\documentclass[12pt]{article}
\pdfoutput=1
\usepackage{pdflscape}
\usepackage{booktabs}
\usepackage{tabulary}
\usepackage{multirow}
\usepackage{rotating}
\usepackage{epsfig}
\usepackage{psfrag}
\usepackage{latexsym}
\usepackage{indentfirst}
\usepackage{fancyhdr}
\usepackage{dsfont}
\usepackage{adjustbox}
\usepackage{amsmath}
\usepackage{amssymb}
\usepackage{amsfonts}
\usepackage{mathrsfs}
\usepackage{amsthm}
\usepackage{pifont}
\usepackage{dsfont}
\usepackage{multirow}
\usepackage{array}
\usepackage{chngpage}
\usepackage{longtable}
\usepackage{cite}
\usepackage{bbold}
\usepackage{color}
\usepackage{braket}
\usepackage{colordvi}
\usepackage{fancybox}
\usepackage[footnotesize]{caption2}
\usepackage{graphicx}
\usepackage[center,footnotesize,hang]{subfigure}
\usepackage{bm}
\usepackage{bbm}
\usepackage{url}
\usepackage[table]{xcolor}
\usepackage[colorlinks, linkcolor=red, anchorcolor=black, citecolor=green]{hyperref}
\usepackage{multirow}
\usepackage{harpoon}
\usepackage{textcomp}  
\usepackage{enumitem}
\usepackage{mathrsfs}  
\usepackage[normalem]{ulem}

\usepackage{tikz}
\usepackage{diagbox}
\usepackage{enumitem}

\newcommand{\PreserveBackslash}[1]{\let\E^c=\\#1\let\\=\E^c}
\newcolumntype{C}[1]{>{\PreserveBackslash\centering}p{#1}}
\newcolumntype{R}[1]{>{\PreserveBackslash\raggedleft}p{#1}}
\newcolumntype{L}[1]{>{\PreserveBackslash\raggedright}p{#1}}
\allowdisplaybreaks  

\newcommand{\bq}{\begin{eqnarray}}
\newcommand{\nq}{\end{eqnarray}}

\newcommand{\ignore}[1]{}

\numberwithin{equation}{section}

\begin{document}
\title{
\begin{flushright}
\hfill\mbox{\small\tt SISSA 17/2024/FISI}  \\
\begin{minipage}{0.2\linewidth}
\normalsize
\end{minipage}
\end{flushright}
{\Large \bf
Non-holomorphic Modular $S_4$ Lepton Flavour Models
\\[2mm]}
\date{}
\author{
Gui-Jun~Ding$^{a}$\footnote{E-mail: {\tt
dinggj@ustc.edu.cn}},  \
Jun-Nan Lu$^{a}$\footnote{E-mail: {\tt
junnanlu@ustc.edu.cn}},  \
S.~T.~Petcov$^{\,b,c,}$\footnote{Also at Institute of Nuclear Research and Nuclear Energy, Bulgarian Academy of Sciences, 1784 Sofia, Bulgaria.},
Bu-Yao Qu$^{a}$\footnote{E-mail: {\tt
qubuyao@mail.ustc.edu.cn}},  \
\\*[20pt]
\centerline{
\begin{minipage}{\linewidth}
\begin{center}
$^a${\it \small Department of Modern Physics, University of Science and Technology of China,\\
Hefei, Anhui 230026, China}\\[2mm]
$^{b}$\,{\it \small SISSA/INFN, Via Bonomea 265, 34136 Trieste, Italy} \\
\vspace{2mm}
$^{c}$\,{\it\small Kavli IPMU (WPI), University of Tokyo,
5-1-5 Kashiwanoha, 277-8583 Kashiwa, Japan} \\
\vspace{2mm}
\end{center}
\end{minipage}}
\\[10mm]}}

\maketitle
\thispagestyle{empty}

\begin{abstract}

In the formalism of the non-supersymmetric modular invariance approach to the flavour problem the elements of the Yukawa coupling and fermion mass matrices are expressed in terms of polyharmonic Maa{\ss} modular forms of level $N$ in addition to the standard modula forms of the same level and a small number of constant parameters. Non-trivial polyharmonic Maa{\ss} forms exist for zero, negative and positive integer modular weights. Employing the finite modula group $S_4$ as a flavour symmetry group and assuming that the three left-handed lepton doublets furnish a triplet irreducible representation of $S_4$, we construct all possible 7- and 8-parameter lepton flavour models in which the neutrino masses are generated either by the Weinberg effective operator or by the type I seesaw mechanism. We identify the phenomenologically viable models and obtain predictions for each of these models for the neutrino mass ordering, the absolute neutrino mass scale, the Dirac and Majorana CP-violation phases and, correspondingly, for the sum of neutrino masses and the neutrinoless double beta decay effective Majorana mass. We comment on how these models can be tested and conclude that they are all falsifiable. Detailed analyses are presented in the case of three representative benchmark lepton flavour scenarios.

\end{abstract}
\thispagestyle{empty}
\vfill

\clearpage

{\hypersetup{linkcolor=black}
\tableofcontents
}

\section{Introduction}

The origin of the flavor structure of quarks and leptons is one of the major challenges in particle physics. The discovery of neutrino
oscillations has brought the dawn for the solution of this puzzle. The tiny
neutrino masses indicate that the origin of neutrino masses may be different
from that of quarks and charged leptons. The atmospheric and solar neutrino
oscillations require two large lepton mixing angles $\theta_{12}$ and
$\theta_{23}$. The reactor mixing angle $\theta_{13}$ is the smallest lepton
mixing angle, and it is of the same order as the quark Cabibbo angle with
$\theta_{13}\sim\theta_C/\sqrt{2}$, where $\theta_C\approx13^{\circ}$ denotes
the Cabibbo angle~\cite{ParticleDataGroup:2024cfk}. A popular approach to
explain the large lepton mixing angles is the  non-Abelian discrete flavour
symmetry~\cite{King:2017guk,Petcov:2017ggy,Feruglio:2019ybq,Xing:2020ijf,Ding:2024ozt}.
There is no exact flavor symmetry at low energy scale, consequently
the non-Abelian discrete flavour symmetry must to be broken. Generally a
large number of scalar fields called flavons as well as auxiliary symmetry is
required and the vacuum expectation values (VEVs) of flavons are the source of flavor symmetry breaking. The alignment of flavon VEVs along specific
directions in flavor space generates the large lepton mixing angles. However,
the dynamics realizing the vacuum alignment of flavon VEVs is quite
sophisticated so that the resulting flavor models are very elaborate.

The modular invariance as flavor symmetry has attracted much attention in the past several years~\cite{Feruglio:2017spp}, see Refs.~\cite{Kobayashi:2023zzc,Ding:2023htn} for reviews. In the paradigm
of modular flavor symmetry, the Yukawa couplings are promoted to dynamical
objects. They are assumed to be modular forms of level $N$, which are
holomorphic functions of a complex scalar field - the  modulus $\tau$,
and they transform as representations of the finite modular groups $\Gamma_N$ or $\Gamma'_N$. The VEV of the modulus $\tau$ is the unique source of modular symmetry breaking in modular models without other flavons, so that there is no need for vacuum alignment anymore, although the VEV of $\tau$ should be
dynamically fixed.

Originally modular symmetry was implemented in the context of supersymmetry which naturally leads to the holomorphicity of modular forms~\cite{Feruglio:2017spp}. Motivated by the modular invariant theory based on automorphic forms~\cite{Ding:2020zxw},  a non-supersymmetric
formulation of the modular flavor symmetry was recently proposed in Ref.~\cite{Qu:2024rns}. The assumption of holomorphicity is superseded by the
harmonic condition, and the modularity condition is preserved. Thus, the Yukawa couplings are polyharmonic Maa{\ss} forms of level $N$,
which can be arranged into multiplets of the finite modular groups $\Gamma_N$ and $\Gamma'_N$~\cite{Qu:2024rns}. The level $N$ polyharmonic Maa{\ss} forms
coincide with the level $N$ holomorphic modular forms at weights $k\geq3$,
however, here exists negative weight polyharmonic Maa{\ss} forms.
At the same time the weights of the standard modular forms must be non-negative. Hence the non-holomorphic modular flavor symmetry
extends the original modular invariance approach due to the presence of
negative weight polyharmonic Maa{\ss} forms, and it provides an interesting
opportunity for constructing models of fermion masses and flavor mixing.
Moreover, this formalism can be consistently combined with the generalized
CP (gCP) symmetry which would reduce, as in the case of supersymmetric modular invarinace approach~\cite{Novichkov:2019sqv}, to the traditional CP symmetry in the basis where both modular generators $S$ and $T$ are represented by unitary and symmetric matrices~\cite{Qu:2024rns}.
The CP transformation of the complex modulus is $\tau \xrightarrow{\mathcal{CP}} -\tau^*$ (see, e.g., \cite{Novichkov:2019sqv,Baur:2019kwi}) up to modular transformations.

Several models for lepton masses and mixing with polyharmonic Maa{\ss} forms based on the finite modular group $\Gamma_3\cong A_4$ have been constructed~\cite{Qu:2024rns,Nomura:2024atp}. In the present work we investigate the non-holomorphic lepton flavor models with $\Gamma_4\cong S_4$ modular symmetry in a systematic way, and study the phenomenological predictions of the models in detail. We focus on the most economical modular invariant models in which no flavon fields are introduced. Both scenarios with gCP and without gCP symmetry are considered. The light neutrinos are assumed to be Majorana particles, and their masses are generated
either via the Weinberg operator or the type I seesaw mechanism. We are aiming at constructing phenomenologically viable models with the smallest number of free parameters. We find that the minimal viable models depend on 7 (8) real parameters including real and imaginary part of $\tau$, if the
gCP symmetry is (isn't) incorporated. The modular $S_4$ symmetry models with holomorphic modular forms have been widely studied~\cite{Penedo:2018nmg,Novichkov:2018ovf,deMedeirosVarzielas:2019cyj,King:2019vhv,
Criado:2019tzk,Ding:2019gof,Wang:2019ovr,Zhao:2021jxg,King:2021fhl,Ding:2021zbg,Qu:2021jdy,Nomura:2021ewm,deMedeirosVarzielas:2023ujt}.
It was found that the minimal phenomenologically viable lepton models involve 7 real parameters as well~\cite{Novichkov:2018ovf,Qu:2021jdy}. The present work extends the previous study of supersymmetric and holomorphic $S_4$ modular models by considering the non-holomorphic polyharmonic Maa{\ss} forms of level 4.

The layout of the remainder of the paper is as follows. In section~\ref{sec:framework}, we briefly review the modular group, polyharmonic Maa{\ss} forms, and the formalism of non-holomorphic modular flavor symmetry. In section~\ref{sec:general-analysis}, we perform a thorough analysis of the possible forms of the charged lepton and neutrino mass terms that are invariant under the $S_4$ modular symmetry. In this section, the corresponding mass matrices of charged leptons and neutrinos are also presented. The method of numerical analysis is outlined in section~\ref{sec:numerical}. We present three example models in section~\ref{sec:benchmark-models}: one with neutrino masses generated by the Weinberg effective operator and two - by the type-I seesaw mechanism with two right-handed (RH) neutrinos. For each of the three models, we derive
the best fit values of the model parameters and of the measured observables (the charged lepton masses, the three neutrino mixing angles and the two neutrino mass squared differences) and obtain  predictions for the neutrino mass ordering, the absolute neutrino mass scale, the Dirac and two Majorana CP-violation (CPV) phases, and, correspondingly, for the sum of neutrino masses and the neutrinoless double beta decay effective Majorana mass.
We draw our conclusion in section~\ref{sec:conclusion}. The group theory of $\Gamma_4\cong S_4$ and the polyharmonic Maa{\ss} forms of level $N=4$ are given in the Appendix~\ref{sec:S4_group}. We give a detailed explanation of the counting of the number of effective parameters in the light neutrino mass
matrix $M_{\nu}$ in Appendix~\ref{sec:effective_parameters}, when the two
right-handed neutrinos of the minimal seesaw model are in singlet
representations of $S_4$. In Appendix~\ref{sec:app_viable_models}, we list  in tables the predictions for the best fit values of the lepton mass and mixing parameters of all phenomenologically viable non-holomorphic $S_4$ modular lepton flavour models with smallest number of free parameters  (seven and eight).

\section{\label{sec:framework}The framework}

The inhomogeneous modular group $\overline{\Gamma}$
is the group of linear fractional transformations acting on the complex modulus $\tau$ in upper-half complex plane as follow:
\begin{equation}
\tau\xrightarrow{\gamma}\gamma\tau=\frac{a\tau+b}{c\tau+d},~~~\text{Im}\tau>0\,,
\end{equation}
where $a$, $b$, $c$, and $d$ are integers satisfying $ad-bc=1$.
Clearly, $\gamma$ and $-\gamma$ give rise to the same action on $\tau$,
therefore $\overline{\Gamma}$ is isomorphic to the projective
special linear group $PSL(2, \mathbb{Z})=SL(2, \mathbb{Z})/\{\pm\mathbb{1}\}$,
where $SL(2, \mathbb{Z})$ is special linear group of $2\times 2$ matrices
with integer elements and unit determinant. The modular group
$\overline{\Gamma}$ is a discrete, infinite and non-compact group, and it can be generated by two elements,
\begin{eqnarray}
\nonumber && S=\begin{pmatrix}
0 ~&~1 \\
-1 ~&~ 0
\end{pmatrix},\quad S\tau=-\frac{1}{\tau}\,,\\
&&T=\begin{pmatrix}
1 ~&~1 \\
0 ~&~ 1
\end{pmatrix},\quad T\tau=\tau+1\,,
\end{eqnarray}
which obey the following relations
\begin{equation}
S^2=(ST)^3=1\,.
\end{equation}
The $SL(2, \mathbb{Z})$ group has a series of infinite normal subgroups
$\Gamma(N)$ with $N=1, 2, \ldots$,
\begin{equation}
\Gamma(N)=\left\{\begin{pmatrix}
a ~&~ b\\
c ~&~ d
\end{pmatrix}\in SL(2, \mathbb{Z})\Big|,~a=d=1~(\text{mod}~N),~b=c=0~(\text{mod}~N)\right\}\,,
\end{equation}
which is the so-called principal congruence subgroup of level $N$.
Note that $T^N$ is an element of $\Gamma(N)$. One can define the
projective principal congruence subgroup
$\overline{\Gamma}(N)=\Gamma(N)/\{\pm\mathbb{1}\}$ for $N=1, 2$,
while $\overline{\Gamma}(N)=\Gamma(N)$ for
$N\geq3$ since $-\mathbb{1}$ does not belong to $\Gamma(N)$.
Taking the quotient $\Gamma_N=\overline{\Gamma}/\overline{\Gamma}(N)$,
one obtains the inhomogeneous finite modular group of level $N$.
Thus, $\Gamma_N$ can be generated by $S$ and $T$ satisfying the
multiplication rule,
\begin{equation}
S^2=(ST)^3=T^N=1,~~~\text{for}~~N\leq5\,.
\end{equation}
It is remarkable that $\Gamma_N$ is isomorphic to the permutation groups,
i.e., $\Gamma_2\cong S_3$,  $\Gamma_3\cong A_4$,  $\Gamma_4\cong S_4$ and
$\Gamma_5\cong A_5$. Additional relations are necessary to render the group
$\Gamma_N$ finite for
 $N\geq6$~\cite{deAdelhartToorop:2011re,Li:2021buv,Ding:2020msi}.
We will be interested in finite modular group $\Gamma_4\cong S_4$
in this work.

Polyharmonic Maa{\ss} forms of weight $k$ and level $N$ are functions $Y(\tau)$ satisfying the following conditions~\cite{Ding:2020zxw,Qu:2024rns}:
\begin{eqnarray}
\nonumber&& Y(\gamma \tau) = \left(c\tau+d\right)^kY(\tau),~~~\gamma\in \Gamma(N)\,, \\
\label{eq:Masss-def}&&\left[-4y^2\dfrac{\partial}{\partial \tau} \dfrac{\partial}{\partial \bar{\tau}} +2iky\dfrac{\partial}{\partial \bar{\tau}}\right] Y(\tau)=0\,,
\end{eqnarray}
where $\tau=x+iy$, and the modular weight $k$ is a generic integer
that can be positive, zero or negative.
The polyharmonic Maa{\ss} forms are implemented with the moderate
growth condition:
$Y(\tau)=\mathcal{O}(y^\alpha)$~~as $y\rightarrow +\infty$ for some $\alpha$.
From $Y(\tau+N)=Y(\tau)$ and the second condition of Eq.~\eqref{eq:Masss-def},
the Fourier expansion of $Y(\tau)$ is determined to be~\cite{Qu:2024rns},
\begin{eqnarray}
\label{eq:poly-maass-form}Y(\tau)=\sum_{\substack{n\in\frac{1}{N}\mathbb{Z} \\ n\geq0}} c^+(n)q^n + c^-(0)y^{1-k}+ \sum_{\substack{n\in\frac{1}{N}\mathbb{Z} \\ n<0}} c^-(n)\Gamma(1-k,-4\pi n y)q^n\,,~~~q\equiv e^{2\pi i\tau} \,,
\end{eqnarray}
where the term $y^{1-k}$ would be $\ln y$ for $k=1$, and $\Gamma(s,x)$ is
the incomplete gamma function defined by Eq.~\eqref{eq:incomplete-gamma}.
All the weight $k$ polyharmonic Maa{\ss} forms of level $N$ span a
linear space of finite dimension. There exists a basis in such a linear space
so that the transformation of a multiplet of polyharmonic Maa{\ss}
forms $Y^{(k)}_{\bm{r}}(\tau)=\left(Y_1(\tau), Y_2(\tau), \ldots\right)^T$
is described by an irreducible representation $\rho_{\bm{r}}$ of the finite
modular group $\Gamma_N$ for even weight $k$~\cite{Qu:2024rns}, i.e.
\begin{equation}
\label{eq:Maass-trans}Y^{(k)}_{\bm{r}}(\gamma\tau)=(c\tau+d)^{k}\rho_{\bm{r}}(\gamma)Y^{(k)}_{\bm{r}}(\tau),~~~\gamma\in\overline{\Gamma}\,.
\end{equation}
where $\gamma$ is a representative element of $\Gamma_N$.
The weight $k$ polyharmonic Maa{\ss} forms can be lifted from the
known modular forms of weight $2-k$~\cite{Qu:2024rns}.
Although the level $N$ polyharmonic Maa{\ss} forms coincide with the level
$N$ holomorphic modular forms at weights $k>2$, there exist non-holomorphic
polyharmonic Maa{\ss} forms at weight $k\leq2$. We are focus on
the level $N=4$ in the present work. The dimension of the linear space of
polyharmonic Maa{\ss} forms of weight $k$ and level 4 is equal to 6
for $k\leq 2$ and $2k+1$ for $k>2$. The expressions of the polyharmonic
Maa{\ss} form multiplets of level 4 are given in Appendix~\ref{sec:S4_group}.

We briefly review the formalism of non-holomorphic modular
flavor symmetry~\cite{Qu:2024rns}. The Lagrangian is required to be invariant
under the modular symmetry and the standard model gauge symmetry
$SU(3)_C\times SU(2)_L\times U(1)_Y$, while supersymmetry is unnecessary.
We are mainly interested in the Yukawa interactions, and we adopt the
two-component spinor notation for fermion fields. The two spinor multiplets in the Yukawa interactions are denoted by $\psi$ and $\psi^c$,
their transformations law is similar to
that of polyharmonic Maa{\ss} forms shown in Eq.~\eqref{eq:Maass-trans},
\begin{eqnarray}
\nonumber&&\psi(x)\rightarrow (c\tau+d)^{-k_{\psi}}\rho_{\psi}(\gamma) \psi(x)\,,\\
&&\psi^c(x)\rightarrow (c\tau+d)^{-k_{\psi^c}}\rho_{\psi^c}(\gamma)\psi^c(x)\,,~~~\gamma=\begin{pmatrix}
a ~&~ b \\
c ~&~ d
\end{pmatrix}\in\overline{\Gamma}\,,
\end{eqnarray}
where $\rho_{\psi}$ and $\rho_{\psi^c}$ are unitary representations of
$\Gamma_N$,  $k_{\psi}$ and $k_{\psi^c}$ are integers. Analogously the modular
transformation of the Higgs field $H(x)$ is given by
\begin{equation}
H(x)\rightarrow (c\tau+d)^{-k_{H}}\rho_{H}(\gamma)H(x)\,,
\end{equation}
where $\rho_{H}$ is a one-dimensional representation of $\Gamma_N$, and
$k_{H}$ is an integer. Then the modular invariant Yukawa interaction
can be written as
\begin{eqnarray}
\label{eq:Yukawa-gener}\mathcal{L}^Y=Y^{(k_Y)}(\tau) \psi^c\psi H+\mathrm{h.c.}\,,
\end{eqnarray}
where the gauge indices are dropped for notational simplicity,
and $Y^{(k_Y)}(\tau)$ is a polyharmonic Maa{\ss} form multiplet of weight
$k_Y$ and level $N$, transforming in the representation
$\rho_{Y}$ of $\Gamma_N$:
\begin{equation}
Y^{(k_Y)}(\gamma\tau)=(c\tau+d)^{k}\rho_{Y}(\gamma)Y^{(k_Y)}(\tau)\,.
\end{equation}
The modular invariance of $\mathcal{L}^Y$ requires the following weight
and representation balance conditions:
\begin{equation}
k_Y=k_{\psi^c}+k_{\psi}+k_H\,,\qquad \rho_{Y}\otimes\rho_{\psi^c}\otimes\rho_{\psi}\otimes\rho_H
\ni\mathbf{1}\,,
\end{equation}
where $\mathbf{1}$ refers to the invariant singlet of $\Gamma_N$.

The non-holomorphic modular flavor symmetry can be extended to include the generalized CP symmetry~\cite{Qu:2024rns}. The action of gCP on a field multiplet $\varphi$ in a representation $\rho_{\bm{r}}$ of $\Gamma_N$ is given by,
\begin{equation}
\varphi\xrightarrow{\mathcal{CP}} X_{\bm{r}}\varphi^{*}\,,
\end{equation}
where the gCP transformation $X_{\bm{r}}$ is a matrix satisfying the
consistency conditions
\begin{equation}
\label{eq:consistency-condition}
X_{\bm{r}}\rho^{*}_{\bm{r}}(S)X^{-1}_{\bm{r}}=\rho^{-1}_{\bm{r}}(S),\qquad X_{\bm{r}}\rho^{*}_{\bm{r}}(T)X^{-1}_{\bm{r}}=\rho^{-1}_{\bm{r}}(T)\,.
\end{equation}
The CP transformation of modulus and polyharmonic Maa{\ss} forms
are determined to be~\cite{Qu:2024rns}
\begin{equation}
\tau \xrightarrow{\mathcal{CP}} -\tau^*\,,\qquad Y^{(k)}_{\bm{r}}(\tau)\stackrel{\mathcal{CP}}{\longrightarrow} Y^{(k)}_{\bm{r}}(-\tau^{*})=X_{\mathbf{r}}Y^{(k)*}_{\bm{r}}(\tau)\,.
\end{equation}
In the basis where the representation matrices $\rho_{\bm{r}}(S)$
and $\rho_{\bm{r}}(T)$ are unitary and symmetric, the consistency condition of Eq.~\eqref{eq:consistency-condition} would be satisfied by
$X_{\mathbf{r}}=\mathbb{1}$. This is exactly the case for our working basis
of $\Gamma_4\cong S_4$ given in Eq.~\eqref{eq:rep-basis-S4}.
As a consequence, the gCP symmetry could enforce all coupling constants
in the modular invariant Lagrangian to be real, and all CP violations would arise
from the vacuum expectation value of $\tau$. These results coincide, apart from the setting, with those obtained in flavour theories based
on the standard (holomorphic) modular invarinace involving
supersymmetry \cite{Novichkov:2019sqv}.

\section{General analysis of model building\label{sec:general-analysis}}

In the present work, we assume that the Higgs field $H$ transforms as a trivial singlet $\mathbf{1}$ of $S_{4}$ with modular weight $k_{H}=0$, and that the neutrinos are Majorana particles. In the following, we shall perform a
general analysis of the possible forms of the charged lepton and neutrino
Yukawa couplings that are invariant under the $S_4$ modular symmetry, and we will present the corresponding charged lepton and neutrino  mass matrices.

\subsection{Charged lepton sector\label{sec:charged_lepton_sector}}

In this section, we investigate the modular invariant Lagrangian of the
charged lepton Yukawa interactions. We assume that the three generations of
lepton $SU(2)_L$ doublets transform as a triplet of $S_{4}$, while the
right-handed (RH) charged leptons transform
as singlets of $S_{4}$:
\begin{equation} \label{eq:assignments}
  L\equiv \begin{pmatrix} L_{1} \\ L_{2} \\ L_{3} \end{pmatrix}\sim \mathbf{3}^{i}\,,\quad E^{c}_{\alpha}\sim \mathbf{1}^{j_{\alpha}}\,,~~\text{with}~~\alpha =1,2,3\,,
\end{equation}
where $i,j_{1,2,3}=0,1$ with $\mathbf{1}\equiv \mathbf{1}^{0}$, $\mathbf{1}'\equiv \mathbf{1}^{1}$, $\mathbf{3}\equiv \mathbf{3}^{0}$ and $\mathbf{3}'\equiv \mathbf{3}^{1}$ for singlet and triplet representations.
$E^{c}_{\alpha}$ stands for $e^{c},\mu^{c},\tau^{c}$ for $\alpha=1,2,3$,
respectively. The exchange of assignments for the three
RH  charged leptons amounts to multiplying the charged-lepton mass matrix on
the right side by permutation matrices. This does not change the lepton mixing and the charged lepton masses.

With these assumptions, we can write down the most general
charged lepton Yukawa interactions as follows:
\begin{equation}
\mathcal{L}^{Y}_\ell=\left[\alpha_{1}(E^c_1LY_{\mathbf{3}^{[j_1+i]}}^{(k_{E^{c}_{1}+k_{L}})})_{\bf{1}}+\alpha_{2}(E^c_2LY_{\mathbf{3}^{[j_2+i]}}^{(k_{E^{c}_{2}+k_{L}})})_{\bf{1}}+\alpha_{3}(E^c_3LY_{\mathbf{3}^{[j_3+i]}}^{(k_{E^{c}_{3}+k_{L}})})_{\bf{1}}\right]H^*
+\text{h.c.}\,,
\end{equation}
where the notation $[j_{\alpha}+i]$ equals to $j_{\alpha}+i$ modulo 2, and $k_{E^{c}_{1,2,3}}$ and $k_{L}$ are the modular weights of $E^{c}_{1,2,3}$ and $L$ respectively. We find that there are three real Yukawa coupling parameters
$\alpha_{1,2,3}$ whose phases can be absorbed by rephasing the
RH charged lepton fields. In the following, we use $\rho_{\psi}$ to represent the representation of field $\psi$ under $S_{4}$.
Considering the possible representation
assignments of $L$ and $E_{\alpha}^{c}$, we obtain the following four
possible forms of $\mathcal{L}^{Y}_{\ell}$.

\begin{itemize}

\item{$(\rho_{L},\rho_{E^{c}_1},\rho_{E^{c}_2},\rho_{E^{c}_3})=(\bf{3},\bf{1},\bf{1},\bf{1})~\text{or}~(\bf{3}',\bf{1}',\bf{1}',\bf{1}')$}

The modular invariant Yukawa interaction is
\begin{eqnarray}
\label{eq:Le_1}
\mathcal{L}_{\ell}^{Y}=\left[\alpha_{1}(E^c_1LY_{\mathbf{3}}^{(k_{E^{c}_{1}}+k_{L})})_{\bf{1}}+\alpha_{2}(E^c_2LY_{\mathbf{3}}^{(k_{E^{c}_{2}}+k_{L})})_{\bf{1}}+\alpha_{3}(E^c_3LY_{\mathbf{3}}^{(k_{E^{c}_{3}}+k_{F})})_{\bf{1}}\right]H^*+\text{h.c.}\,.
\end{eqnarray}
The resulting charged lepton mass matrix reads:
\begin{eqnarray}
M_{\ell}=\left(\begin{matrix}
\alpha_{1} Y_{\mathbf{3},1}^{(k_{E^{c}_{1}}+k_{L})}~&~\alpha_{1} Y_{\mathbf{3},3}^{(k_{E^{c}_{1}}+k_{L})}~&~\alpha_{1} Y_{\mathbf{3},2}^{(k_{E^{c}_{1}}+k_{L})}\\
\alpha_{2} Y_{\mathbf{3},1}^{(k_{E^{c}_{2}}+k_{L})}~&~\alpha_{2} Y_{\mathbf{3},3}^{(k_{E^{c}_{2}}+k_{L})}~&~\alpha_{2} Y_{\mathbf{3},2}^{(k_{E^{c}_{2}}+k_{L})}\\
\alpha_{3} Y_{\mathbf{3},1}^{(k_{E^{c}_{3}}+k_{L})}~&~\alpha_{3} Y_{\mathbf{3},3}^{(k_{E^{c}_{3}}+k_{L})}~&~\alpha_{3} Y_{\mathbf{3},2}^{(k_{E^{c}_{3}}+k_{L})}
\end{matrix}\right)v
\label{eq:Me_1}\,,
\end{eqnarray}
where $v=\langle H\rangle$ denotes the VEV of Higgs field.

\item{$(\rho_{L},\rho_{E^{c}_1},\rho_{E^{c}_2},\rho_{E^{c}_3})=(\mathbf{3},\mathbf{1}, \mathbf{1}, \mathbf{1}')~\text{or}~(\mathbf{3}',\mathbf{1}', \mathbf{1}', \mathbf{1})$}

The charged lepton Yukawa coupling
is given by:
\begin{eqnarray}
\label{eq:Le_2}
\mathcal{L}_{\ell}^{Y}=\left[\alpha_{1}(E^c_1LY_{\mathbf{3}}^{(k_{E^{c}_{1}}+k_{L})})_{\bf{1}}+\alpha_{2}(E^c_2LY_{\mathbf{3}}^{(k_{E^{c}_{2}}+k_{L})})_{\bf{1}}+\alpha_{3}(E^c_3LY_{\mathbf{3}'}^{(k_{E^{c}_{3}}+k_{F})})_{\bf{1}}\right]H^*+\text{h.c.}\,,
\end{eqnarray}
which leads to
\begin{eqnarray}
M_{\ell}=\left(\begin{matrix}
\alpha_{1} Y_{\mathbf{3},1}^{(k_{E^{c}_{1}}+k_{L})}~&~\alpha_{1} Y_{\mathbf{3},3}^{(k_{E^{c}_{1}}+k_{L})}~&~\alpha_{1} Y_{\mathbf{3},2}^{(k_{E^{c}_{1}}+k_{L})}\\
\alpha_{2} Y_{\mathbf{3},1}^{(k_{E^{c}_{2}}+k_{L})}~&~\alpha_{2} Y_{\mathbf{3},3}^{(k_{E^{c}_{2}}+k_{L})}~&~\alpha_{2} Y_{\mathbf{3},2}^{(k_{E^{c}_{2}}+k_{L})}\\
\alpha_{3} Y_{\mathbf{3}',1}^{(k_{E^{c}_{3}}+k_{L})}~&~\alpha_{3} Y_{\mathbf{3}',3}^{(k_{E^{c}_{3}}+k_{L})}~&~\alpha_{3} Y_{\mathbf{3}',2}^{(k_{E^{c}_{3}}+k_{L})}
\end{matrix}\right)v
\label{eq:Me_2}\,.
\end{eqnarray}

\item{$(\rho_{L},\rho_{E^{c}_1},\rho_{E^{c}_2},\rho_{E^{c}_3})=(\bf{3},\bf{1},\bf{1}',\bf{1}')~\text{or}~(\bf{3}',\bf{1}',\bf{1},\bf{1})$}

The Yukawa interaction takes the following form:
\begin{eqnarray}
\label{eq:Le_3}
\mathcal{L}_{\ell}^{Y}=\left[\alpha_{1}(E^c_1LY_{\mathbf{3}}^{(k_{E^{c}_{1}}+k_{L})})_{\bf{1}}+\alpha_{2}(E^c_2LY_{\mathbf{3}'}^{(k_{E^{c}_{2}}+k_{L})})_{\bf{1}}+\alpha_{3}(E^c_3LY_{\mathbf{3}'}^{(k_{E^{c}_{3}}+k_{F})})_{\bf{1}}\right]H^*+\text{h.c.}\,,
\end{eqnarray}
which implies
\begin{eqnarray}
M_{\ell}=\left(\begin{matrix}
\alpha_{1} Y_{\mathbf{3},1}^{(k_{E^{c}_{1}}+k_{L})}~&~\alpha_{1} Y_{\mathbf{3},3}^{(k_{E^{c}_{1}}+k_{L})}~&~\alpha_{1} Y_{\mathbf{3},2}^{(k_{E^{c}_{1}}+k_{L})}\\
\alpha_{2} Y_{\mathbf{3}',1}^{(k_{E^{c}_{2}}+k_{L})}~&~\alpha_{2} Y_{\mathbf{3}',3}^{(k_{E^{c}_{2}}+k_{L})}~&~\alpha_{2} Y_{\mathbf{3}',2}^{(k_{E^{c}_{2}}+k_{L})}\\
\alpha_{3} Y_{\mathbf{3}',1}^{(k_{E^{c}_{3}}+k_{L})}~&~\alpha_{3} Y_{\mathbf{3}',3}^{(k_{E^{c}_{3}}+k_{L})}~&~\alpha_{3} Y_{\mathbf{3}',2}^{(k_{E^{c}_{3}}+k_{L})}
\end{matrix}\right)v
\label{eq:Me_3}\,.
\end{eqnarray}

\item{$(\rho_{L},\rho_{E^{c}_1},\rho_{E^{c}_2},\rho_{E^{c}_3})=(\bf{3},\bf{1}',\bf{1}',\bf{1}')~\text{or}~(\bf{3}',\bf{1},\bf{1},\bf{1})$}

The charged lepton Yukawa coupling invariant under $S_4$ modular symmetry
is given by:
\begin{eqnarray}
\label{eq:Le_4}
\mathcal{L}_{\ell}^{Y}=\left[\alpha_{1}(E^c_1LY_{\mathbf{3}'}^{(k_{E^{c}_{1}}+k_{L})})_{\bf{1}}+\alpha_{2}(E^c_2LY_{\mathbf{3}'}^{(k_{E^{c}_{2}}+k_{L})})_{\bf{1}}+\alpha_{3}(E^c_3LY_{\mathbf{3}'}^{(k_{E^{c}_{3}}+k_{F})})_{\bf{1}}\right]H^*+\text{h.c.}\,,~~~~~
\end{eqnarray}
from which we can read out
\begin{eqnarray}
M_{\ell}=\left(\begin{matrix}
\alpha_{1} Y_{\mathbf{3}',1}^{(k_{E^{c}_{1}}+k_{L})}~&~\alpha_{1} Y_{\mathbf{3}',3}^{(k_{E^{c}_{1}}+k_{L})}~&~\alpha_{1} Y_{\mathbf{3}',2}^{(k_{E^{c}_{1}}+k_{L})}\\
\alpha_{2} Y_{\mathbf{3}',1}^{(k_{E^{c}_{2}}+k_{L})}~&~\alpha_{2} Y_{\mathbf{3}',3}^{(k_{E^{c}_{2}}+k_{L})}~&~\alpha_{2} Y_{\mathbf{3}',2}^{(k_{E^{c}_{2}}+k_{L})}\\
\alpha_{3} Y_{\mathbf{3}',1}^{(k_{E^{c}_{3}}+k_{L})}~&~\alpha_{3} Y_{\mathbf{3}',3}^{(k_{E^{c}_{3}}+k_{L})}~&~\alpha_{3} Y_{\mathbf{3}',2}^{(k_{E^{c}_{3}}+k_{L})}
\end{matrix}\right)v\label{eq:Me_4}\,.
\end{eqnarray}

\end{itemize}

The possible representation assignments to the $L$ and $E^{c}_{\alpha}$, along with the Yukawa couplings $\mathcal{L}^{Y}_{\ell}$ and charged lepton mass matrix are summarized in Table~\ref{Tab:L_e}. In this work, we focus on the modular forms $Y^{(k)}_{\mathbf{r}}$ with weights $-4\leq k \leq 4$. The higher weight modular forms generally lead to more free parameters which can weaken the predictive power of models. From Table~\ref{Tab:LeveL4_MM} we see that there is always a modular form multiplet transforming as $\mathbf{3}$ of $S_{4}$ at these weights. However, the modular forms in the triplet representation $\mathbf{3}'$ begin to appear at weight $k_{Y}=4$. As a result, one row of the charged lepton mass matrix
would vanish if the Yukawa couplings are modular form $Y^{(k_Y)}_{\mathbf{3}'}$ with $k_Y<4$. We require the charged-lepton mass matrix to have rank three in order to accommodate the three nonzero masses of the electron, muon and tau.
In what follows, we will give possible values of the modular weights of
lepton fields for each representation assignment of $L$ and $E^{c}_{\alpha}$.

\begin{table}[t!]
\centering
\begin{tabular}{|c|c|c|c|c|c|c|c|}
\hline  \hline
 $\mathcal{L}^{Y}_{\ell}$ & $M_{\ell}$ & $\rho_{L}$ & $\rho_{E^{c}_1}$ & $\rho_{E^{c}_2}$ & $\rho_{E^{c}_3}$ & Constraints & rank($M_{\ell}$)\\ \hline
 \multirow{2}{*}{eq.\eqref{eq:Le_1}} & \multirow{2}{*}{eq.\eqref{eq:Me_1}} & $\mathbf{3}$ & $\mathbf{1}$ & $\mathbf{1}$ & $\mathbf{1}$ & \multirow{2}{*}{$k_{E^{c}_{1}}\neq k_{E^{c}_{2}} \neq k_{E^{c}_{3}}$} & \multirow{2}{*}{3} \\
 & & $\mathbf{3}'$ & $\mathbf{1}'$ & $\mathbf{1}'$ & $\mathbf{1}'$ & &\\ \hline
 \multirow{2}{*}{eq.\eqref{eq:Le_2}} & \multirow{2}{*}{eq.\eqref{eq:Me_2}} & $\mathbf{3}$ & $\mathbf{1}$ & $\mathbf{1}$ & $\mathbf{1}'$ & \multirow{2}{*}{$k_{E^{c}_{1}}\neq k_{E^{c}_{2}}$} & \multirow{2}{*}{3}\\
 & & $\mathbf{3}'$ & $\mathbf{1}'$ & $\mathbf{1}'$ & $\mathbf{1}$ & & \\ \hline
 \multirow{2}{*}{eq.\eqref{eq:Le_3}} & \multirow{2}{*}{eq.\eqref{eq:Me_3}} & $\mathbf{3}$ & $\mathbf{1}$ & $\mathbf{1}'$ & $\mathbf{1}'$ & \multirow{2}{*}{$k_{E^{c}_{2}} \neq k_{E^{c}_{3}}$} & \multirow{2}{*}{2}\\
 && $\mathbf{3}'$ & $\mathbf{1}'$ & $\mathbf{1}$ & $\mathbf{1}$ & & \\ \hline
 \multirow{2}{*}{eq.\eqref{eq:Le_4}} & \multirow{2}{*}{eq.\eqref{eq:Me_4}} & $\mathbf{3}$ & $\mathbf{1}'$ & $\mathbf{1}'$ & $\mathbf{1}'$& \multirow{2}{*}{$k_{E^{c}_{1}}\neq k_{E^{c}_{2}} \neq k_{E^{c}_{3}}$} & \multirow{2}{*}{1} \\
 && $\mathbf{3}'$ & $\mathbf{1}$ & $\mathbf{1}$ & $\mathbf{1}$ & &\\
  \hline \hline
\end{tabular}
\caption{
\label{Tab:L_e}
The possible assignments of irreps for $L$ and $E^{c}_{\alpha}$ with $\alpha=1,2,3$.}
\end{table}

\begin{itemize}

\item{$(\rho_{L},\rho_{E^{c}_1},\rho_{E^{c}_2},\rho_{E^{c}_3})=(\bf{3},\bf{1},\bf{1},\bf{1})~\text{or}~(\bf{3}',\bf{1}',\bf{1}',\bf{1}')$}

The three generations of the RH charged lepton fields transform as the same singlet representation of $S_{4}$, but they are distinguished by different modular weights. The charged lepton mass matrix of rank 3 can be obtained for the following values of the modular weights:
\begin{equation}
    k_{E^{c}_{1}}\neq k_{E^{c}_{2}} \neq k_{E^{c}_{3}}\,,\quad k_{E^{c}_{1,2,3}}+k_{L}\in \{-4,-2,0,2,4\}\,.
\end{equation}
In this case, there are $C_{5}^{3}=10$ allowed combinations of
$(k_{E^{c}_{1}}+k_{L},k_{E^{c}_{2}}+k_{L},k_{E^{c}_{3}}+k_{L})$,
  \begin{eqnarray}\nonumber
    &&(-4,-2,0)\,,(-4,-2,2)\,,(-4,-2,4)\,,(-4,0,2)\,,(-4,0,4)\,,\\
    &&(-4,2,4)\,,(-2,0,2)\,,(-2,0,4)\,,(-2,2,4)\,,(0,2,4)\,.
  \end{eqnarray}

\item{$(\rho_{L},\rho_{E^{c}_1},\rho_{E^{c}_2},\rho_{E^{c}_3})=(\bf{3},\bf{1},\bf{1},\bf{1}')~\text{or}~(\bf{3}',\bf{1}',\bf{1}',\bf{1})$}

In the considered case, the first two RH charged lepton fields $E^{c}_1$ and $E^{c}_2$ should carry two distinct modular weights to avoid that two rows of the charged lepton mass matrix are proportional. The modular multiplet that couple to $L$ and $E^{c}_{3}$ should transform as $\mathbf{3}'$ of $S_{4}$. The modular weights of the charged leptons satisfy:
\begin{equation}
    k_{E^{c}_{1}}\neq k_{E^{c}_{2}}\,,\quad k_{E^{c}_{1,2}}+k_{L}\in \{-4,-2,0,2,4\}\,,\quad k_{E^{c}_3}+k_{L}=4\,.
\end{equation}
As a result, we  obtain $C_{5}^{2}=10$ allowed combinations of
$(k_{E^{c}_{1}}+k_{L},k_{E^{c}_{2}}+k_{L},k_{E^{c}_{3}}+k_{L})$:
  \begin{eqnarray}\nonumber
    &&(-4,-2,4)\,,(-4,0,4)\,,(-4,2,4)\,,(-4,4,4)\,,(-2,0,4)\,,\\
    &&(-2,2,4)\,,(-2,4,4)\,,(0,2,4)\,,(0,4,4)\,,(2,4,4)\,.
  \end{eqnarray}

\item{$(\rho_{L},\rho_{E^{c}_1},\rho_{E^{c}_2},\rho_{E^{c}_3})=(\bf{3},\bf{1},\bf{1}',\bf{1}')~\text{or}~(\bf{3}',\bf{1}',\bf{1},\bf{1})$}

To distinguish the second and third RH charged lepton fields $E_{2}^{c}$ and $E_{3}^{c}$, we cannot assign the same modular weights to $E_{2}^{c}$ and $E_{3}^{c}$. However, the second and third rows of $M_{\ell}$ do not vanish only if $k_{E^{c}_{2}}+k_{L}=k_{E^{c}_{3}}+k_{L}=4$. In this case, the maximal possible rank of the charged lepton mass matrix is 2, and we will not consider such assignments further.

\item{$(\rho_{L},\rho_{E^{c}_1},\rho_{E^{c}_2},\rho_{E^{c}_3})=(\bf{3},\bf{1}',\bf{1}',\bf{1}')~\text{or}~(\bf{3}',\bf{1},\bf{1},\bf{1})$}

  For the  relevant  Maa{\ss} form of weight $-4\leq k \leq 4$,
the following relation must hold to avoid
having a  vanishing row of $M_{\ell}$:
\begin{equation}
    k_{E^{c}_{1}}+k_{L}=k_{E^{c}_{2}}+k_{L}=k_{E^{c}_{3}}+k_{L}=4\,.
\end{equation}
As a consequence, the three
RH charged lepton fields $E^{c}_{1,2,3}$
 are un-distinguishable and the rank of $M_{\ell}$ is 1.
\end{itemize}
Requiring that the charged-lepton mass matrix
has rank three, we summarized in Table~\ref{tab:chraged_lepton_model}
all possible charged lepton models
 for which the modular weights of the involved modular forms
satisfy $|k_{E_{\alpha}^{c}}+k_L|\leq 4$.
\begin{table}[hptb!]
\centering
\begin{tabular}{c}
\begin{tabular}{|c|c|c||c|c|c|}  \hline\hline
  Models & $(k_{E_{1}^{c}}',k_{E_{2}^{c}}',k_{E_{3}^{c}}')$ & $(\rho_{L},\rho_{E^{c}_1},\rho_{E^{c}_2},\rho_{E^{c}_3})$ &         Models & $(k_{E_{1}^{c}}',k_{E_{2}^{c}}',k_{E_{3}^{c}}')$ & $(\rho_{L},\rho_{E^{c}_1},\rho_{E^{c}_2},\rho_{E^{c}_3})$ \\
  \hline
  $\mathcal{C}_{1}$ & $(-4,-2,0)$ &  & $\mathcal{C}_{11}$ & $(-4,-2,4)$ &  \\
  \cline{1-2} \cline{4-5}
  $\mathcal{C}_{2}$ & $(-4,-2,2)$ &  & $\mathcal{C}_{12}$ & $(-4,0,4)$ & \\
  \cline{1-2} \cline{4-5}
  $\mathcal{C}_{3}$ & $(-4,-2,4)$ &  & $\mathcal{C}_{13}$ & $(-4,2,4)$ & \\
  \cline{1-2} \cline{4-5}
  $\mathcal{C}_{4}$ & $(-4,0,2)$ &  & $\mathcal{C}_{14}$ & $(-4,4,4)$ & \\
  \cline{1-2} \cline{4-5}
  $\mathcal{C}_{5}$ & $(-4,0,4)$ & $(\bf{3},\bf{1},\bf{1},\bf{1})$ & $\mathcal{C}_{15}$ & $(-2,0,4)$ & $(\bf{3},\bf{1},\bf{1},\bf{1}')$\\
  \cline{1-2} \cline{4-5}
  $\mathcal{C}_{6}$ & $(-4,2,4)$ & $\text{or}~(\bf{3}',\bf{1}',\bf{1}',\bf{1}')$ & $\mathcal{C}_{16}$ & $(-2,2,4)$ & $\text{or}~(\bf{3}',\bf{1}',\bf{1}',\bf{1})$\\
  \cline{1-2} \cline{4-5}
  $\mathcal{C}_{7}$ & $(-2,0,2)$ &  & $\mathcal{C}_{17}$ & $(-2,4,4)$ & \\
  \cline{1-2} \cline{4-5}
  $\mathcal{C}_{8}$ & $(-2,0,4)$ &  & $\mathcal{C}_{18}$ & $(0,2,4)$ & \\
  \cline{1-2} \cline{4-5}
  $\mathcal{C}_{9}$ & $(-2,2,4)$ &  & $\mathcal{C}_{19}$ & $(0,4,4)$ & \\
  \cline{1-2} \cline{4-5}
  $\mathcal{C}_{10}$ & $(0,2,4)$ &  & $\mathcal{C}_{20}$ & $(2,4,4)$ & \\
  \hline\hline
\end{tabular}
\end{tabular}
\caption{\label{tab:charged_lepton_model}
List of the charged lepton models $\mathcal{C}_{1}$, $\mathcal{C}_2$,\ldots,
$\mathcal{C}_{20}$, where $k_{E_{\alpha}^{c}}'\equiv k_{E_{\alpha}^{c}}+k_L$,
$\alpha=1,2,3$, are the modular weights
of the modular forms in the charged lepton Yukawa coupling.
}
\end{table}

\subsection{Neutrino masses via Weinberg operator\label{sec:neutrino_sector_WO}}

In this section, we consider the case that the neutrino masses arise from
the Weinberg operator. Given the assignments of left-handed lepton fields as in Eq.~\eqref{eq:assignments}, the Lagrangian for the Weinberg operator can be written as:
\begin{equation}
    \mathcal{L}^{M}_{\nu}= \frac{1}{2\Lambda}\left[ g_{1}\left((LL)_{\mathbf{1}}Y_{\mathbf{1}}^{(2k_{L})}\right)_{\mathbf{1}} +g_{2}\left((LL)_{\mathbf{2}}Y_{\mathbf{2}}^{(2k_{L})}\right)_{\mathbf{1}}+g_{3}\left((LL)_{\mathbf{3}'}Y_{\mathbf{3}'}^{(2k_{L})}\right)_{\mathbf{1}}\right]HH+\text{h.c.}\,.
\end{equation}
The corresponding light neutrino Majorana mass matrix is:
\begin{eqnarray}
\nonumber
M_\nu&=&
 \frac{g_{1}}{\Lambda}Y_{\mathbf{1}}^{(2k_{L})}\left(\begin{matrix}
1~&~0 ~&~0\\
0 ~&~0 ~&~1 \\
0 ~&~1 ~&~0
\end{matrix}\right)v^{2}
+\frac{g_{2}}{\Lambda} \left(\begin{matrix}
2Y_{\mathbf{2},1}^{(2k_{L})}&0 &0\\
0 &\sqrt{3}Y_{\mathbf{2},2}^{(2k_{L})} &-Y_{\mathbf{2},1}^{(2k_{L})} \\
0 &-Y_{\mathbf{2},1}^{(2k_{L})} & \sqrt{3}Y_{\mathbf{2},2}^{(2k_{L})}
\end{matrix}\right)v^{2}
\\
&&~+
\frac{g_{3}}{\Lambda} \left(\begin{matrix}
0&Y_{\mathbf{3}',2}^{(2k_{L})} &-Y_{\mathbf{3}',3}^{(2k_{L})}\\
Y_{\mathbf{3}',2}^{(2k_{L})} &Y_{\mathbf{3}',1}^{(2k_{L})} &0 \\
-Y_{\mathbf{3}',3}^{(2k_{L})} &0 &-Y_{\mathbf{3}',1}^{(2k_{L})}
\end{matrix}\right)v^{2} \,.
\end{eqnarray}
Considering the allowed values of $k_{L}$, we list 5 possible structures of
the light neutrino mass matrix in Table~\ref{tab:Weinberg_operator_model}.
From the summary of polyharmonic Maa{\ss} forms of level $N=4$ in
Table~\ref{Tab:LeveL4_MM}, we find that if $k_{L}\in\{-2,-1,0,1\}$,
there will be two complex Yukawa couplings $g_{1}$ and $g_{2}$ in $M_{\nu}$.
The phase of $g_{1}$ can be removed by rephasing $L$, thus there are $3$ real
coupling constant parameters in $M_{\nu}$. In the case of $k_{L}=2$,
$M_{\nu}$ depends on $5$ real parameters (real $g_{1}$, complex $g_{2}$ and
$g_{3}$) and the complex $\tau$. If the gCP symmetry is imposed,
all Yukawa coupling parameters will be real \cite{Novichkov:2019sqv}.

As we have discussed,
there are always 3 real Yukawa coupling
parameters in the charged lepton mass matrix $M_{\ell}$.
Thus, the total number of free parameters in the charged
lepton mass matrix $M_{\ell}$ and the light neutrino Majorana mass matrix $M_{\nu}$ satisfies
\begin{eqnarray}\nonumber
  &&k_{L}\in\{-2,-1,0,1\}:\qquad 8(7)~\text{real parameters}\,,\\
  &&k_{L}=2: \qquad 10(8)~\text{real parameters}\,,\label{eq:number_WO}
\end{eqnarray}
where the number given in parenthesis correspond to the case
that gCP symmetry is imposed.

\begin{table}[t!]
\centering
\resizebox{1.0\textwidth}{!}{
\begin{tabular}{|c|c|c|c|} \hline\hline

Model  & $k_L$ & $\rho_{L}$ &  Neutrino mass matrix  \\
  \hline
$\mathcal{W}_{1,2,3,4}$ & $-2,-1,0,1$ & \multirow{4}{*}{$\mathbf{3}$ or $\mathbf{3}'$}  & {\scriptsize $ M_\nu = \frac{g_{1}}{\Lambda} Y_{\mathbf{1}}^{(2k_{L})}\left(\begin{matrix}
1&0 &0\\
0 &0 &1 \\
0 &1 &0
\end{matrix}\right)v^{2}
+\frac{g_{2}}{\Lambda} \left(\begin{matrix}
2Y_{\mathbf{2},1}^{(2k_{L})}&0 &0\\
0 &\sqrt{3}Y_{\mathbf{2},2}^{(2k_{L})} &-Y_{\mathbf{2},1}^{(2k_{L})} \\
0 &-Y_{\mathbf{2},1}^{(2k_{L})} & \sqrt{3}Y_{\mathbf{2},2}^{(2k_{L})}
\end{matrix}\right)v^{2}$}  \\
 \cline{1-2} \cline{4-4}
$\mathcal{W}_5$ & $2$ & & {\scriptsize $   M_\nu =  \frac{g_{1}}{\Lambda} Y_{\mathbf{1}}^{(4)}\left(\begin{matrix}
1&0 &0\\
0 &0 &1 \\
0 &1 &0
\end{matrix}\right)v^{2}
+\frac{g_{2}}{\Lambda} \left(\begin{matrix}
2Y_{\mathbf{2},1}^{(4)}&0 &0\\
0 &\sqrt{3}Y_{\mathbf{2},2}^{(4)} &-Y_{\mathbf{2},1}^{(4)} \\
0 &-Y_{\mathbf{2},1}^{(4)} & \sqrt{3}Y_{\mathbf{2},2}^{(4)}
\end{matrix}\right)v^{2}+\frac{g_{3}}{\Lambda} \left(\begin{matrix}
0&Y_{\mathbf{3}',2}^{(4)} &-Y_{\mathbf{3}',3}^{(4)}\\
Y_{\mathbf{3}',2}^{(4)} &Y_{\mathbf{3}',1}^{(4)} &0 \\
-Y_{\mathbf{3}',3}^{(4)} &0 &-Y_{\mathbf{3}',1}^{(4)}
\end{matrix}\right)v^{2}$}  \\
  \hline
  \hline
\end{tabular} }
\caption{
\label{tab:Weinberg_operator_model}
The predictions for the neutrino mass matrices
for the models $\mathcal{W}_{1,2,3,4,5}$
in which the neutrino masses are generated through the Weinberg operator.
}
\end{table}

\subsection{Neutrino masses via Type-I seesaw mechanism\label{sec:neutrino_sector_SS}}

If neutrino masses are generated through the type-I seesaw mechanism,
at least two RH neutrino fields are required to accommodate the present neutrino oscillation data, namely, three nonzero lepton mixing angles and
two non-zero mass-squared differences. In this section, we will consider
the seesaw models with two and three RH neutrinos separately.

\subsubsection{\label{sec:two_right_handed_neutrinos}Two right-handed neutrinos}
We first consider the case of two heavy RH neutrinos. The two
RH neutrino fields $N^{c}=\left(N^{c}_{1}, N^{c}_{2}\right)^{T}$ are assumed to transform as doublet or a direct sum of
two one-dimensional representations of $S_{4}$.

\begin{itemize}

\item{$\rho_{N^{c}}=\bf{2}$}

  In this case the Dirac and Majorana neutrino mass terms
 can be written as:
  \begin{equation}
    \mathcal{L}_{\nu}=\mathcal{L}^{Y}_{\nu_{D}}+\mathcal{L}^{M}_{N^c}\,,
  \end{equation}
 where
 \begin{eqnarray}\nonumber
    \mathcal{L}^{Y}_{\nu_{D}}&=&\left[\beta_{1}\left((N^cL)_{\mathbf{3}^{i}}Y_{\mathbf{3}^{i}}^{(k_{N^{c}}+k_{L})}\right)_{\bf{1}}+\beta_{2}\left((N^cL)_{\mathbf{3}^{[i+1]}}Y_{\mathbf{3}^{[i+1]}}^{(k_{N^{c}}+k_{L})}\right)_{\bf{1}}\right]H+\text{h.c.}\\
    \mathcal{L}^{M}_{N^c}&=&\Big[g_{1}\left((N^cN^{c})_{\mathbf{1}}Y_{\mathbf{1}}^{(2k_{N^{c}})}\right)_{\bf{1}}+g_{2}\left((N^cN^{c})_{\mathbf{2}}Y_{\mathbf{2}}^{(2k_{N^{c}})}\right)_{\bf{1}}\Big]\Lambda +\text{h.c.}\,,
  \end{eqnarray}
where $k_{N^{c}}$ is the modular weight of $N^{c}$.
  The neutrino Dirac mass matrix and the heavy RH neutrino Majorana mass matrix are given by:
\begin{eqnarray}
\nonumber
  M_{\nu_D}&=&\beta_{1} \left(\begin{matrix}
2Y_{\mathbf{3}^{i},1}^{(k_{N^{c}}+k_{L})}~&~ -Y_{\mathbf{3}^{i},3}^{(k_{N^{c}}+k_{L})}~&~ -Y_{\mathbf{3}^{i},2}^{(k_{N^{c}}+k_{L})}\\
0~&~ \sqrt{3}Y_{\mathbf{3}^{i},2}^{(k_{N^{c}}+k_{L})}~&~ \sqrt{3}Y_{\mathbf{3}^{i},3}^{(k_{N^{c}}+k_{L})}\\
\end{matrix}\right)v \\
&&+\beta_{2} \left(\begin{matrix}
0& \sqrt{3}Y_{\mathbf{3}^{[i+1]},2}^{(k_{N^{c}}+k_{L})}~&~ \sqrt{3}Y_{\mathbf{3}^{[i+1]},3}^{(k_{N^{c}}+k_{L})}\\
-2Y_{\mathbf{3}^{[i+1]},1}^{(k_{N^{c}}+k_{L})} ~&~ Y_{\mathbf{3}^{[i+1]},3}^{(k_{N^{c}}+k_{L})}~&~ Y_{\mathbf{3}^{[i+1]},2}^{(k_{N^{c}}+k_{L})}\\
\end{matrix}\right)v\,,\\
\label{eq:MN_two}
 M_{N^c}&=&\left(\begin{matrix}
g_{1}Y_{\mathbf{1}}^{(2k_{N^{c}})}-g_{2}Y_{\mathbf{2},1}^{(2k_{N^{c}})}~&~g_{2}Y_{\mathbf{2},2}^{(2k_{N^{c}})}\\
g_{2}Y_{\mathbf{2},2}^{(2k_{N^{c}})}~&~g_{1}Y_{\mathbf{1}}^{(2k_{N^{c}})}+g_{2}Y_{\mathbf{2},1}^{(2k_{N^{c}})}\\
\end{matrix}\right)
\Lambda \,.
\end{eqnarray}
Here the possible assignments for $\rho_{L}$ can be
$\bf{3}$ or $\bf{3}'$, and the allowed values of $k_{L}+k_{N^{c}}$ are $k_{L}+k_{N^{c}}=-4,-2,0,2,4$. From Table~\ref{Tab:LeveL4_MM}, we find that for the cases $k_{L}+k_{N^{c}}=-4,-2,0,2$, there is one modular form multiplet that transforms as $\mathbf{3}$ of $S_{4}$, while there is no modular form in the triplet representation $\mathbf{3}'$. For $k_{L}+k_{N^{c}}=4$, there are two modular form triplets $Y^{(4)}_{\mathbf{3}}$ and $Y^{(4)}_{\mathbf{3}'}$,  which transform as $\mathbf{3}$ and $\mathbf{3}'$, respectively. As a consequence, we obtain four general forms of $M_{\nu_{D}}$.
\begin{eqnarray}
\label{eq:MD_1}
 \rho_{L}=\mathbf{3}\,, k_{L}+k_{N^{c}}< 4:~ M_{\nu_{D}}&=&\beta_{1} \left(\begin{matrix}
2Y_{\mathbf{3},1}^{(k_{N^{c}}+k_{L})} ~&~ -Y_{\mathbf{3},3}^{(k_{N^{c}}+k_{L})}~&~ -Y_{\mathbf{3},2}^{(k_{N^{c}}+k_{L})}\\
0~&~ \sqrt{3}Y_{\mathbf{3},2}^{(k_{N^{c}}+k_{L})}~&~ \sqrt{3}Y_{\mathbf{3},3}^{(k_{N^{c}}+k_{L})}\\
\end{matrix}\right)v\,,\\ \nonumber
\rho_{L}=\mathbf{3}\,, k_{L}+k_{N^{c}}= 4:~ M_{\nu_{D}}&=&\beta_{1} \left(\begin{matrix}
2Y_{\mathbf{3},1}^{(k_{N^{c}}+k_{L})}~&~ -Y_{\mathbf{3},3}^{(k_{N^{c}}+k_{L})}~&~ -Y_{\mathbf{3},2}^{(k_{N^{c}}+k_{L})}\\
0~&~ \sqrt{3}Y_{\mathbf{3},2}^{(k_{N^{c}}+k_{L})}~&~ \sqrt{3}Y_{\mathbf{3},3}^{(k_{N^{c}}+k_{L})}\\
\end{matrix}\right)v\\ \label{eq:MD_2}
&&\hskip-0.1in+\beta_{2}\left(\begin{matrix}
0~&~ \sqrt{3}Y_{\mathbf{3}',2}^{(k_{N^{c}}+k_{L})}~&~ \sqrt{3}Y_{\mathbf{3}',3}^{(k_{N^{c}}+k_{L})}\\
-2Y_{\mathbf{3}',1}^{(k_{N^{c}}+k_{L})} ~&~ Y_{\mathbf{3}',3}^{(k_{N^{c}}+k_{L})}~&~ Y_{\mathbf{3}',2}^{(k_{N^{c}}+k_{L})}\\
\end{matrix}\right)v\,,\\ \label{eq:MD_3}
\rho_{L}=\mathbf{3}'\,, k_{L}+k_{N^{c}}< 4:~   M_{\nu_{D}}&=&\beta_{1} \left(\begin{matrix}
0~&~ \sqrt{3}Y_{\mathbf{3},2}^{(k_{N^{c}}+k_{L})}~&~ \sqrt{3}Y_{\mathbf{3},3}^{(k_{N^{c}}+k_{L})}\\
-2Y_{\mathbf{3},1}^{(k_{N^{c}}+k_{L})} ~&~ Y_{\mathbf{3},3}^{(k_{N^{c}}+k_{L})}~&~ Y_{\mathbf{3},2}^{(k_{N^{c}}+k_{L})}\\
\end{matrix}\right)v\,,\\ \nonumber
\rho_{L}=\mathbf{3}'\,, k_{L}+k_{N^{c}}= 4:~ M_{\nu_{D}}&=&\beta_{1} \left(\begin{matrix}
2Y_{\mathbf{3}',1}^{(k_{N^{c}}+k_{L})}~&~ -Y_{\mathbf{3}',3}^{(k_{N^{c}}+k_{L})}~&~ -Y_{\mathbf{3}',2}^{(k_{N^{c}}+k_{L})}\\
0~&~ \sqrt{3}Y_{\mathbf{3}',2}^{(k_{N^{c}}+k_{L})}~&~ \sqrt{3}Y_{\mathbf{3}',3}^{(k_{N^{c}}+k_{L})}\\
\end{matrix}\right)v\\ \label{eq:MD_4}
&&\hskip-0.1in +\beta_{2} \left(\begin{matrix}
0~&~ \sqrt{3}Y_{\mathbf{3},2}^{(k_{N^{c}}+k_{L})}~&~ \sqrt{3}Y_{\mathbf{3},3}^{(k_{N^{c}}+k_{L})}\\
-2Y_{\mathbf{3},1}^{(k_{N^{c}}+k_{L})} ~&~ Y_{\mathbf{3},3}^{(k_{N^{c}}+k_{L})}~&~ Y_{\mathbf{3},2}^{(k_{N^{c}}+k_{L})}\\
\end{matrix}\right)v\,.
\end{eqnarray}
We note that the assignments $\rho_{L}=\mathbf{3}$, $\rho_{N^{c}}=\mathbf{2}$
and $\rho_{L}=\mathbf{3}'$, $\rho_{N^{c}}=\mathbf{2}$  lead to the same
predictions for neutrino masses and mixing.
If we change the representation assignment
  \begin{equation}
    \rho_{L}=\mathbf{3}\,,\rho_{N^{c}}=\mathbf{2}\quad \rightarrow \quad \rho_{L}=\mathbf{3}'\,,\rho_{N^{c}}=\mathbf{2}\,,
  \end{equation}
for any given modular weights, the neutrino Dirac and the heavy RH
neutrino Majorana mass matrices would turn into:
  \begin{equation}
    M_{\nu_{D}}\rightarrow M'_{\nu_{D}}=\begin{pmatrix} 0 ~&~ 1 \\ -1 ~&~ 0 \end{pmatrix}M_{\nu_{D}} \,,\quad M_{N^{c}}\rightarrow M'_{N^{c}}=M_{N^{c}}\,.
  \end{equation}
The effective light neutrino Majorana mass matrix $M_{\nu}$ is given by
the well known seesaw expression:
\begin{equation}
\label{eq:seesaw}
  M_{\nu}=-M_{\nu_{D}}^{T}M_{N^c}^{-1}M_{\nu_{D}}\,.
\end{equation}
As a consequence, the light neutrino mass matrix $M_{\nu}$ would change as follows:
\begin{equation}
  M_{\nu}\rightarrow M_{\nu}'=-(M'_{\nu_{D}})^{T}(M'_{N^c})^{-1}M'_{\nu_{D}}=-M_{\nu_{D}}^{T}M_{N^c}^{-1}M_{\nu_{D}}\left( g_{2}\rightarrow -g_{2}\right)=M_{\nu}\left( g_{2}\rightarrow -g_{2}\right)\,.
\end{equation}
Since $g_{2}$ is a free parameter, the matrices $M_{\nu}$ and $M_{\nu}'$
can always yield same predictions for neutrino masses and mixing.
Therefore we will consider only the case $\rho_{L}=\mathbf{3}$
in the following analysis.

Considering the allowed values of $k_{N^{c}}+k_{L}$,
we list the 5 possible structures of the neutrino Dirac mass matrix in
Table~\ref{tab:seesaw_neutrino_SSN2}. The explicit form of the
heavy RH neutrino Majorana mass matrix $M_{N^{c}}$ depends on $k_{N^{c}}$
and it can take five possible forms, as shown in Table~\ref{tab:seesaw_neutrino_SSN2}. The combinations of structures of $M_{\nu_{D}}$ and $M_{N^{c}}$ are summarized in Table~\ref{tab:seesaw_neutrino_SSN2}.

The effective light neutrino mass matrix can be obtained by using seesaw
expression given in Eq.~\eqref{eq:seesaw}.
In the case of $k_{N^{c}}+k_{L}<4$, $M_{\nu}$ depends on the overall factor
$\frac{\beta_{1}^{2}v^{2}}{g_{1}\Lambda}$ and a coupling $g_{2}$,
besides the complex modulus $\tau$.
If $k_{N^{c}}+k_{L}=4$, there will be one additional complex parameter
$\beta_{2}$ in $M_{\nu}$. Combining the charge lepton and neutrino sectors, the
number of free parameters in the charged lepton mass matrix $M_{\ell}$ and the
light neutrino Majorana mass matrix $M_{\nu}$ satisfies:
\begin{eqnarray} \nonumber
  &&k_{N^{c}}+k_{L}\in \{-4,-2,0,2\}\,,2k_{N^{c}}\in\{-4,-2,0,2,4\}:\quad 8(7)~\text{real parameters}\,,\\
  &&k_{N^{c}}+k_{L}=4\,,2k_{N^{c}}\in\{-4,-2,0,2,4\}:\quad 10(8)~\text{real parameters}\,. \label{eq:number_SSN2}
\end{eqnarray}
\begin{table}[t!]
\centering
\begin{tabular}{c}
\begin{tabular}{|c|c|c|c|c|c|}  \hline\hline
  \multicolumn{2}{|c|}{Models}& $k_{L}$  & $(\rho_{L},\rho_{N^{c}})$ & $M_{\nu_{D}}$ & $M_{N^{c}}$ \\
  \hline
  $\mathcal{D}_{1}$ & \multirow{5}{*}{$\mathcal{N}_{1,2,3,4,5}$} & $-4-k_{N^{c}}$  & \multirow{5}{*}{$(\bf{3},\bf{2})$ or $(\bf{3}',\bf{2})$} & \multirow{4}{*}{eq.\eqref{eq:MD_1}} & \multirow{5}{*}{eq.\eqref{eq:MN_two}}  \\
  \cline{1-1}
  $\mathcal{D}_{2}$ &  & $-2-k_{N^{c}}$ &  & & \\
  \cline{1-1}
  $\mathcal{D}_{3}$ &  & $-k_{N^{c}}$ &  & & \\
  \cline{1-1}
  $\mathcal{D}_{4}$ &  & $2-k_{N^{c}}$ &  & & \\
  \cline{1-1} \cline{5-5}
  $\mathcal{D}_{5}$ &  & $4-k_{N^{c}}$ & & eq.\eqref{eq:MD_2} & \\
  \hline\hline
\end{tabular}
\end{tabular}
\caption{\label{tab:seesaw_neutrino_SSN2} List of
the neutrino Dirac and Majorana
mass matrices in the case of $N^{c}\sim \mathbf{2}$.
Here $k_{N^{c}}=-2,-1,0,1,2$ for the heavy Majorana neutrino models $\mathcal{N}_{1}$, $\mathcal{N}_{2}$, $\mathcal{N}_{3}$, $\mathcal{N}_{4}$, $\mathcal{N}_{5}$ respectively. }
\end{table}

\item{$\rho_{N^c} =\mathbf{1}^{j_1}\oplus\mathbf{1}^{j_2}$}

  If the RH heavy neutrinos are assumed to transform as singlet
representations of $S_{4}$, the general  Dirac and Majorana neutrino
mass terms can be written as:
  \begin{equation}
    \mathcal{L}_{\nu}=\mathcal{L}^{Y}_{\nu_{D}}+\mathcal{L}^{M}_{N^c}\,,
  \end{equation}
where
 \begin{eqnarray}\nonumber
   \hskip -0.5in \mathcal{L}^{Y}_{\nu_{D}}&=&\left[\beta_{1}(N^c_1LY_{\mathbf{3}^{[j_1+i]}}^{(k_{N^{c}_{1}}+k_{L})})_{\bf{1}}+\beta_{2}(N^c_2LY_{\mathbf{3}^{[j_{2}+i]}}^{(k_{N^{c}_{2}}+k_{L})})_{\bf{1}}\right]H+\text{h.c.}\,,\\
   \hskip -0.5in \mathcal{L}^{M}_{N^c}&=&\Big[g_{1}(N^c_{1}N^{c}_{1})_{\mathbf{1}}Y_{\mathbf{1}}^{(2k_{N_{1}^{c}})}+g_{2}(N^c_{2}N^{c}_{2})_{\mathbf{1}}Y_{\mathbf{1}}^{(2k_{N_{2}^{c}})}+2g_{3}\left((N^c_{1}N^{c}_{2})_{\mathbf{1}^{[j_{1}+j_{2}]}}Y_{\mathbf{1}^{[j_{1}+j_{2}]}}^{(k_{N_{1}^{c}}+k_{N_{2}^{c}})}\right)_{\bf{1}}\Big]\Lambda +\text{h.c.}\,,
  \end{eqnarray}
where $k_{N^{c}_{1,2}}$ are the modular weights of $N^{c}_{1,2}$.
The corresponding neutrino Dirac mass matrix and the heavy RH
neutrino Majorana mass matrix read:
\begin{eqnarray}\label{eq:N11_MD}
  M_{\nu_{D}}&=&\left(\begin{matrix}
\beta_{1} Y_{\mathbf{3}^{[j_{1}+i]},1}^{(k_{N^{c}_{1}}+k_{L})}~&~\beta_{1} Y_{\mathbf{3}^{[j_{1}+i]},3}^{(k_{N^{c}_{1}}+k_{L})}~&~\beta_{1} Y_{\mathbf{3}^{[j_{1}+i]},2}^{(k_{N^{c}_{1}}+k_{L})}\\
\beta_{2} Y_{\mathbf{3}^{[j_{2}+i]},1}^{(k_{N^{c}_{2}}+k_{L})}~&~\beta_{2} Y_{\mathbf{3}^{[j_{2}+i]},3}^{(k_{N^{c}_{2}}+k_{L})}~&~\beta_{2} Y_{\mathbf{3}^{[j_{2}+i]},2}^{(k_{N^{c}_{2}}+k_{L})}\\
\end{matrix}\right)v\\
 M_{N^c}&=&\left(\begin{matrix}
g_{1}Y_{\mathbf{1}}^{(2k_{N^c_1})} ~&g_{3}Y_{\mathbf{1}^{[j_1+j_2]}}^{(k_{N^c_1}+k_{N^c_2})} \\
g_{3}Y_{\mathbf{1}^{[j_1+j_2]}}^{(k_{N^c_1}+k_{N^c_2})} ~& g_{2}Y_{\mathbf{1}}^{(2k_{N^c_2})} \\
\end{matrix}\right)\Lambda\,.
\end{eqnarray}

Similar to the singlet RH charged lepton fields,
the two RH neutrino fields must be distinguishable from each other by their
modular weights and/or representations. Notice that exchanging the assignments of the RH neutrinos effectively multiplies the neutrino Dirac mass matrix $M_{\nu_{D}}$ by certain permutation matrices on the left-hand side, and the heavy neutrino mass matrix $M_{N^{c}}$ by the same matrices on both sides. Consequently, the resulting effective light neutrino mass matrix $M_{\nu}$ remains unchanged. Without loss of generality, we assume $k_{N^{c}_{1}}\leq k_{N^{c}_{2}}$.  We have demanded that the modular weight of the involved modular forms should satisfy $|k_{Y}|\leq 4$, i.e.,
\begin{equation}
  k_{L}+k_{N^{c}_{1}}\,,k_{L}+k_{N^{c}_{2}}\,,k_{N^{c}_{1}}+k_{N^{c}_{2}}\,,2k_{N^{c}_{1}}\,,2k_{N^{c}_{2}}\in \{-4,-2,0,2,4\}\,,
\end{equation}
which leads to
\begin{equation}
\label{eq:constraint_weight_11}
  k_{N^{c}_{1}}\,,k_{N^{c}_{2}}\in\{-2,-1,0,1,2\}\,,
\qquad k_{N^{c}_{2}}-k_{N^{c}_{1}}\in \{0,2,4\}\,.
\end{equation}
To distinguish the two RH neutrino fields $N^{c}_{1}$ and $N^{c}_{2}$,
we should have $k_{N^{c}_{1}}< k_{N^{c}_{2}}$ if $N^{c}_{1}$ and $N^{c}_{2}$ transform as the same singlet representation of $S_{4}$,
i.e., if  $\rho_{N^{c}_{1}}=\rho_{N^{c}_{2}}$. In the case where $\rho_{N^{c}_{1}}\neq \rho_{N^{c}_{2}}$, we require that $k_{N^{c}_{1}}\leq k_{N^{c}_{2}}$. Analyzing the possible representation assignments of $L$ and $N^{c}_{1,2}$ , we can derive the following possible forms of $M_{\nu_{D}}$.
\begin{itemize}

\item{$(\rho_{L},\rho_{N^{c}_{1}},\rho_{N^{c}_2})=(\bf{3},\bf{1},\bf{1})~\text{or}~(\bf{3}',\bf{1}',\bf{1}')$}

In this case, the neutrino Dirac mass matrix reads:
\begin{equation}
    \label{eq:MD_s_1}
  M_{\nu_{D}}=\left(\begin{matrix}
\beta_{1} Y_{\mathbf{3},1}^{(k_{N^{c}_{1}}+k_{L})}~&~\beta_{1} Y_{\mathbf{3},3}^{(k_{N^{c}_{1}}+k_{L})}~&~\beta_{1} Y_{\mathbf{3},2}^{(k_{N^{c}_{1}}+k_{L})}\\
\beta_{2} Y_{\mathbf{3},1}^{(k_{N^{c}_{2}}+k_{L})}~&~\beta_{2} Y_{\mathbf{3},3}^{(k_{N^{c}_{2}}+k_{L})}~&~\beta_{2} Y_{\mathbf{3},2}^{(k_{N^{c}_{2}}+k_{L})}\\
\end{matrix}\right)v\,.
  \end{equation}
Since $\rho_{N^{c}_{1}}=\rho_{N^{c}_{2}}$, the modular weights of $N^{c}_{1}$
and $N^{c}_{2}$ should satisfy $k_{N^{c}_{1}}< k_{N^{c}_{2}}$, as well as
Eq.~\eqref{eq:constraint_weight_11}.
The allowed values of $(k_{N^{c}_{1}}+k_{L},k_{N^{c}_{2}}+k_{L})$ are:
\begin{equation}
    (-4,-2)\,,(-4,0)\,,(-2,0)\,,(-2,2)\,,(0,2)\,,(0,4)\,,(2,4)\,.
\end{equation}

\item{$(\rho_{L},\rho_{N^{c}_{1}},\rho_{N^{c}_2})=(\bf{3},\bf{1},\bf{1}')~\text{or}~(\bf{3}',\bf{1}',\bf{1})$}

The neutrino Dirac mass matrix is given by:
\begin{equation}
    \label{eq:MD_s_2}
  M_{\nu_{D}}=\left(\begin{matrix}
\beta_{1} Y_{\mathbf{3},1}^{(k_{N^{c}_{1}}+k_{L})}~&~\beta_{1} Y_{\mathbf{3},3}^{(k_{N^{c}_{1}}+k_{L})}~&~\beta_{1} Y_{\mathbf{3},2}^{(k_{N^{c}_{1}}+k_{L})}\\
\beta_{2} Y_{\mathbf{3}',1}^{(k_{N^{c}_{2}}+k_{L})}~&~\beta_{2} Y_{\mathbf{3}',3}^{(k_{N^{c}_{2}}+k_{L})}~&~\beta_{2} Y_{\mathbf{3}',2}^{(k_{N^{c}_{2}}+k_{L})}\\
\end{matrix}\right)v\,.
  \end{equation}
In order to contract a trivial singlet $\mathbf{1}$ of $S_{4}$, the modular multiplet which couple to $L$ and $N^{c}_{2}$ should transform as
$\mathbf{3}'$ of $S_{4}$ in the case of $\rho_{L}\otimes \rho_{N^{c}_{2}}=\mathbf{3}'$. The possible combinations of $(k_{N^{c}_{1}}+k_{L},k_{N^{c}_{2}}+k_{L})$ are:
\begin{equation}
 (0,4)\,,(2,4)\,,(4,4)\,.
\end{equation}

\item{$(\rho_{L},\rho_{N^{c}_{1}},\rho_{N^{c}_2})=(\bf{3},\bf{1}',\bf{1}')~\text{or}~(\bf{3}',\bf{1},\bf{1})$}

The  neutrino Dirac mass matrix fas the form:
\begin{equation}
\label{eq:MD_s_3}
M_{\nu_{D}}=\left(\begin{matrix}
\beta_{1} Y_{\mathbf{3}',1}^{(k_{N^{c}_{1}}+k_{L})}~&~\beta_{1} Y_{\mathbf{3}',3}^{(k_{N^{c}_{1}}+k_{L})}~&~\beta_{1} Y_{\mathbf{3}',2}^{(k_{N^{c}_{1}}+k_{L})}\\
\beta_{2} Y_{\mathbf{3}',1}^{(k_{N^{c}_{2}}+k_{L})}~&~\beta_{2} Y_{\mathbf{3}',3}^{(k_{N^{c}_{2}}+k_{L})}~&~\beta_{2} Y_{\mathbf{3}',2}^{(k_{N^{c}_{2}}+k_{L})}\\
\end{matrix}\right)v\,.
  \end{equation}
The row of $M_{\nu_{D}}$ does not vanish only if $k_{N^{c}_{1}}+k_{L}=k_{N^{c}_{2}}+k_{L}=4$. This leads to $k_{N^{c}_{1}}=k_{N^{c}_{2}}$. The two rows of $M_{\nu_{D}}$ are proportional and the rank of $M_{\nu_{D}}$ is 1. We do not consider this assignment of the neutrino fields further.
\end{itemize}
We list the possible structures of $M_{\nu_{D}}$ in Table~\ref{tab:seesaw_neutrino_SSN11}. Now, let's consider the structures of the heavy right-handed (RH) neutrino Majorana mass matrix $M_{N^{c}}$.

\begin{itemize}

\item{$(\rho_{N^{c}_{1}},\rho_{N^{c}_2})=(\bf{1},\bf{1})~\text{or}~(\bf{1}',\bf{1}')$}

In this case the RH neutrino Majorana mass matrix reads:
\begin{equation}
\label{eq:MN_2}
    M_{N^c}=\left(\begin{matrix}
g_{1}Y_{\mathbf{1}}^{(2k_{N^c_1})} ~&~g_{3}Y_{\mathbf{1}}^{(k_{N^c_1}+k_{N^c_2})} \\
g_{3}Y_{\mathbf{1}}^{(k_{N^c_1}+k_{N^c_2})} ~&~ g_{2}Y_{\mathbf{1}}^{(2k_{N^c_2})} \\
\end{matrix}\right)\Lambda\,.
\end{equation}
The allowed values of $(k_{N^{c}_{1}},k_{N^{c}_{2}})$ are:
\begin{equation}
  (-2,0)\,,(-2,2)\,,(-1,1)\,,(0,2)\,.
\end{equation}

\item{$(\rho_{N^{c}_{1}},\rho_{N^{c}_2})=(\bf{1},\bf{1}')~\text{or}~(\bf{1}',\bf{1})$}

With this assignment $M_{N^c}$ has the form:
\begin{equation} \label{eq:MN_1}
    M_{N^c}=\left(\begin{matrix}
g_{1}Y_{\mathbf{1}}^{(2k_{N^c_1})} ~&0 \\
0 ~& g_{2}Y_{\mathbf{1}}^{(2k_{N^c_2})} \\
\end{matrix}\right)\Lambda\,.
\end{equation}

There are 9 allowed values of $(k_{N^{c}_{1}},k_{N^{c}_{2}})$:
\begin{equation}
  (-2,-2)\,,(-1,-1)\,,(0,0)\,,(1,1)\,,(2,2)\,,(-2,0)\,,(-2,2)\,,
(-1,-1)\,,(0,2)\,.
\end{equation}

\end{itemize}

We list the possible structures of $M_{N^{c}}$ in Table~\ref{tab:seesaw_neutrino_SSN11}. By combining the results for $M_{\nu_{D}}$ and $M_{N^{c}}$ that we have derived, we also list the allowed combinations of $M_{\nu_{D}}$ and $M_{N^{c}}$ in Table~\ref{tab:seesaw_neutrino_SSN11}.

Using the seesaw expression given in Eq.~\eqref{eq:seesaw}, we find that there are 3 and 5 real parameters in $M_{\nu}$ for $[j_{1}+j_{2}]$ equals to $1$ and $0$, respectively. A detailed explanation is presented in Appendix~\ref{sec:effective_parameters}. As a result, the number of free parameters in the charged lepton mass matrix $M_{\ell}$ and the light neutrino Majorana mass matrix $M_{\nu}$ satisfy:
\begin{eqnarray}\nonumber
  &&[j_{1}+j_{2}]=1:\quad 8~(7)~\text{real parameters}\,,\\
  &&[j_{1}+j_{2}]=0:\quad 10~(8)~\text{real parameters}\,,\label{eq:number_SSN11}
\end{eqnarray}
where the the numbers in the brackets correspond to the case
of imposed gCP symmetry.
\end{itemize}
\begin{table}[h!]
\centering
\begin{tabular}{c}
\begin{tabular}{|c|c|c|c|c|c|}  \hline\hline
  \multicolumn{2}{|c|}{Models}&  $(k_{L},k_{N_{1}^{c}},k_{N_{2}^{c}})$ &  $(\rho_{L},\rho_{N^{c}_{1}},\rho_{N^{c}_{2}})$ & $M_{\nu_{D}}$ & $M_{N^{c}}$  \\
  \hline
  \multirow{3}{*}{$\mathfrak{D}_{1}$} & $\mathfrak{N}_{1}$ & $(-2,-2,0)$ & \multirow{15}{*}{$\begin{array}{c}(\bf{3},\bf{1},\bf{1})\,,\text{or}\\(\bf{3}',\bf{1}',\bf{1}') \end{array}$}  & \multirow{15}{*}{eq.\eqref{eq:MD_s_1}} & \multirow{15}{*}{eq.\eqref{eq:MN_2}} \\ \cline{2-3}
   & $\mathfrak{N}_{3}$  & $(-3,-1,1)$ & & & \\ \cline{2-3}
   & $\mathfrak{N}_{4}$  & $(-4,0,2)$ & & & \\ \cline{1-3}
  $\mathfrak{D}_{2}$ & $\mathfrak{N}_{2}$ & $(-2,-2,2)$ & & & \\ \cline{1-3}
  \multirow{3}{*}{$\mathfrak{D}_{3}$} & $\mathfrak{N}_{1}$ & $(0,-2,0)$ &   &  &\\ \cline{2-3}
   & $\mathfrak{N}_{3}$  & $(-1,-1,1)$ & &  &\\ \cline{2-3}
   & $\mathfrak{N}_{4}$  & $(-2,0,2)$ & &  &\\ \cline{1-3}
  $\mathfrak{D}_{4}$ & $\mathfrak{N}_{2}$ & $(0,-2,2)$ & & & \\ \cline{1-3}
  \multirow{3}{*}{$\mathfrak{D}_{5}$} & $\mathfrak{N}_{1}$ & $(2,-2,0)$ &   & & \\ \cline{2-3}
   & $\mathfrak{N}_{3}$  & $(1,-1,1)$ & & & \\ \cline{2-3}
   & $\mathfrak{N}_{4}$  & $(0,0,2)$ & & & \\ \cline{1-3}
  $\mathfrak{D}_{6}$ & $\mathfrak{N}_{2}$ & $(2,-2,2)$ & &&  \\ \cline{1-3}
  \multirow{3}{*}{$\mathfrak{D}_{7}$} & $\mathfrak{N}_{1}$ & $(4,-2,0)$ &   & & \\ \cline{2-3}
   & $\mathfrak{N}_{3}$  & $(3,-1,1)$ & &&  \\ \cline{2-3}
   & $\mathfrak{N}_{4}$  & $(2,0,2)$ & &&  \\
  \hline
  $\mathfrak{D}_{8}$ & $\mathfrak{N}_{13}$ & $(2,-2,2)$ & \multirow{9}{*}{$\begin{array}{c}(\bf{3},\bf{1},\bf{1}')\,,\text{or}\\(\bf{3}',\bf{1}',\bf{1})\end{array}$} & \multirow{9}{*}{eq.\eqref{eq:MD_s_2}} & \multirow{9}{*}{ eq.\eqref{eq:MN_1}}  \\
  \cline{1-3}
  \multirow{3}{*}{$\mathfrak{D}_{9}$} & $\mathfrak{N}_{10}$ & $(4,-2,0)$ &   & & \\ \cline{2-3}
   & $\mathfrak{N}_{11}$  & $(3,-1,1)$ & & & \\ \cline{2-3}
   & $\mathfrak{N}_{12}$  & $(2,0,2)$ & &  &\\ \cline{1-3}
  \multirow{5}{*}{$\mathfrak{D}_{10}$} & $\mathfrak{N}_{5}$ & $(6,-2,-2)$ &   & & \\ \cline{2-3}
   & $\mathfrak{N}_{6}$  & $(5,-1,-1)$ & & & \\ \cline{2-3}
   & $\mathfrak{N}_{7}$  & $(4,0,0)$ & & & \\ \cline{2-3}
   & $\mathfrak{N}_{8}$  & $(3,1,1)$ & & & \\ \cline{2-3}
   & $\mathfrak{N}_{9}$  & $(2,2,2)$ & & & \\
  \hline\hline
\end{tabular}
\end{tabular}
\caption{
\label{tab:seesaw_neutrino_SSN11} List of the
 neutrino Dirac and Majorana mass matrices in case of
$N^{c}\sim \mathbf{1}^{j_{1}}\oplus\mathbf{1}^{j_{2}}$.}
\end{table}

\subsubsection{Three right-handed neutrinos}

In this section, we discuss seesaw models with three RH neutrinos. The three
RH neutrino fields $N^{c}=\left(N^{c}_{1}, N^{c}_{2}, N^{c}_{3}\right)^{T}$ are assumed to transform as a triplet or a direct sum of one-dimensional and two-dimensional representations of $S_{4}$. In what follows, we consider distinct assignments of the RH neutrino fields and the resulting neutrino mass matrices.

\begin{itemize}

\item{$\rho_{L}=\mathbf{3}^{i}\,,\rho_{N^{c}}= \mathbf{3}^{j}$}

In this case, the Dirac and Majorana neutrino
mass terms can be written as:
  \begin{equation}
    \mathcal{L}_{\nu}=\mathcal{L}^{Y}_{\nu_{D}}+\mathcal{L}^{M}_{N^c}\,,
  \end{equation}
where
  \begin{eqnarray}\nonumber
   \hskip -0.5in \mathcal{L}^{Y}_{\nu_{D}}&=&\beta_{1}\left((N^{c}L)_{\mathbf{1}^{[i+j]}}Y_{\mathbf{1}^{[i+j]}}^{(k_{N^c}+k_{L})}\right)_{\mathbf{1}}H +\beta_{2}\left((N^{c}L)_{\mathbf{2}}Y_{\mathbf{2}}^{(k_{N^c}+k_{L})}\right)_{\mathbf{1}}H\\ \nonumber
&~&~+\beta_{3}\left((N^{c}L)_{\mathbf{3}^{[i+j]}}Y_{\mathbf{3}^{[i+j]}}^{(k_{N^c}+k_{L})}\right)_{\mathbf{1}}H+\beta_{4} \left((N^{c}L)_{\mathbf{3}^{[i+j+1]}}Y_{\mathbf{3}^{[i+j+1]}}^{(k_{N^c}+k_{L})}\right)_{\mathbf{1}}H+\text{h.c.}\,,\\
   \hskip -0.5in \mathcal{L}^{M}_{N^c}&=&\left[ g_{1}\left((N^{c}N^{c})_{\mathbf{1}}Y_{\mathbf{1}}^{(2k_{N^{c}})}\right)_{\mathbf{1}} +g_{2}\left((N^{c}N^{c})_{\mathbf{2}}Y_{\mathbf{2}}^{(2k_{N^{c}})}\right)_{\mathbf{1}}+g_{3}\left((N^{c}N^{c})_{\mathbf{3}'}Y_{\mathbf{3}'}^{(2k_{N^{c}})}\right)_{\mathbf{1}}\right]\Lambda +\text{h.c.} \,.
  \end{eqnarray}
The neutrino Dirac mass matrix and the heavy RH neutrino Majorana mass
matrix are given by:
  \begin{eqnarray} \small \nonumber
M_{\nu_{D}}&=&{}
\beta_{1} Y_{\mathbf{1}}^{(k_{N^{c}}+k_{L})}\left(\begin{matrix}
1~&~0 ~&~0\\
0 ~&~0 ~&~1 \\
0 ~&~1 ~&~0
\end{matrix}\right)v
+\beta_{2} \left(\begin{matrix}
2Y_{\mathbf{2},1}^{(k_{N^{c}}+k_{L})}&0 &0\\
0 &\sqrt{3}Y_{\mathbf{2},2}^{(k_{N^{c}}+k_{L})} &-Y_{\mathbf{2},1}^{(k_{N^{c}}+k_{L})} \\
0 &-Y_{\mathbf{2},1}^{(k_{N^{c}}+k_{L})} & \sqrt{3}Y_{\mathbf{2},2}^{(k_{N^{c}}+k_{L})}
\end{matrix}\right)v
\\ \nonumber
&+&
\beta_{3} \left(\begin{matrix}
0&-Y_{\mathbf{3}^{[i+j]},3}^{(k_{N^{c}}+k_{L})} &Y_{\mathbf{3}^{[i+j]},2}^{(k_{N^{c}}+k_{L})}\\
Y_{\mathbf{3}^{[i+j]},3}^{(k_{N^{c}}+k_{L})} &0 &-Y_{\mathbf{3}^{[i+j]},1}^{(k_{N^{c}}+k_{L})} \\
-Y_{\mathbf{3}^{[i+j]},2}^{(k_{N^{c}}+k_{L})} &Y_{\mathbf{3}^{[i+j]},1}^{(k_{N^{c}}+k_{L})} &0
\end{matrix}\right)v
+
\beta_{4} \left(\begin{matrix}
0&Y_{\mathbf{3}^{[i+j+1]},2}^{(k_{N^{c}}+k_{L})} &-Y_{\mathbf{3}^{[i+j+1]},3}^{(k_{N^{c}}+k_{L})}\\
Y_{\mathbf{3}^{[i+j+1]},2}^{(k_{N^{c}}+k_{L})} &Y_{\mathbf{3}^{[i+j+1]},1}^{(k_{N^{c}}+k_{L})} &0 \\
-Y_{\mathbf{3}^{[i+j+1]},3}^{(k_{N^{c}}+k_{L})} &0 &-Y_{\mathbf{3}^{[i+j+1]},1}^{(k_{N^{c}}+k_{L})}
\end{matrix}\right)v \,,\label{eq:MD_33_1}\\ \nonumber
M_{N^c} &=&  \frac{1}{\Lambda} Y_{\mathbf{1}}^{(2k_{N^{c}})}\left(\begin{matrix}
1~&~0 ~&~0\\
0 ~&~0 ~&~1 \\
0 ~&~1 ~&~0
\end{matrix}\right)v^{2}
+\frac{g_{2}}{\Lambda} \left(\begin{matrix}
2Y_{\mathbf{2},1}^{(2k_{N^{c}})}&0 &0\\
0 &\sqrt{3}Y_{\mathbf{2},2}^{(2k_{N^{c}})} &-Y_{\mathbf{2},1}^{(2k_{N^{c}})} \\
0 &-Y_{\mathbf{2},1}^{(2k_{N^{c}})} & \sqrt{3}Y_{\mathbf{2},2}^{(2k_{N^{c}})}
\end{matrix}\right)v^{2}\\
    &&+\frac{g_{3}}{\Lambda} \left(\begin{matrix}
0&Y_{\mathbf{3}',2}^{(2k_{N^{c}})} &-Y_{\mathbf{3}',3}^{(2k_{N^{c}})}\\
Y_{\mathbf{3}',2}^{(2k_{N^{c}})} &Y_{\mathbf{3}',1}^{(2k_{N^{c}})} &0 \\
-Y_{\mathbf{3}',3}^{(2k_{N^{c}})} &0 &-Y_{\mathbf{3}',1}^{(2k_{N^{c}})}
\end{matrix}\right)v^{2}\,.
 \end{eqnarray}

The number of the coupling constants depends on the values of the modular
weight of $L$ and $N^{c}$, i.e.,
 \begin{eqnarray}
\nonumber
   &&k_{L}+k_{N^{c}}<4\,,k_{N^{c}}<2: \quad  \beta_{1}\,,\beta_{2}\,,\beta_{3} (\beta_{4})\,,g_{1}\,,g_{2}\,,\quad \text{for}\quad [i+j]=0(1)\,,\\ \nonumber
   &&k_{L}+k_{N^{c}}<4\,,k_{N^{c}}=2: \quad  \beta_{1}\,,\beta_{2}\,,\beta_{3} (\beta_{4})\,,g_{1}\,,g_{2}\,,g_{3}\,,\quad \text{for}\quad [i+j]=0(1)\,,\\
   &&k_{L}+k_{N^{c}}=4\,,k_{N^{c}}=2: \quad  \beta_{1}\,,\beta_{2}\,,\beta_{3}\,,\beta_{4}\,,g_{1}\,,g_{2}\,,g_{3}\,. \label{eq:number_SSN3}
 \end{eqnarray}
 We find that in the minimal case where $k_{L}+k_{N^{c}}<4\,,k_{N^{c}}<2$,
the Yukawa coupling parameters are $\beta_{1}\,,\beta_{2}\,,\beta_{3}(\beta_{4})\,,g_{1}\,,g_{2}$. Using the seesaw formula in Eq.~\eqref{eq:seesaw},
we obtain that there are one overall factor parameter and three complex
coupling parameters in $M_{\nu}$. If gCP symmetry is imposed, all coupling parameters would be real, resulting in four real parameters in $M_{\nu}$. Including the charged lepton sector, the minimal number of free real parameters of the lepton model is $7+3+2=12$ without gCP symmetry, or $4+3+2=9$ with gCP symmetry.

\item{$\rho_{L}=\mathbf{3}^{i}\,,\rho_{N^{c}}=\mathbf{2}\oplus \mathbf{1}^{j}$}

  In this case the Dirac and Majorana neutrino mass terms can be written
as:
  \begin{equation}
    \mathcal{L}_{\nu}=\mathcal{L}^{Y}_{\nu_{D}}+\mathcal{L}^{M}_{N^c}\,,
  \end{equation}
where
 \begin{eqnarray}\nonumber
    \mathcal{L}^{Y}_{\nu_{D}}&=&\left[\beta_{1}\left((N^c_DL)_{\mathbf{3}^{j}}Y_{\mathbf{3}^{j}}^{(k_{1})}\right)_{\bf{1}}+\beta_{2}\left((N^c_DL)_{\mathbf{3}^{[j+1]}}Y_{\mathbf{3}^{[j+1]}}^{(k_{1})}\right)_{\bf{1}}+\beta_{3}(N^c_3LY_{\mathbf{3}^{[i+j]}}^{k_{2}})_{\bf{1}}\right]H+\text{h.c.}\,,\\
  \nonumber  \mathcal{L}^{M}_{N^c}&=&\Big[g_{1}\left((N^c_DN^{c}_D)_{\mathbf{1}}Y_{\mathbf{1}}^{(2k_{N^{c}_{D}})}\right)_{\bf{1}}+g_{2}\left((N^c_DN^{c}_D)_{\mathbf{2}}Y_{\mathbf{2}}^{(2k_{N^{c}_{D}})}\right)_{\bf{1}}\\
  &~&~+2g_{3}\left(N^c_DN^{c}_3Y_{\mathbf{2}}^{(k_{N^{c}_{D}}+k_{N^{c}_{3}})}\right)_{\bf{1}}+g_{4} \left(N^c_3N^{c}_3Y_{\mathbf{1}}^{(2k_{N^{c}_{3}})}\right)_{\mathbf{1}}\Big]\Lambda +\text{h.c.}~~\,.
  \end{eqnarray}
The  neutrino Dirac mass matrix and the heavy RH neutrino Majorana
mass matrix are given by:
\begin{eqnarray} \nonumber
M_{\nu_{D}}&=& \left(\begin{matrix}
2\beta_{1}Y_{\mathbf{3}^{i},1}^{(k_{N_{D}^{c}}+k_{L})}& -\beta_{1}Y_{\mathbf{3}^{i},3}^{(k_{N_{D}^{c}}+k_{L})}+\sqrt{3}\beta_{2}Y_{\mathbf{3}^{[i+1]},2}^{(k_{N_{D}^{c}}+k_{L})}& -\beta_{1}Y_{\mathbf{3}^{i},2}^{(k_{N_{D}^{c}}+k_{L})}+\sqrt{3}\beta_{2}Y_{\mathbf{3}^{[i+1]},3}^{(k_{N_{D}^{c}}+k_{L})}\\
-2\beta_{2}Y_{\mathbf{3}^{[i+1]},1}^{(k_{N_{D}^{c}}+k_{L})}& \sqrt{3}\beta_{1}Y_{\mathbf{3}^{i},2}^{(k_{N_{D}^{c}}+k_{L})}+\beta_{2}Y_{\mathbf{3}^{[i+1]},3}^{(k_{N_{D}^{c}}+k_{L})}& \sqrt{3}\beta_{1}Y_{\mathbf{3}^{i},3}^{(k_{N_{D}^{c}}+k_{L})}+\beta_{2}Y_{\mathbf{3}^{[i+1]},2}^{(k_{N_{D}^{c}}+k_{L})}\\
\beta_{3}Y_{\mathbf{3}^{[i+j]},1}^{(k_{N_{3}^{c}}+k_{L})}& \beta_{3}Y_{\mathbf{3}^{[i+j]},3}^{(k_{N_{3}^{c}}+k_{L})}& \beta_{3}Y_{\mathbf{3}^{[i+j]},2}^{(k_{N_{3}^{c}}+k_{L})}
\end{matrix}\right)v
  \label{eq:2+1 and 3}\,,\\
M_{N^c}&=&\left(\begin{matrix}
g_{1}Y_{\mathbf{1}}^{(2k_{N^{c}_{D}})}-g_{2}Y_{\mathbf{2},1}^{(2k_{N^{c}_{D}})}&g_{2}Y_{\mathbf{2},2}^{(2k_{N^{c}_{D}})}&g_{3} Y_{\mathbf{2},1}^{(k_{N^{c}_{D}}+k_{N^{c}_{3}})}\\
g_{2}Y_{\mathbf{2},2}^{(2k_{N^{c}_{D}})}&g_{1}Y_{\mathbf{1}}^{(2k_{N^{c}_{D}})}+g_{2}Y_{\mathbf{2},1}^{(2k_{N^{c}_{D}})}&g_{3} Y_{\mathbf{2},2}^{(k_{N^{c}_{D}}+k_{N^{c}_{3}})}\\
g_{3} Y_{\mathbf{2},1}^{(k_{N^{c}_{D}}+k_{N^{c}_{3}})}&g_{3} Y_{\mathbf{2},2}^{(k_{N^{c}_{D}}+k_{N^{c}_{3}})} &g_{4} Y_{\mathbf{1}}^{(2k_{N^{c}_{3}})}
\end{matrix}\right)\Lambda\label{eq:Majorana2+1}\,.
\end{eqnarray}
The number of the coupling constants depends on the values of modular weight of $L$, $N^{c}_{D}$ and $N^{c}_{3}$:
\begin{eqnarray} \nonumber
   &&k_{L}+k_{N_{D}^{c}}<4: \quad  \beta_{1}(\beta_{2})\,,\beta_{3}\,,g_{1}\,,g_{2}\,,g_{3}\,,g_{4}\,,\quad \text{for}\quad i=0(1)\,,\\
   &&k_{L}+k_{N_{D}^{c}}=4: \quad  \beta_{1}\,,\beta_{2}\,,\beta_{3}\,,g_{1}\,,g_{2}\,,g_{3}\,,g_{4}\,, \label{eq:number_SSN21}
 \end{eqnarray}
where we have required that the rank of $M_{\nu}$ is $3$.
In minimal scenario where $k_{L}+k_{N_{D}^{c}}<4$,
the Yukawa coupling parameters are $\beta_{1}(\beta_{2})\,,\beta_{3}\,,g_{1}\,,g_{2}\,,g_{3}\,,g_{4}$. Applying the seesaw formula in Eq.~\eqref{eq:seesaw}, we get that there are one overall factor parameter and four complex coupling parameters in $M_{\nu}$. If the gCP holds, all coupling parameters become real leaving $5$ parameters in $M_{\nu}$. Including the charged lepton sector, the minimal number of the free real parameters of the lepton flavour model is $9+2+3=14$ or $5+2+3=10$ for the cases without or with gCP symmetry, respectively.

\end{itemize}


\subsection{Summary of Models}

In sections~\ref{sec:charged_lepton_sector}, \ref{sec:neutrino_sector_WO} and section~\ref{sec:neutrino_sector_SS}, we have separately discussed the assignments and resulting mass matrices of the charged leptons and neutrinos. In concrete lepton models, the assignments of representations and modular weights of $L$ in the charged lepton sector must be consistent with those in the neutrino sector. In this work, we focus on the case where the left-handed leptons transform as a triplet of $S_{4}$, identical in both the charged lepton and neutrino sectors. The modular weight $k_{L}$ is fixed for different neutrino mass matrices as shown in Table~\ref{tab:Weinberg_operator_model},
Table~\ref{tab:seesaw_neutrino_SSN2} and Table~\ref{tab:seesaw_neutrino_SSN11}. In the charged lepton part, $k_{L}$
is less constrained, while $k_{E^{c}_{\alpha}}+k_{L}$ is fixed as shown in
Table~\ref{tab:charged_lepton_model}. Consequently, all the charged
lepton mass matrices provided in Table~\ref{tab:charged_lepton_model}
can be combined with the neutrino mass matrices given in
Tables~\ref{tab:Weinberg_operator_model},~\ref{tab:seesaw_neutrino_SSN2}
and Table~\ref{tab:seesaw_neutrino_SSN11}.

As mentioned in section~\ref{sec:charged_lepton_sector}, we are concerned with the modular forms $Y^{(k)}_{\mathbf{r}}$ with weights $-4\leq k \leq 4$. There are three real Yukawa coupling parameters in $M_{\ell}$. The total number of free parameters in lepton models for different neutrino mass generation mechanisms have been given in Eqs.~\eqref{eq:number_WO}, ~\eqref{eq:number_SSN2},~\eqref{eq:number_SSN11},~\eqref{eq:number_SSN3}, and
Eq.~\eqref{eq:number_SSN21}. Here we summarize them in
Table~\ref{Tab:number_lepton_model}. We choose to perform analyses for the
``minimal'' models, i.e., the models with the smallest number of constant
parameters. From Table~\ref{Tab:number_lepton_model}, we can find that
the ``minimal'' models contain 7 (8) real parameters including $\text{Re}(\tau)$ and $\text{Im}(\tau)$ if the gCP symmetry
is (not) imposed. As for the ``minimal'' models, if neutrino masses are
generated via the Weinberg operator, we can get $20\times 4 =80$ pairs of
$(M_{\ell}, M_{\nu})$, as given in Table~\ref{tab:charged_lepton_model}
and Table~\ref{tab:Weinberg_operator_model}. For the case of the
type-I seesaw mechanism, from Table~\ref{tab:seesaw_neutrino_SSN2} and Table~\ref{tab:seesaw_neutrino_SSN11}, we find that there are
$20\times 4\times 5 =400$ and $20\times 9 =180$ combinations of $(M_{\ell}, M_{\nu})$ corresponding to $\rho_{N^{c}}=\mathbf{2}$ and $\rho_{N^{c}}=\mathbf{1}^{j_{1}}\oplus \mathbf{1}^{j_{2}}$ respectively.
Thus, there are a total of $660$ lepton models containing 7(8) free real
parameters if the gCP symmetry  is (not) imposed. With these constructed
``minimal'' lepton models, we will perform a numerical analysis in the next
section.

\begin{table}[h!]
\centering
\begin{tabular}{|c|c|c|c|}
\hline  \hline
& Representation  & Constraint & Number of free parameters\\ \hline
  \multirow{2}{*}{WO} & \multirow{2}{*}{$\rho_{L}=\mathbf{3}^{i}$} & $k_{L}<2$ & 8 (7) \\ \cline{3-4}
  & & $k_{L}=2$ & 10 (8) \\ \hline
  \multirow{8}{*}{SS} & \multirow{2}{*}{$\rho_{L}=\mathbf{3}^{i}$, $\rho_{N^{c}}=\mathbf{2}$} & $k_{N^{c}}+k_{L}<4$ & 8 (7) \\ \cline{3-4}
     & & $k_{N^{c}}+k_{L}=4$ & 10 (8) \\ \cline{2-4}
     & \multirow{2}{*}{$\rho_{L}=\mathbf{3}^{i}$, $\rho_{N^{c}}=\mathbf{1}^{j_1}\oplus\mathbf{1}^{j_2}$} & $[j_{1}+j_{2}]=1$ & 8 (7)\\ \cline{3-4}
     & & $[j_{1}+j_{2}]=0$ & 10 (8) \\ \cline{2-4}
   & \multirow{3}{*}{$\rho_{L}=\mathbf{3}^{i}$, $\rho_{N^{c}}= \mathbf{3}^{j}$} & $k_{N^{c}}+k_{L}<4\,,k_{N^{c}}<2$ & 12 (9) \\ \cline{3-4}
  & & $k_{N^{c}}+k_{L}<4\,,k_{N^{c}}=2$ & 14 (10) \\ \cline{3-4}
  & & $k_{N^{c}}+k_{L}=4\,,k_{N^{c}}=2$ & 16 (11) \\ \cline{2-4}
     & \multirow{2}{*}{$\rho_{L}=\mathbf{3}^{i}$, $\rho_{N^{c}}=\mathbf{2}\oplus\mathbf{1}^{j}$} & $k_{N_{D}^{c}}+k_{L}<4$ & 14 (10)\\ \cline{3-4}
     & & $k_{N_{D}^{c}}+k_{L}=4$ & 16 (11)\\
  \hline \hline
\end{tabular}
\caption{\label{Tab:number_lepton_model}Number of free independent real parameters in models containing modular forms of weights $|k|\leq 4$. Here ``WO'' denotes the cases that neutrino mass is described by Weinberg operator, and ``SS'' refers to these cases neutrino mass is generated by seesaw mechanism. }
\end{table}
\section{\label{sec:numerical}Numerical analysis method}

We have systematically constructed lepton flavour models
based on $S_{4}$ modular symmetry. In this section, we will perform a numerical analysis of some of these models. We choose to perform such analyses for the ``minimal'' models, i.e., the models with the smallest number of constant parameters. It turns out that the minimal phenomenologically viable models depend on 7 (8) real parameters including $\text{Re}(\tau)$ and $\text{Im}(\tau)$ if gCP is (not) imposed. These models lead to experimenatlly testable predictions for the neutrino observables which have not been experimentally determined yet: the neutrino mass ordering, the value of the lightest neutrino mass, the Dirac and Majorana CP violation (CPV) phases, and correspondingly, for the sum of neutrino masses $\sum_i m_i$ and neutrinoless double beta decay effective Majorana mass $m_{\beta\beta}$. Thus, the models we will consider are falsifiable.

For each lepton flavor model, it is necessary to verify whether the model can reproduce the input data within the experimental uncertainties. To achieve this, we conduct a $\chi^2$ analysis of the proposed fermion models, considering both normal ordering (NO) and inverted ordering (IO) for the neutrino mass spectrum.
The $\chi^{2}$ function is adopted in its standard form:
\begin{equation}
\chi^2=\sum^{n}_{i=1}\left(\frac{P_i(x)-\mu_i}{\sigma_i}\right)^2\,,
\end{equation}
where the vector $x$ contains the model parameters, $P_{i}(x)$ are the model
predictions for the observables, $\mu_{i}$ and $\sigma_{i}$ denote the central
values and standard deviations of the corresponding quantities obtained
from experimental data, see Table~\ref{tab:lepton-data}.
For lepton models, we fit seven dimensionless physical observables:
$\theta_{12}$, $\theta_{13}$, $\theta_{23}$, $\delta_{CP}$,
$m_{e}/m_{\mu}$, $m_{\mu}/m_{\tau}$, and $\Delta m_{21}^{2}/\Delta m_{31}^{2}$.
The mass of the electron $m_{e}$ and the solar neutrino mass squared difference $\Delta m_{21}^{2}$ can be fixed by the overall mass scale parameters of
respectively the charged lepton and neutrino mass matrices.

We adopt the standard parametrization of the
Pontecorvo, Maki, Nakagawa and Sakata (PMNS)
lepton mixing matrix~\cite{ParticleDataGroup:2024cfk}:
\begin{equation}\label{eq:PMNS}
U=\left(\begin{array}{ccc}
c_{12}c_{13}  &   s_{12}c_{13}   &   s_{13}e^{-i\delta_{CP}}  \\
-s_{12}c_{23}-c_{12}s_{13}s_{23}e^{i\delta_{CP}}   &  c_{12}c_{23}-s_{12}s_{13}s_{23}e^{i\delta_{CP}}  &  c_{13}s_{23}  \\
s_{12}s_{23}-c_{12}s_{13}c_{23}e^{i\delta_{CP}}   & -c_{12}s_{23}-s_{12}s_{13}c_{23}e^{i\delta_{CP}}  &  c_{13}c_{23}
\end{array}\right)\text{diag}(1,e^{i\frac{\alpha_{21}}{2}},e^{i\frac{\alpha_{31}}{2}})\,,
\end{equation}
where $c_{ij}\equiv \cos\theta_{ij}$, $s_{ij}\equiv \sin\theta_{ij}$, $\delta_{CP}$ is the Dirac CP violation (CPV) phase, and $\alpha_{21,31}$ are Majorana CPV phases \cite{Bilenky:1980cx}.
The CPV phases are CP-conserving if they are multiples of $\pi$.
 The Dirac phase $\delta_{CP}$, as is well known,
can cause CP-violating effects in
neutrino oscillations, i.e., a difference between the probabilities of the
$\nu_l \rightarrow \nu_{l'}$ and $\bar{\nu}_l \rightarrow \bar{\nu}_{l'}$
oscillations, $l\neq l' = e,\mu,\tau$.
The magnitude of the CPV effects in neutrino oscillations
depends, in particular,
on the rephasing invariant $J^{lep}_{CP}$ of the PMNS matrix
associated with $\delta_{CP}$ \cite{Krastev:1988yu}.
In the standard parameterisation of the PMNS matrix
the $J^{lep}_{CP}$ invariant has the form:
\begin{equation}
 J^{lep}_{CP} =
\dfrac{1}{8}\,\sin2\theta_{12}\,\sin2\theta_{23}\,\sin2\theta_{13}\,
\cos\theta_{13}\,\sin\delta_{CP}\,.
\label{eq:JCP}
\end{equation}
The $J^{lep}_{CP}$ invariant is a leptonic analog of the
invariant in the quark sector introduced by
Jarlskog \cite{Jarlskog:1985ht}~\footnote{Note, however, that the CPV effects in neutrino oscillations
depend on additional factor, which has oscillatory dependence on the energy of neutrinos, and the distance traveled, and involves the neutrino mass squared differences~\cite{Krastev:1988yu}. No analogous factor is present in the quark CPV observables. }.

It follows from Table~\ref{tab:lepton-data} that  $\delta_{CP}$ is poorly constrained by the existing data. We note also that the $3\sigma$ C.L. interval of allowed values of $\sin^2\theta_{23}$ is relatively wide. If the lightest neutrino is massless, there is a single Majorana phase
$\phi$, and the diagonal phase matrix in the above equation can be replaced
by $\text{diag}(1,e^{i\phi/2},1)$. Information on the Majorana phases could
potentially be provided by the neutrinoless double beta decay ($0\nu\beta\beta$) experiments (see, e.g., \cite{Pascoli:2005zb}). If the $0\nu\beta\beta$ decay is generated by the exchange of the three virtual light Majorana neutrinos, the decay amplitude is proportional to the effective Majorana neutrino mass $m_{\beta\beta}$,
\begin{equation}
  m_{\beta\beta}=|m_{1}\cos^{2}\theta_{12}\cos^{2}\theta_{13}+m_{2}\sin^{2}\theta_{12}\cos^{2}\theta_{13}e^{i\alpha_{21}}+m_{3}\sin^{2}\theta_{13}e^{i(\alpha_{31}-2\delta_{CP})}|\,,
\end{equation}
which in the case massless lightest neutrino reduces to
\begin{equation}
  m_{\beta\beta}=\left\{ \begin{aligned} &|m_{2}\sin^{2}\theta_{12}\cos^{2}\theta_{13}e^{i\phi}+m_{3}\sin^{2}\theta_{13}e^{-2i\delta_{CP}}|,& m_{1}=0\;({\rm NO})\,,
      \\ &|m_{1}\cos^{2}\theta_{12}\cos^{2}\theta_{13}+m_{2}\sin^{2}\theta_{12}\cos^{2}\theta_{13}e^{i\phi}|\,, & m_{3}=0\;({\rm IO})\,.\end{aligned} \right.
\end{equation}
We will consider also the kinematical mass $m_{\beta}$, information about which is obtained in the  beta decay experiments. It is defined as:
\begin{equation}
m_{\beta}=\left(m_{1}^{2}\cos^{2}\theta_{12}\cos^{2}\theta_{13}+m_{2}^{2}\sin^{2}\theta_{12}\cos^{2}\theta_{13}+m_{3}^{2}\sin^{2}\theta_{13}\right)^{1/2}\,.
\end{equation}
Given that in the considered lepton flavour models the
number of parameters describing the neutrino sector
is smaller than the number of the described observables
and that all observables depend on the VEV of the modulus $\tau$,
there are unusual correlations between observables that are unique to flavour theories  based on modular invariance \cite{Novichkov:2018ovf}.
More specifically, the predicted values of the Dirac and
Majorana CPV phases $\delta$ and $\alpha_{21,31}$
and of the effective Majorana mass $m_{\beta\beta}$, may be correlated with $\sin^2\theta_{23}$,
the prediction for the sum of neutrino masses $\sum_i m_i$
may be correlated with the predicted value of the Dirac
CPV phase $\delta$~\footnote{Note, for example, that $m_{\beta\beta}$ does not depend explicitly on $\sin^2\theta_{23}$, and that the CPV phases,
$\sum_i m_i$ and $\sin^2\theta_{23}$ are physically very different observables.}, etc. We will show examples of such unusual correlations between the neutrino observables in each of the statistically analyzed models.

\begin{table}[t!]
\centering
\begin{tabular}{| c | c | c |} \hline \hline
Observable & Central value and $1\sigma$ error & $3\sigma$ range \\ \hline
$m_e/m_\mu $ & $0.004737 $ & $-$  \\
$m_\mu/m_\tau$ & $0.05882$ & $-$  \\
$m_e/{\rm MeV}$ & $0.469652$ & $-$ \\
$\Delta m_{21}^2 / 10^{-5}\text{eV}^2$ & $7.41^{+0.21}_{-0.20}$ & $[6.81\,, 8.03]$   \\
$\Delta m_{31}^2 / 10^{-3}\text{eV}^2$(NO) & $2.505^{+0.024}_{-0.026}$ & $[2.426\,, 2.586]$  \\
$\Delta m_{32}^2 / 10^{-3}\text{eV}^2$(IO) & $-2.487^{+0.027}_{-0.024}$ & $[-2.566\,, -2.407]$     \\ \hline
$\delta_{CP}/\pi$(NO) & $1.289^{+0.217}_{-0.139}$ & $[0.772\,, 1.944]$  \\
$\delta_{CP}/\pi$(IO) & $1.517^{+0.144}_{-0.133}$ & $[1.083\,, 1.900]$  \\
$\sin^2\theta_{12}$ $(\text{NO} \;\&\; \text{IO})$ & $0.307^{+0.012}_{-0.011}$ & $[0.275\,, 0.344]$   \\
$\sin^2\theta_{13}$(NO) & $0.02224^{+0.00056}_{-0.00057}$ & $[0.02047\,, 0.02397]$  \\
$\sin^2\theta_{13}$(IO) & $0.02222^{+0.00069}_{-0.00057}$ & $[0.02049\,, 0.02420]$   \\
$\sin^2\theta_{23}$(NO) & $0.454^{+0.019}_{-0.016}$ & $[0.411\,, 0.606]$  \\
$\sin^2\theta_{23}$(IO) & $0.568^{+0.016}_{-0.021}$ & $[0.412\,, 0.611]$  \\ \hline \hline
\end{tabular}
\caption{
\label{tab:lepton-data}
The central values and the $1\sigma$ errors of the  mass ratios, mixing angles and Dirac CP violation phase in the lepton sector. The central
values of the charged lepton mass ratios are taken from~\cite{Xing:2007fb}.
When scanning the parameter space of our models we set the uncertainties of the charged lepton mass ratios to be $0.1\%$ of their central value. We adopt the values of the lepton mixing parameters from NuFIT v5.3 with
Super-Kamiokanda atmospheric data for normal ordering (NO) and inverted ordering (IO) of neutrino masses~\cite{Esteban:2020cvm}. }
\end{table}

The minimization of the $\chi^{2}$ function is performed using the
CERN package \texttt{TMinuit}~\cite{minuit}. The parameter space for the
Yukawa couplings $g_{i}$ is constrained as follows: $|g_{i}| \in [0, 10^{5}]$
and $\text{arg}(g_{i}) \in [0, 2\pi)$. The complex modulus $\tau$ is
restricted to the fundamental domain $\mathcal{F}$,
defined by $|\text{Re}\,\tau| \leq \frac{1}{2}$, $\text{Im}\,\tau > 0$,
and $|\tau| \geq 1$. A lepton model is considered phenomenologically viable
if the predictions for the neutrino masses and mixing parameters at
the $\chi^{2}$ minimum fall within the corresponding $3\sigma$ ranges
listed in Table~\ref{tab:lepton-data}. We impose the bound on the neutrino mass sum $m_1+m_2+m_3<0.12$ eV from Planck collaboration~\cite{Planck:2018vyg}. Additionally, we require that the
predicted charged lepton masses do not deviate from the experimental central
values by more than $0.3\%$. By performing a $\chi^{2}$ analysis on all
660 "minimal" lepton flavor models, we can identify a substantial
but significantly smaller number of phenomenologically viable models.
All viable models and their corresponding best-fit results for lepton
observables are summarized in Tables~\ref{tab:lepton_res_par7_WO_NO},~\ref{tab:lepton_res_par7_SSN2_NO},~\ref{tab:lepton_res_par7_SSN2_IO},~\ref{tab:lepton_res_par7_SSN11_NO} and Table~\ref{tab:lepton_res_par7_SSN11_IO} in Appendix~\ref{sec:app_viable_models}.

\section{\label{sec:benchmark-models}Example models
for lepton masses and mixing}

By performing a $\chi^{2}$ analysis on the constructed lepton flavour models, we can obtain
a large number of phenomenologically viable models based on the polyharmonic Maa{\ss} forms of level 4, with the corresponding finite modular group being $\Gamma_4\cong S_4$.
It is beyond the scope of our study to explore all the viable cases in detail and to present a complete graphical treatment of each model's predictions. In what follows we consider three representative cases in which the quality of the results can be thoroughly appreciated. No flavons are introduced in these models. The VEV of the modulus $\tau$ is the only source of breaking of the flavour (modular) symmetry. We also investigate the possibility that it is the sole source of CP symmetry breaking\cite{Novichkov:2019sqv}.

\subsection{\label{subsec:model-Weinberg}Neutrino masses from Weinberg operator}

The light neutrino masses are generated by the effective Weinberg
operator in this model. The assumed modular weight and representation assignments of lepton
fields are summarized as follows:
\begin{eqnarray}
\rho_{E_1^c}=\bm{1}\,,\,\rho_{E_2^c}=\bm{1}\,,\,\rho_{E_3^c}=\bm{1}\,,\,\rho_L=\bm{3}\,,~~~k_{E_1^c}=-4\,,\,k_{E_2^c}=2\,,\,k_{E_3^c}=4\,,\,k_{L}=0\,,
\end{eqnarray}
which corresponds to the combination
$\mathcal{C}_{6}-\mathcal{W}_{3}$, where
$\mathcal{C}_{6}$ and $\mathcal{W}_{3}$ are defined in
Table~\ref{tab:charged_lepton_model} and
Table~\ref{tab:Weinberg_operator_model},
respectively.
With these assignments, the modular-invariant Lagrangian for the charged
lepton Yukawa interaction and the Weinberg operator takes the following form:
\begin{eqnarray}
\nonumber -\mathcal{L}^{Y}_{\ell} &=& \alpha (E^c_1 L Y^{(-4)}_{\bm{3}}H^*)_{\bm{1}}  + \beta (E^c_2 L Y^{(2)}_{\bm{3}}H^*)_{\bm{1}}  + \gamma (E^c_3 L Y^{(4)}_{\bm{3}}H^*)_{\bm{1}}  + \text{h.c.}\,, \\
\label{eq:Lag-Weinberg}
\mathcal{L}^{M}_\nu &=& \dfrac{1}{2\Lambda} (LLHH Y^{(0)}_{\bm{1}})_{\bm{1}} + \dfrac{g}{2\Lambda} (LL HH Y^{(0)}_{\bm{2}})_{\bm{1}} + \text{h.c.}\,.
\end{eqnarray}
The phases of the constant parameters
$\alpha$, $\beta$, $\gamma$, and $\Lambda$
can be absorbed by redefining the lepton fields and consequently they can
be taken as real without loss of generality, while the coupling $g$ is complex,
in general. From Eq. (\ref{eq:Lag-Weinberg}), we can read out the charged lepton and neutrino mass matrices:
\begin{eqnarray}
\nonumber M_e&=&\begin{pmatrix}
\alpha Y_{\bm{3},1}^{(-4)} ~& \alpha Y_{\bm{3},3}^{(-4)} ~& \alpha Y_{\bm{3},2}^{(-4)} \\
\beta Y_{\bm{3},1}^{(2)} ~& \beta Y_{\bm{3},3}^{(2)} ~& \beta Y_{\bm{3},2}^{(2)} \\
\gamma Y_{\bm{3},1}^{(4)} ~& \gamma Y_{\bm{3},3}^{(4)} ~& \gamma Y_{\bm{3},2}^{(4)}
\end{pmatrix}v\,,\\ \label{eq:M_nu_WO}
M_{\nu}&=&\begin{pmatrix}
Y_{\bm{1}}^{(0)} + 2g Y_{\bm{2},1}^{(0)} ~& 0 ~& 0 \\
0 ~& \sqrt{3}gY_{\bm{2},2}^{(0)} ~&  Y_{\bm{1}}^{(0)} - g Y_{\bm{2},1}^{(0)}\\
0 ~& Y_{\bm{1}}^{(0)} - g Y_{\bm{2},1}^{(0)} ~& \sqrt{3}gY_{\bm{2},2}^{(0)}
\end{pmatrix}\dfrac{v^2}{\Lambda}\,,
\end{eqnarray}
where $v=\langle H^{0}\rangle$ is the VEV of the Standard Model Higgs field,
$v = 174$ GeV.

\begin{figure}[h!]
\centering
\includegraphics[width=0.92\textwidth]{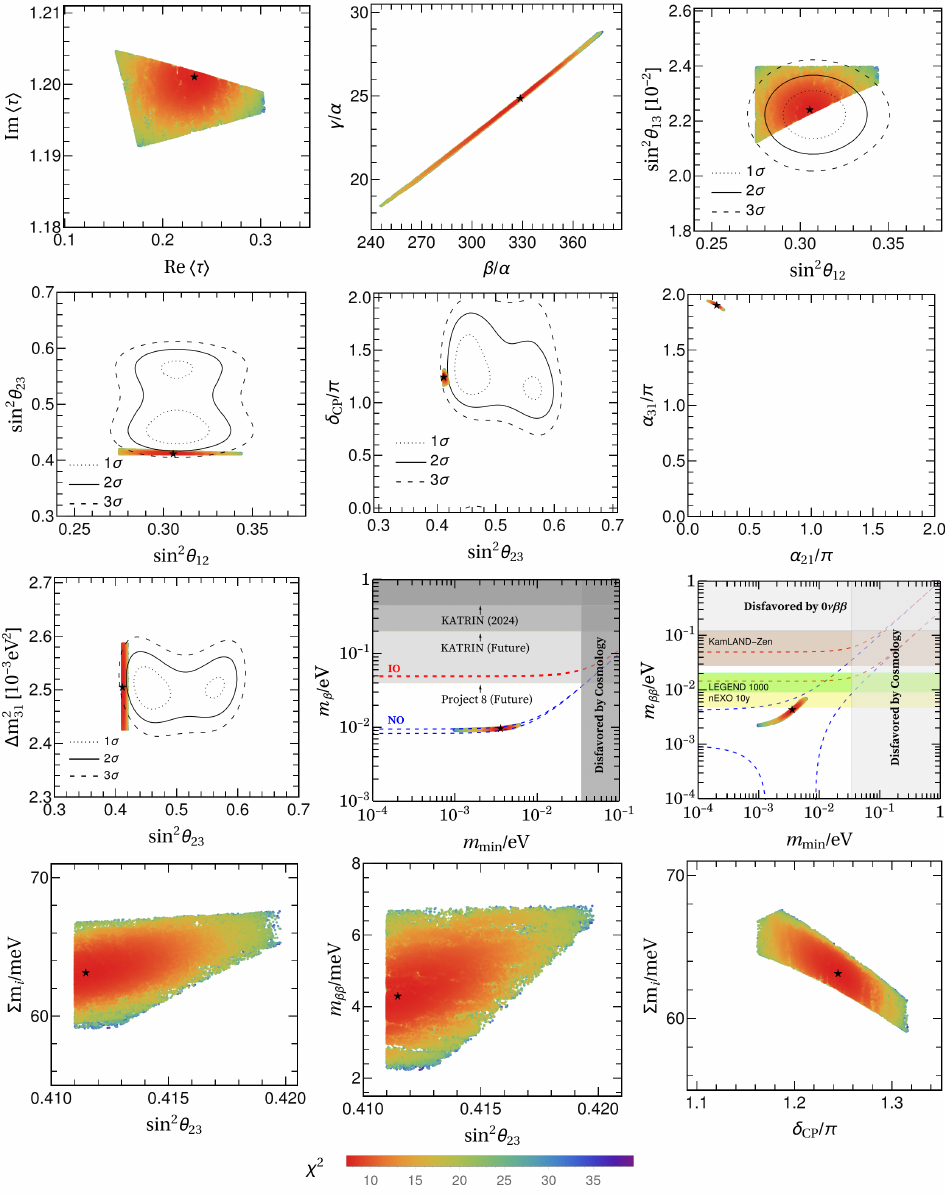}
\caption{\label{fig:model_lepton_7para_WO}
Predictions for correlations between the input free constant parameters, neutrino mixing angles, CP violation phases and neutrino masses in the lepton model where the neutrino masses are generated via the effective Weinberg operator and the gCP symmetry is imposed. The best fit values of the input parameters and lepton observables are indicated by black stars. The gray shaded regions represent the current KATRIN upper bound ($m_{\beta}<0.45\,\text{eV}$ at $90\%$ CL)~\cite{Katrin:2024tvg}, future KATRIN sensitivity ($m_{\beta}<0.2\,\text{eV}$ at $90\%$ CL)~\cite{KATRIN:2021dfa} and Project 8 future sensitivity ($m_{\beta}<0.04\,\text{eV}$)~\cite{Project8:2022wqh} respectively. In the two panels for $m_{\beta}$ and $m_{\beta\beta}$, the blue (red) dashed lines represent the most general allowed regions for NO (IO) neutrino mass spectrum, where the neutrino oscillation parameters are varied within their $3\sigma$ ranges~\cite{Esteban:2020cvm}. The vertical band disfavored by cosmology arises from the neutrino mass sum $\sum_i m_{i}<0.12$ eV by Planck~\cite{Planck:2018vyg}.}
\end{figure}

The charged lepton mass matrix $M_e$ involves three real constants
$\alpha$, $\beta$, and $\gamma$, which can be adjusted to reproduce the
charged lepton masses. The neutrino mass matrix $M_{\nu}$ depends on
the complex coupling $g$ and an overall scale factor $v^2/\Lambda$,
in addition to the complex modulus $\tau$. If gCP symmetry is imposed, the parameter $g$ will be constrained to be real. We search for the minimum of the $\chi^2$ function constructed with the data in Table~\ref{tab:lepton-data}, and we find that the experimental data on lepton masses and mixing angles can only be accommodated by the NO mass spectrum. The best fit values of the input parameters and lepton flavor observables are found to be:
\begin{eqnarray}
\label{eq:bf_WO_WCP}
\nonumber &&\langle\tau\rangle = 0.2323 + 1.2011 i \,,\,\beta/\alpha = 328.6763\,,\,\gamma/\alpha = 24.8490\,,\\
\nonumber && g = 2.6594\,,\, \alpha v= 3.8895\,{\rm MeV}\,,\, \dfrac{v^2}{\Lambda}= 18.6332\,{\rm meV}\,, \\
\nonumber &&\sin^2\theta_{12}=0.305\,,~~\sin^2\theta_{13}=0.02241\,,~~\sin^2\theta_{23}=0.411\,, \\
\nonumber &&\delta_{CP}=1.245\pi\,,~~\alpha_{21}=0.234\pi\,,~~\alpha_{31}=1.904\pi\,,\\
&& m_1=3.619\,{\rm meV}\,,~~m_2=9.338\,{\rm meV}\,,~~m_3=50.181\,{\rm meV}\,,
\end{eqnarray}
with $\chi^2_{\text{min}}= 7.28$. These predictions are in excellent agreement with experimental data. The central values of the charged lepton masses are again exactly reproduced. Using the best fit values given in Eq.~\eqref{eq:bf_WO_WCP} we find the following values for $\sum_i m_i$, $J^{lep}_{CP}$ and  $m_{\beta\beta}$ in this case:
\begin{equation}
\sum_i m_i = 63.137~{\rm meV}\,,~~
J^{lep}_{CP} = -\,0.023\,,~~
m_{\beta\beta} = 4.297~{\rm meV}\,.
\label{eq:summbb4.5}
\end{equation}
The corresponding 3$\sigma$ allowed intervals of $\sum_i m_i$, $J^{lep}_{CP}$
and $m_{\beta\beta}$ are:
\begin{equation}
\label{eq:nu-masses-Weinberg}
\hskip -0.1in\sum_i m_i \in \left[59.332\text{meV},67.571\text{meV}\right]\,,
J^{lep}_{CP}\in \left[-0.0296, -0.0153\right]\,,
m_{\beta\beta} \in \left[2.233\text{meV},6.817\text{meV}\right]\,.
  \end{equation}
In Figure~\ref{fig:model_lepton_7para_WO}, we show correlations between
the input free constant parameters, the neutrino masses, and the neutrino
mixing observables predicted in this model.

\begin{figure}[h!]
\centering
\includegraphics[width=0.98\textwidth]{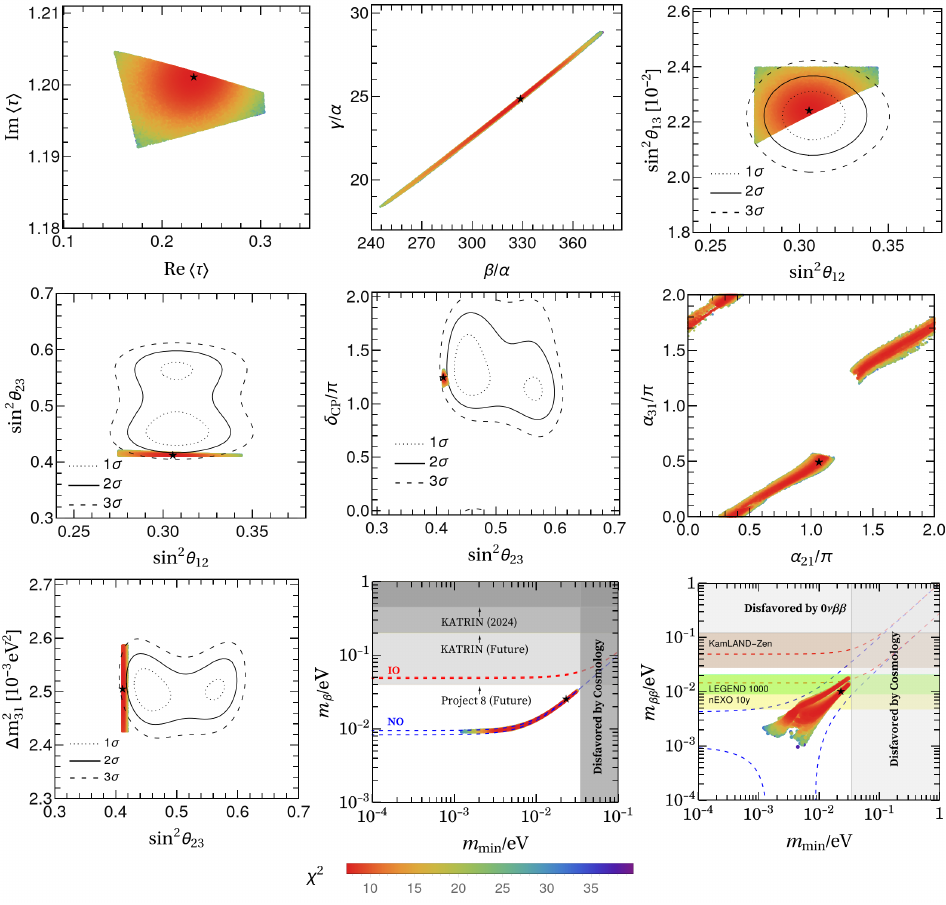}
\caption{\label{fig:model_lepton_7para_WO_WOCP}
The same as in Figure~\ref{fig:model_lepton_7para_WO} but for the case without gCP symmetry. }
\end{figure}

In the case where the gCP symmetry does not hold, the coupling $g$ is a complex parameter. We find that a much better description of the data can be achieved for the NO mass spectrum than the IO case. At the best-fit point for the IO neutrino masses spectrum, the neutrino mass sum $\sum_i m_i\simeq1.856$ eV which significantly exceeds the upper limit 0.12 eV from Planck~\cite{Planck:2018vyg}, and we obtain values of solar and atmospheric mixing angles $\sin^{2}\theta_{12}=0.500$ and $\sin^{2}\theta_{23}=0.403$, which are outside the corresponding $3\sigma$ ranges of experimental data as
given in Table~\ref{tab:lepton-data}.
The best fit values of the input parameters and lepton flavor observables
for the NO neutrino mass spectrum are found to be:
\begin{eqnarray}
\label{eq:bf_WO_WOCP}
\nonumber &&\langle\tau\rangle = 0.2323 + 1.2011 i \,,\,\beta/\alpha = 328.6761\,,\,\gamma/\alpha = 24.8489\,,\\
\nonumber && |g| = 1.6711\,,\, \text{arg}(g) =0.2837\,\pi\,,\, \alpha v= 3.8895\,{\rm MeV}\,,\, \dfrac{v^2}{\Lambda}= 29.6532\,{\rm meV}\,, \\
\nonumber &&\sin^2\theta_{12}=0.305\,,~~\sin^2\theta_{13}=0.02241\,,~~\sin^2\theta_{23}=0.411\,, \\
\nonumber &&\delta_{CP}=1.245\pi\,,~~\alpha_{21}=1.062\pi\,,~~\alpha_{31}=0.491\pi\,,\\
\label{eq:fit-model1-NO}&& m_1=23.350\,{\rm meV}\,,~~m_2=24.886\,{\rm meV}\,,~~m_3=55.228\,{\rm meV}\,,
\end{eqnarray}
%
with $\chi^2_{\text{min}}= 7.28$.
These predictions are in agreement with experimental data as well.
The central values of the charged lepton masses are exactly reproduced.
As discussed in the end of this section, there are many other (actually infinite number of) values of $|g|$ and $\text{arg}(g)$ leading to the same $\chi^2_{\text{min}}$, so the values given in Eq.~\eqref{eq:fit-model1-NO}
are representative of all the best fit points having the same $\chi^2$. Using the best fit values of neutrino masses, mixing angles and
Dirac and Majorana CP violation phases from Eq.~\eqref{eq:fit-model1-NO}, we get the following predictions for the sum of neutrino masses
$\sum_i m_i$,
the $J^{lep}_{CP}$ factor and the neutrinoless double beta decay effective Majorana mass $m_{\beta\beta}$:
\begin{equation}
\sum_i m_i = 103.464~{\rm meV}\,,~~J^{lep}_{CP} = -\,0.023\,,~~
m_{\beta\beta} = 9.907~{\rm meV}\,.
\label{eq:summbb4.4}
\end{equation}
By using the sampler \texttt{MultiNest}~\cite{Feroz:2007kg,Feroz:2008xx}
to scan the parameter space and considering the Planck bound $\sum_i m_i<0.12$ eV~\cite{Planck:2018vyg}, we find that the predicted 3$\sigma$ allowed intervals of $\sum_i m_i$, $J^{lep}_{CP}$, and $m_{\beta\beta}$ are:
\begin{equation}\hskip-0.1in
\sum_i m_i \in \left[59.332\,\text{meV}, 120\,\text{meV}\right]\,,
J^{lep}_{CP}\in \left[-0.0296, -0.0153\right]\,,
m_{\beta\beta} \in \left[0.961\,\text{meV}, 18.013\,\text{meV}\right]\,.
\end{equation}
The current experimental bound on $m_{\beta\beta}$, provided by the KamLAND-Zen experiment, is $m_{\beta\beta}<(28\sim122)\,$meV~\cite{KamLAND-Zen:2024eml}. Future large-scale $0\nu\beta\beta$-decay experiments, such as LEGEND-1000~\cite{LEGEND:2021bnm}, aim to improve the sensitivity to $m_{\beta\beta} < (9 \sim 21)\,$meV, while nEXO~\cite{nEXO:2021ujk} expects to achieve $m_{\beta\beta} < (4.7 \sim 20.3)\,$meV. These forthcoming experiments have the potential to test the predictions of this model (for a review of the potential of the future planned neutrinoless double beta decay experiments see, e.g.,~\cite{bbonuReview:Neutrino2024}). In Figure~\ref{fig:model_lepton_7para_WO_WOCP}, we show correlations between the
model free constant parameters, the neutrino masses and the neutrino
mixing observables predicted in this model. The best-fit values of the input parameters and lepton observables are indicated by black stars.


We note that the values of $\chi^{2}_{\text{min}}$ are almost the same in
the two considered versions of the model, both without gCP symmetry and with
gCP symmetry. However, the best-fit values of the three neutrino masses
and the Majorana CPV phases differ significantly in the two cases.
As a consequence, the values of $\sum_i m_i$ and  $m_{\beta\beta}$ predicted
in the two cases, also differ significantly. These differences in the
predicted values of the two observables, especially the difference in the predicted values of $\sum_i m_i$, can be used to distinguish experimentally
between the two cases. The reason that the $\chi^{2}_{\text{min}}$ values are almost the same in the cases without and with gCP symmetry is that the
diagonalization matrix of the neutrino mass matrix $M_{\nu}$
has a special form. The light neutrino mass matrix $M_{\nu}$ given in Eq.~\eqref{eq:M_nu_WO} can be diagonalized as
\begin{equation}\label{eq:diag_m_nu}
  U_{\nu}^{T}M_{\nu}U_{\nu}=\widehat{M}_{\nu}=\text{diag}(m_{1},m_{2},m_{3})\,,
\end{equation}
%
with the three light neutrino masses are given as
\begin{eqnarray} \nonumber
  &&m_{1}=\frac{v^{2}}{\Lambda}\left|\sqrt{3}gY_{\mathbf{2},2}^{(0)}-\eta\left(Y_{\mathbf{1}}^{(0)}-gY_{\mathbf{2},1}^{(0)}\right)\right|\,,\\ \nonumber
  &&m_{2}=\frac{v^{2}}{\Lambda}\left|\sqrt{3}gY_{\mathbf{2},2}^{(0)}+\eta\left(Y_{\mathbf{1}}^{(0)}-gY_{\mathbf{2},1}^{(0)}\right)\right|\,,\\ \label{eq:neutrino_mass}
  &&m_{3}=\frac{v^{2}}{\Lambda}\left|Y_{\mathbf{1}}^{(0)}+2gY_{\mathbf{2},1}^{(0)}\right|\,.
\end{eqnarray}
where $\eta\equiv\text{sign}\left(\text{Re}\left[gY_{\mathbf{2},2}^{(0)}\left(Y_{\mathbf{1}}^{(0)}-gY_{\mathbf{2},1}^{(0)}\right)^{*}\right]\right)$.
The diagonalization matrix $U_{\nu}$ is given as
\begin{equation}
  U_{\nu}=\left( \begin{array}{ccc} 0 ~&~ 0 ~&~ 1 \\
                    \frac{-\eta}{\sqrt{2}} ~&~ \frac{\eta}{\sqrt{2}} ~&~0\\
                    \frac{1}{\sqrt{2}} ~&~ \frac{1}{\sqrt{2}} ~&~0\\ \end{array} \right).\begin{pmatrix}e^{-i\rho_{1}/2} & 0 & 0\\
                 0 & e^{-i\rho_{2}/2} & 0\\
                 0 & 0 & e^{-i\rho_{3}/2}
               \end{pmatrix}\,.
\label{eq:UnuW}
\end{equation}
where
\begin{eqnarray}
\nonumber&&\rho_{1}=\text{arg}\left(\sqrt{3}gY_{\mathbf{2},2}^{(0)}-\eta\left(Y_{\mathbf{1}}^{(0)}-gY_{\mathbf{2},1}^{(0)}\right)\right)\,,\\
\nonumber&&\rho_{2}=\text{arg}\left(\sqrt{3}gY_{\mathbf{2},2}^{(0)}+\eta\left(Y_{\mathbf{1}}^{(0)}-gY_{\mathbf{2},1}^{(0)}\right)\right)\,,\\
&&\rho_{3}=\text{arg}\left(Y_{\mathbf{1}}^{(0)}+2gY_{\mathbf{2},1}^{(0)}\right)\,.
\end{eqnarray}
It is not difficult to show using the $q$-expansion of $Y_{\bm{2},1}^{(0)}(\tau)$ and $Y_{\bm{2},2}^{(0)}(\tau)$ that
i) both $Y_{\bm{2},1}^{(0)}(\tau)$ and $Y_{\bm{2},2}^{(0)}(\tau)$
are real functions (see Eq.~\eqref{eq:Y02a}), and that
ii) for any $\tau =x+i\,y$, up to corrections $\mathcal{O}(few\times 10^{-4})$, $Y_{\bm{2},1}^{(0)}(\tau)$ and $Y_{\bm{2},2}^{(0)}(\tau)$ are given by the simple expressions shown in Eq.~\eqref{eq:Y02b}. This follows from the explicit forms of the $q$-expansions of $Y_{\bm{2},1}^{(0)}(\tau)$ and $Y_{\bm{2},2}^{(0)}(\tau)$ and the fact that in the fundamental domain of the modular group $y\geq \sqrt{3}/2$. Since $Y_{\bm{1}}^{(0)} = 1$ and
both $Y_{\bm{2},1}^{(0)}(\tau)$ and $Y_{\bm{2},2}^{(0)}(\tau)$ are real,
in the case of a real constant $g$, the CP-violation in the PMNS neutrino mixing matrix originates from the unitary matrix $U_e$ diagonalising the charged lepton mass matrix, which in turn is generated by the CP-violating
VEV of $\tau$. Moreover, given the form of $U_\nu$, the contribution of $U_e$ to the PMNS matrix $U = U^\dagger_e\,U_\nu$ is crucial (both in the cases of complex and real $g$) for obtaining in the fit the correct values of
the three neutrino mixing angles, as well as the predicted CP-violating value of the Dirac phase $\delta_{CP}$. The requirement of reproducing correctly the values of these observables fixes the value of the VEV of $\tau$.
This implies, in particular, that there should be correlations between
the values of some of the mixing angles, and between some of the angles
and $\delta_{CP}$. Indeed, such correlations are shown to
take place in Figure~\ref{fig:model_lepton_7para_WO_WOCP}.

In the case when the gCP symmetry holds, $g$ is real, $Y_{\bm{2},1}^{(0)}(\tau)$ and $Y_{\bm{2},1}^{(0)}(\tau)$ are also real
and the phases $\rho_{1}$, $\rho_{2}$, $\rho_{3}$ are equal to 0 or $\pi$, note that $\rho_{1}=\rho_{2}=\rho_{3}=0$ at the best fit point. Given the value of $\langle\tau\rangle$, the value of $g$ is determined by the measured value of the ratio $\Delta m^2_{21}/\Delta m^2_{31}$. For the best-fit values of $\langle\tau\rangle$ in Eq.~\eqref{eq:bf_WO_WCP} and of the ratio in Table~ \ref{tab:lepton-data}, using the fact that $Y_{\bm{1}}^{(0)} = 1$ and
calculating the values of $Y_{\bm{2},1}^{(0)}(\tau)$ and $Y_{\bm{2},2}^{(0)}(\tau)$ at $\langle\tau\rangle$ from Eq.~\eqref{eq:Y02b},
we get $g=2.6594$. The Majorana phases $\alpha_{21}$ and $\alpha_{31}$
get relatively small contributions from $U_e$, which shifts them somewhat from the CP-conserving values $0$ and $\pi$.

\begin{figure}[ht!]
\centering
\includegraphics[width=0.55\textwidth]{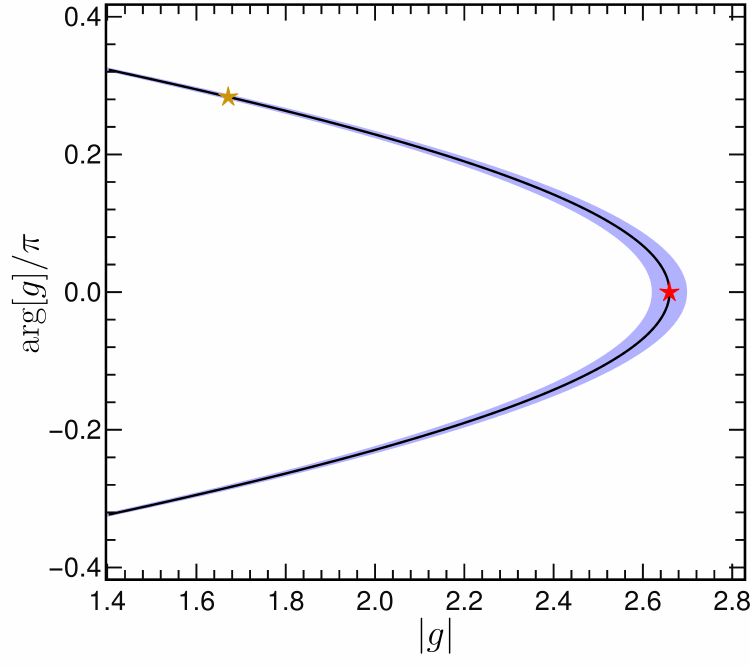}
\caption{\label{fig:region}
The region of the complex $g$ compatible with experimental data in the
case where the gCP symmetry does not hold. The other input parameters are
fixed at their best-fit values given in Eq.~\eqref{eq:bf_WO_WOCP}.
The black line indicates the values of the complex parameter
$g$ that lead to $\Delta m_{21}^{2}/m_{31}^{2} = 0.02958$
and have the same $\chi^2_{\text{min}}=7.28$ as the representative best fit value of $g$ quoted in Eq.~\eqref{eq:bf_WO_WOCP}. The yellow star corresponds
to this representative best-fit complex value of $g$. The red star indicates the best-fit point in the case of imposed gCP symmetry. On all the points on the contour the Planck constraint $m_1+m_2+m_3 < 0.12$ eV is satisfied.
The width of the contour accounts for the uncertainties in the determination of the values of $g$. See text for further details.}
\end{figure}

As it follows from Eq.~\eqref{eq:UnuW}, in the case of complex $g$
the diagonalization matrix $U_{\nu}$ of the neutrino mass matrix $M_{\nu}$ depends on the sign factor $\eta$ and a phase matrix $\rho=\text{diag}(e^{-i\rho_{1}/2},e^{-i\rho_{2}/2},e^{-i\rho_{3}/2})$.
The phase matrix $\rho$  influences only the values of the two
Majorana CP-violation phases $\alpha_{21}$ and $\alpha_{31}$,
but not the three lepton mixing angles and the Dirac CP-violation phase
$\delta_{\text{CP}}$. For the $\chi^{2}$ analysis, we fit seven dimensionless
physical observables: $\theta_{12}$, $\theta_{13}$, $\theta_{23}$,
$\delta_{\text{CP}}$, $m_{e}/m_{\mu}$, $m_{\mu}/m_{\tau}$ and
$\Delta m_{21}^{2}/\Delta m_{31}^{2}$. The variable $\eta$ has two discrete values $1$ and $-1$: $\eta={\rm sign} (\text{Re}\left[gY_{\mathbf{2},2}^{(0)}\left(Y_{\mathbf{1}}^{(0)}-gY_{\mathbf{2},1}^{(0)}\right)^{*}\right])$.
Given the best fit values of $\braket{\tau}$, $\beta/\alpha$ and $\gamma/\alpha$ in Eq.~\eqref{eq:bf_WO_WOCP}, we find that the predicted values of $\theta_{12}$, $\theta_{13}$, $\theta_{23}$, $\delta_{\text{CP}}$, $m_{e}/m_{\mu}$ and $m_{\mu}/m_{\tau}$ given in Eq.~\eqref{eq:bf_WO_WOCP} can always be obtained as long as $\eta=1$. This fact indicates that the value of $g$ will not influence the determination of $\theta_{12}$, $\theta_{13}$, $\theta_{23}$ and $\delta_{\text{CP}}$. However, different values of $g$ will generate distinct predictions of neutrino masses $m_{1}$, $m_{2}$, $m_{3}$,
and the Majorana CP-violation phases $\alpha_{21}$, $\alpha_{31}$, as
Eqs.~\eqref{eq:bf_WO_WOCP} and \eqref{eq:bf_WO_WCP} show.
In order to illustrate the impact of $g$ on the physical observable,
we plot in Figure~\ref{fig:region} the allowed  values of the complex $g$, which are compatible with the experimental data at $3\sigma$ C.L. Using the approximate values of $Y_{\bm{2},1}^{(0)}(\tau)$ and $Y_{\bm{2},2}^{(0)}(\tau)$  given by
the simple expressions shown in Eq.~\eqref{eq:Y02b} at the best value of $\langle\tau\rangle$ obtained in the fit, $\langle\tau\rangle=0.2323 + 1.2011 i$, and the experimental value of the ratio $\Delta m^2_{31}/\Delta m^2_{21}$ from Table \ref{tab:lepton-data}, we find the following constraint on the
complex constant $g = |g|e^{i\,\phi}$: $2.659\cos\phi-|g| = 0$.
This constraint (including the relevant uncertainties) is shown in
Figure~\ref{fig:region}, where the bound on the sum of neutrino masses
$m_1+m_2+m_3 < 0.12$ eV from Planck collaboration ~\cite{Planck:2018vyg}
has been included. The black solid line is the contour plot for the minimal
$\chi^2_{\text{min}}=7.28$ in the plane $\text{arg}(g)$ versus $|g|$, on which $m_1+m_2+m_3 < 0.12$ eV is satisfied. All the points on the black contour line give the same predictions for
the lepton mixing angles and the Dirac CPV phase
and reproduce correctly the best-fit value of the experimentally
determined ratio $\Delta m_{21}^{2}/\Delta m_{31}^{2} = 0.02958$.
However, the predictions for the light neutrino masses
and the Majorana CPV phases change with the point,
with the sum of neutrino masses and the effective Majorana mass
varying in the ranges
$\sum_i m_i \in\left[63.138\,\text{meV}, 120\,\text{meV}\right]$ and
$m_{\beta\beta} \in \left[3.658\,\text{meV}, 17.509\,\text{meV}\right]$.

We note finally that the model, in both its considered versions, provides a very specific prediction for $\sin^2 \theta_{23}$, namely, $\sin^2 \theta_{23} \in \left[0.411, 0.420\right]$ at $3\sigma$ C.L., as well as very specific, rather strong correlations between the values of $\sin^2 \theta_{12}$ and $\sin^2 \theta_{13}$, $\sin^2 \theta_{23}$ and $\delta_{\text{CP}}$, as well as weaker ones between $\sum_i m_i$, $m_{\beta\beta}$, and $\sin^2 \theta_{23}$, and $\sum_i m_i$ and $\delta_{\text{CP}}$. Critical tests of the viability of the model will be provided by the planned high-precision measurements of $\sin^2 \theta_{23}$ as well as of $\delta_{\text{CP}}$ at the T2HK~\cite{Hyper-Kamiokande:2018ofw} and DUNE~\cite{DUNE:2020ypp} experiments under construction and at the discussed ESS$\nu$SB experiment~\cite{Alekou:2022emd}. The predicted value of $\sin^2 \theta_{23}$ might also be probed at the currently operative T2K~\cite{T2K:2023smv} and NO$\nu$A~\cite{NOvA:2021nfi} experiments.

\subsection{\label{subsec:model-seesaw_N11}
Neutrino masses from
minimal seesaw with RH neutrinos transforming as singlets of $S_4$ }

As a benchmark model for the case of Majorana neutrino masses generated
via the type-I seesaw mechanism, we take the following
$S_{4}$ representation and modular-weight assignments:
\begin{eqnarray}
\nonumber && \rho_{E^c_1}=\bm{1}\,,~~ \rho_{E^c_2}=\bm{1}\,,~~ \rho_{E^c_3}=\bm{1}' \,,~~ \rho_L=\bm{3} \,,~~ \rho_{N^c_1}=\bm{1}\,,~~ \rho_{N^c_2}=\bm{1}'\,,\\
&& k_{E^c_1}=-6 \,,~~k_{E^c_2}=-4 \,,~ k_{E^c_3}=2 \,, ~~ k_L=2\,, ~~ k_{N^c_1}=0\,, ~~ k_{N^c_2}= 2\,,
\end{eqnarray}
which corresponds to the combination $\mathcal{C}_{11}-\mathfrak{D}_{9}-\mathfrak{N}_{12}$, where $\mathcal{C}_{11}$ and $\mathfrak{D}_{9}-\mathfrak{N}_{12}$ are defined in Table~\ref{tab:charged_lepton_model} and Table~\ref{tab:seesaw_neutrino_SSN11} respectively.
The modular invariant Lagrangian for the lepton masses is given by:
\begin{eqnarray}
\nonumber -\mathcal{L}^{Y}_{\ell} &=& \alpha (E^c_1 L Y^{(-4)}_{\bm{3}}H^*)_{\bm{1}}  + \beta (E^c_2 L Y^{(-2)}_{\bm{3}}H^*)_{\bm{1}}  + \gamma (E^c_3 L Y^{(4)}_{\bm{3}'}H^*)_{\bm{1}}  + \text{h.c.}\,, \\
-\mathcal{L}_\nu &=& g_1 (N^c_1 L H Y^{(2)}_{\bm{3}})_{\bm{1}}   + g_2 (N^c_2 L Y^{(4)}_{\bm{3}}H)_{\bm{1}}  + \frac{1}{2}\Lambda_1 N^c_1 N^c_1 Y^{(0)}_{\bm{1}} + \dfrac{1}{2} \Lambda_2 N^c_2 N^c_2 Y^{(4)}_{\bm{1}} +\text{h.c.}~\,.
\end{eqnarray}
The charged lepton mass matrix, the neutrino Dirac mass matrix and
the heavy Majorana neutrino mass matrix read:
\begin{eqnarray}
\nonumber M_e &=& \begin{pmatrix}
\alpha Y^{(-4)}_{\bm{3},1} ~& \alpha Y^{(-4)}_{\bm{3},3} ~& \alpha Y^{(-4)}_{\bm{3},2} \\
\beta Y^{(-2)}_{\bm{3},1} ~& \beta Y^{(-2)}_{\bm{3},3} ~& \beta Y^{(-2)}_{\bm{3},2} \\
\gamma Y^{(4)}_{\bm{3}',1} ~& \gamma Y^{(4)}_{\bm{3}',3} ~& \gamma Y^{(4)}_{\bm{3}',2}
\end{pmatrix}v \,,  ~~
M_N =\begin{pmatrix}
\Lambda_1 Y^{(0)}_{\bm{1}} ~& 0 \\
0 ~& \Lambda_2 Y^{(4)}_{\bm{1}}
\end{pmatrix}\,, \\
M_D &=& \begin{pmatrix}
g_1 Y^{(2)}_{\bm{3},1} ~&~ g_1 Y^{(2)}_{\bm{3},3} ~& g_1 Y^{(2)}_{\bm{3},2}  \\
g_2 Y^{(4)}_{\bm{3}',1} ~&~ g_2 Y^{(4)}_{\bm{3}',3} ~& g_2 Y^{(4)}_{\bm{3}',2}
\end{pmatrix} v\,.
\end{eqnarray}
The light neutrino mass matrix is given by the seesaw expression:
\begin{small}
\begin{eqnarray}
\nonumber M_\nu &=& - M_D^T M_N^{-1} M_D  \\
 &=& -\frac{g_{1}^{2}v^{2}}{\Lambda_{1}Y_{\mathbf{1}}^{(0)}}\left(\begin{matrix}
Y_{\mathbf{3} ,1}^{(2)}Y_{\mathbf{3} ,1}^{(2)}~&~Y_{\mathbf{3} ,1}^{(2)}Y_{\mathbf{3} ,3}^{(2)}~&~Y_{\mathbf{3} ,1}^{(2)}Y_{\mathbf{3} ,2}^{(2)}\\
Y_{\mathbf{3} ,3}^{(2)}Y_{\mathbf{3} ,1}^{(2)}~&~Y_{\mathbf{3} ,3}^{(2)}Y_{\mathbf{3} ,3}^{(2)}~&~Y_{\mathbf{3} ,3}^{(2)}Y_{\mathbf{3} ,2}^{(2)}\\
Y_{\mathbf{3} ,2}^{(2)}Y_{\mathbf{3} ,1}^{(2)}~&~Y_{\mathbf{3} ,2}^{(2)}Y_{\mathbf{3} ,3}^{(2)}~&~Y_{\mathbf{3} ,2}^{(2)}Y_{\mathbf{3} ,2}^{(2)}\\
\end{matrix}\right)-\frac{g_{2}^{2}v^{2}}{\Lambda_{2}Y_{\mathbf{1}}^{(4)}}\left(\begin{matrix}
Y_{\mathbf{3}' ,1}^{(4)}Y_{\mathbf{3}' ,1}^{(4)}~&~Y_{\mathbf{3}' ,1}^{(4)}Y_{\mathbf{3}' ,3}^{(4)}~&~Y_{\mathbf{3}' ,1}^{(4)}Y_{\mathbf{3}' ,2}^{(4)}\\
Y_{\mathbf{3}' ,3}^{(4)}Y_{\mathbf{3}' ,1}^{(4)}&Y_{\mathbf{3}' ,3}^{(4)}Y_{\mathbf{3}' ,3}^{(4)}&Y_{\mathbf{3}' ,3}^{(4)}Y_{\mathbf{3}' ,2}^{(4)}\\
Y_{\mathbf{3}' ,2}^{(4)}Y_{\mathbf{3}' ,1}^{(4)}&Y_{\mathbf{3}' ,2}^{(4)}Y_{\mathbf{3}' ,3}^{(4)}&Y_{\mathbf{3}' ,2}^{(4)}Y_{\mathbf{3}' ,2}^{(4)}\\
\end{matrix}\right)\,.~~~
\end{eqnarray}
\end{small}
We see that the light neutrino mass matrix $M_{\nu}$ depends on two
combinations of parameters $\frac{g_{2}^{2}\Lambda_{1}}{g_{1}^{2}\Lambda_{2}}$,
$\frac{g_{1}^{2}v^{2}}{\Lambda_{1}}$ as well as on the modulus $\tau$. The overall phase in each mass matrix is unphysical, consequently one may choose both $\alpha$ and $g^2_1/\Lambda_1$ to be real without loss of genrality. Thus this model describes all the lepton masses and mixing observables in terms of
8 real parameters including
$\text{Re}\,\tau$ and $\text{Im}\,\tau$.
A correct description of the experimental data can only be achieved for the NO neutrino mass spectrum. The best-fit values of the input parameters and the lepton flavor observables are determined to be:
%
\begin{eqnarray}
\nonumber && \langle \tau \rangle =   0.3875 + 1.2615 i\,,
~\beta/\alpha = 8.7553\,, ~\gamma/\alpha = 0.0152\,,  \\
\nonumber && \Big|\frac{g_{2}^{2}\Lambda_{1}}{g_{1}^{2}\Lambda_{2}}\Big| = 3.9618\,,~\text{arg}\left(\frac{g_{2}^{2}\Lambda_{1}}{g_{1}^{2}\Lambda_{2}}\right)=0.3263\,\pi \,,~\alpha v = 0.2538\,{\rm GeV}\,, ~ \frac{g_{1}^{2}v^{2}}{\Lambda_{1}}= 5.6129\,{\rm meV}\,, \\
\nonumber && \sin^2\theta_{12}=0.306 \,,~ \sin^2\theta_{13}=0.02224\,,~ \sin^2\theta_{23}=0.454\,,~ \delta_{CP}=1.385 \pi\,,~ \phi =0.489\pi\,, \\
&&m_1 = 0\,{\rm meV}\,,m_2 = 8.608\,{\rm meV}\,,~ m_2 = 50.048\,{\rm meV}\,,~ \chi^2_{\text{min}}=0.207\,. \label{eq:seesaw1}
\end{eqnarray}

In the case where the gCP symmetry holds, all couplings would be
constrained to be real.
In this case the lepton flavours are described by 7 real parameters: 3 real constants $\alpha$, $\beta$, $\gamma$ for the 3 charged lepton masses, 2 real constants $g^2_1/\Lambda_1$, $g^2_2/\Lambda_2$ together with $\text{Re}\,\tau$ and $\text{Im}\,\tau$ describe the 9 observables
in the neutrino sector.
We find the experimental data of lepton masses and mixing angles
can only be accommodated for NO mass spectrum in this case. The best fit values of the input constant parameters and the lepton flavor observables are found to be:
\begin{eqnarray}
\nonumber && \langle \tau \rangle =  0.3954 + 1.2427 i\,,~\beta/\alpha = 8.8011\,, ~\gamma/\alpha = 0.0143\,,  \\
\nonumber && \frac{g_{2}^{2}\Lambda_{1}}{g_{1}^{2}\Lambda_{2}} = 3.8866 \,,~\alpha v = 0.2614\,{\rm GeV}\,, ~ \frac{g_{1}^{2}v^{2}}{\Lambda_{1}}= 5.4314\,{\rm meV}\,, \\
\nonumber && \sin^2\theta_{12}=0.303 \,,~ \sin^2\theta_{13}=0.02227\,,~ \sin^2\theta_{23}=0.455\,, ~ \delta_{CP}=1.406 \pi\,,~ \phi =0.827\pi\,,\\
&& m_1 = 0\,{\rm meV}\,, ~ m_2 = 8.608\,{\rm meV}\,,~ m_3 = 50.045\,{\rm meV}\,,~ \chi^2_{\text{min}}=0.403\,.
\label{eq:seesaw1gCP}
\end{eqnarray}
%
The sum of neutrino masses $\sum_i m_i$,
the $J^{lep}_{CP}$ invariant and the effective Majorana mass
$m_{\beta\beta}$, corresponding to the best-fit values of
the relevant observables quoted
in Eq.~(\ref{eq:seesaw1}) and Eq.~(\ref{eq:seesaw1gCP}) are given in the model by:
  \begin{equation}
\sum_i m_i = 58.657~(58.653){\rm meV}\,,
J^{lep}_{CP} = -\,0.0313~(-0.0319)\,,
m_{\beta\beta} = 1.982~(3.188){\rm meV}\,,
\label{eq:summbbseesaw1}
\end{equation}
where the values (values in brackets) correspond to the case of not imposed (imposed) gCP symmetry.

\begin{figure}[h!]
\centering
\includegraphics[width=0.85\textwidth]{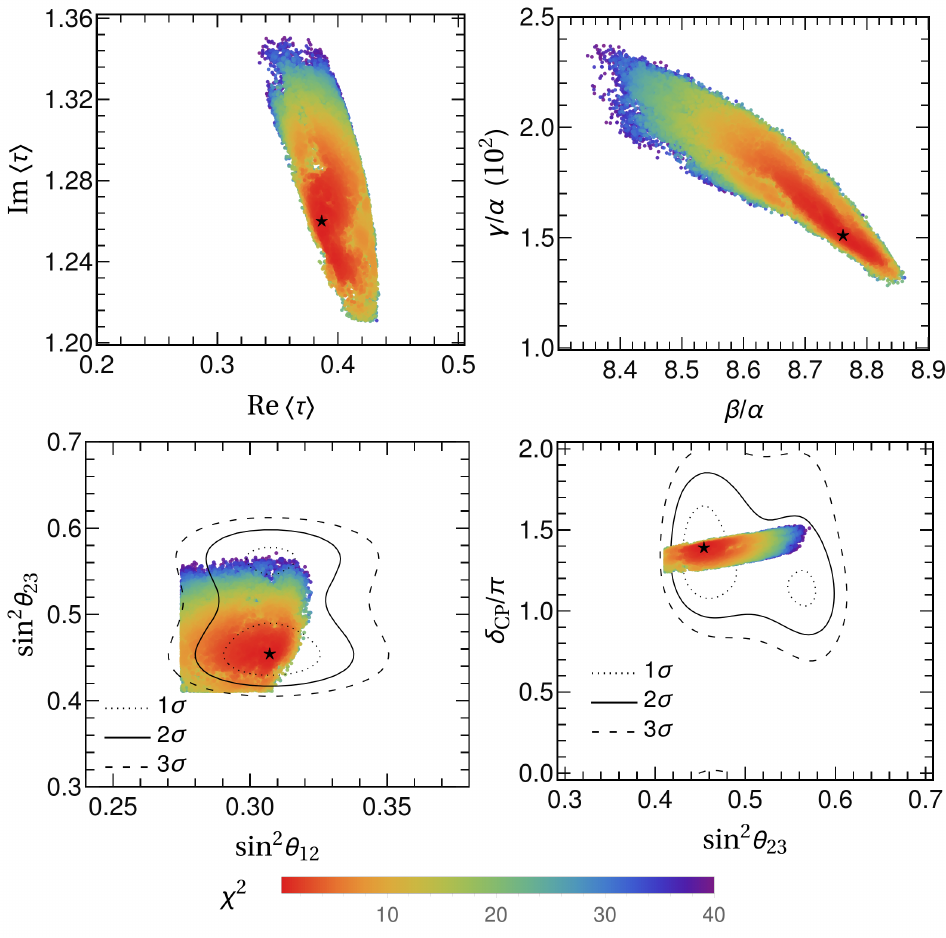}
\caption{\label{fig:model_lepton_7para_SSN11_WOCP}
Predictions for correlations between the input free constant
parameters, the neutrino mixing angles, CPV phases and neutrino masses
in the lepton model where the neutrino masses are generated via the type
I seesaw mechanism and have NO type of spectrum, and the gCP symmetry does not hold. The best fitting values of the input parameters and lepton observables are indicated by black stars. Here we consider the case of
$N^{c}\sim \mathbf{1}\oplus \mathbf{1}'$. }
\end{figure}

\begin{figure}[h!]
\centering
\includegraphics[width=0.98\textwidth]{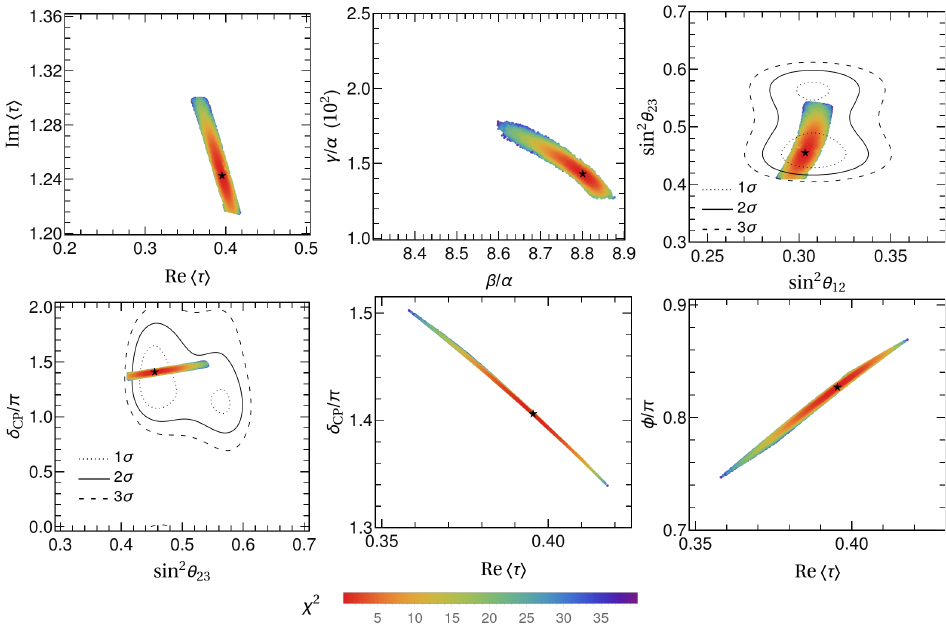}
\caption{\label{fig:model_lepton_7para_SSN11}
The same as in Figure \ref{fig:model_lepton_7para_SSN11_WOCP}
but for the version of the model with imposed gCP symmetry.}
\end{figure}

It is clear from the results shown in Eqs. (\ref{eq:seesaw1}),
(\ref{eq:seesaw1gCP}), and
(\ref{eq:summbbseesaw1}) that
distinguishing between the two versions without and with gCP symmetry of the model would be extremely difficult in the NO case.
The predicted values of the Dirac
CPV phase $\delta_{CP}$, of the $J^{lep}_{CP}$ factor and  especially of the
allowed values of $\sin^2\theta_{23}$  in the model under discussion differ from those in the model considered in the preceding subsection.
Thus, sufficiently high-precision measurements of $\sin^2\theta_{23}$
and of $\delta_{CP}$, $J^{lep}_{CP}$ as well as of $\sum_i m_i$,
will allow testing the two models and possibly distinguishing
between them. In Figures~\ref{fig:model_lepton_7para_SSN11_WOCP}
and~\ref{fig:model_lepton_7para_SSN11} we show correlations
between the input free constant  parameters,
the neutrino masses and neutrino mixing observables, predicted
for NO spectrum by the discussed model, in the cases,
respectively, without and with gCP symmetry.

\subsection{\label{subsec:model-seesaw_N2}Neutrino masses from
minimal seesaw with RH neutrinos transforming as doublet of $S_4$ }

As a benchmark model for the case of Majorana neutrino masses generated via
the type-I seesaw mechanism, we take the $S_{4}$ representation and
modular-weight assignments:
\begin{eqnarray}
\nonumber && \rho_{E^c_1}=\bm{1}\,,~~ \rho_{E^c_2}=\bm{1}\,,~~ \rho_{E^c_3}=\bm{1} \,,~~ \rho_L=\bm{3} \,,~~ \rho_{N^c}=\bm{2}\,,\\
&& k_{E^c_1}=2 \,,~~k_{E^c_2}=4 \,,~ k_{E^c_3}=6 \,, ~~ k_L=-2\,, ~~ k_{N^c}=-2\,,
\end{eqnarray}
%
which corresponds to the combination $\mathcal{C}_{10}-\mathcal{D}_{1}-\mathcal{N}_{1}$, where $\mathcal{C}_{10}$ and $\mathcal{D}_{1}-\mathcal{N}_{1}$ are defined in Table~\ref{tab:charged_lepton_model} and Table~\ref{tab:seesaw_neutrino_SSN2}, respectively. The modular-invariant charged lepton and neutrino Yukawa couplings are  given by:
\begin{eqnarray}
\nonumber -\mathcal{L}^{Y}_{\ell} &=& \alpha (E^c_1 L Y^{(0)}_{\bm{3}}H^*)_{\bm{1}}  + \beta (E^c_2 L Y^{(2)}_{\bm{3}}H^*)_{\bm{1}}  + \gamma (E^c_3 L Y^{(4)}_{\bm{3}}H^*)_{\bm{1}}  + \text{h.c.}\,, \\
-\mathcal{L}_\nu &=& g (N^c L H Y^{(-4)}_{\bm{3}})_{\bm{1}}+ \frac{1}{2}\Lambda_1 \left( N^c N^c \right)_{\bm{1}} Y^{(-4)}_{\bm{1}} + \dfrac{1}{2} \Lambda_2 \left(\left( N^c N^c\right)_{\bm{2}} Y^{(-4)}_{\bm{2}}\right)_{\bm{1}} +\text{h.c.}~\,.
\end{eqnarray}
%
Correspondingly, the charged lepton, the neutrino Dirac and the heavy Majorana neutrino mass matrices read:
\begin{eqnarray}
\nonumber M_e &=& \begin{pmatrix}
\alpha Y^{(0)}_{\bm{3},1} ~& \alpha Y^{(0)}_{\bm{3},3} ~& \alpha Y^{(0)}_{\bm{3},2} \\
\beta Y^{(2)}_{\bm{3},1} ~& \beta Y^{(2)}_{\bm{3},3} ~& \beta Y^{(2)}_{\bm{3},2} \\
\gamma Y^{(4)}_{\bm{3},1} ~& \gamma Y^{(4)}_{\bm{3},3} ~& \gamma Y^{(4)}_{\bm{3},2}
\end{pmatrix}v \,,  ~~
M_N =\begin{pmatrix}
\Lambda_1 Y^{(-4)}_{\bm{1}}-\Lambda_{2}Y^{(-4)}_{\bm{2},1}  ~& \Lambda_{2}Y^{(-4)}_{\bm{2},2} \\
\Lambda_{2}Y^{(-4)}_{\bm{2},2} ~& \Lambda_1 Y^{(-4)}_{\bm{1}}+\Lambda_{2}Y^{(-4)}_{\bm{2},1}
\end{pmatrix}\,, \\
M_D &=& \begin{pmatrix}
2g Y^{(-4)}_{\bm{3},1} ~&~ -gY^{(-4)}_{\bm{3},3} ~& -gY^{(-4)}_{\bm{3},2}\\
0 ~&~ \sqrt{3}g Y^{(-4)}_{\bm{3},2} ~& \sqrt{3}g Y^{(-4)}_{\bm{3},3}
\end{pmatrix} v\,.
\end{eqnarray}
%
The phases of $\alpha$, $\beta$, $\gamma$, $g$, $\Lambda_1$ can be removed by field redefinition, while $\Lambda_2/\Lambda_1$ is a complex parameter if gCP symmetry is not considered. This model describes successfully all the lepton masses and mixing parameters in terms of 8 real parameters including $\text{Re}\tau$ and $\text{Im}\tau$.

\begin{figure}[t!]
\centering \includegraphics[width=0.85\textwidth]{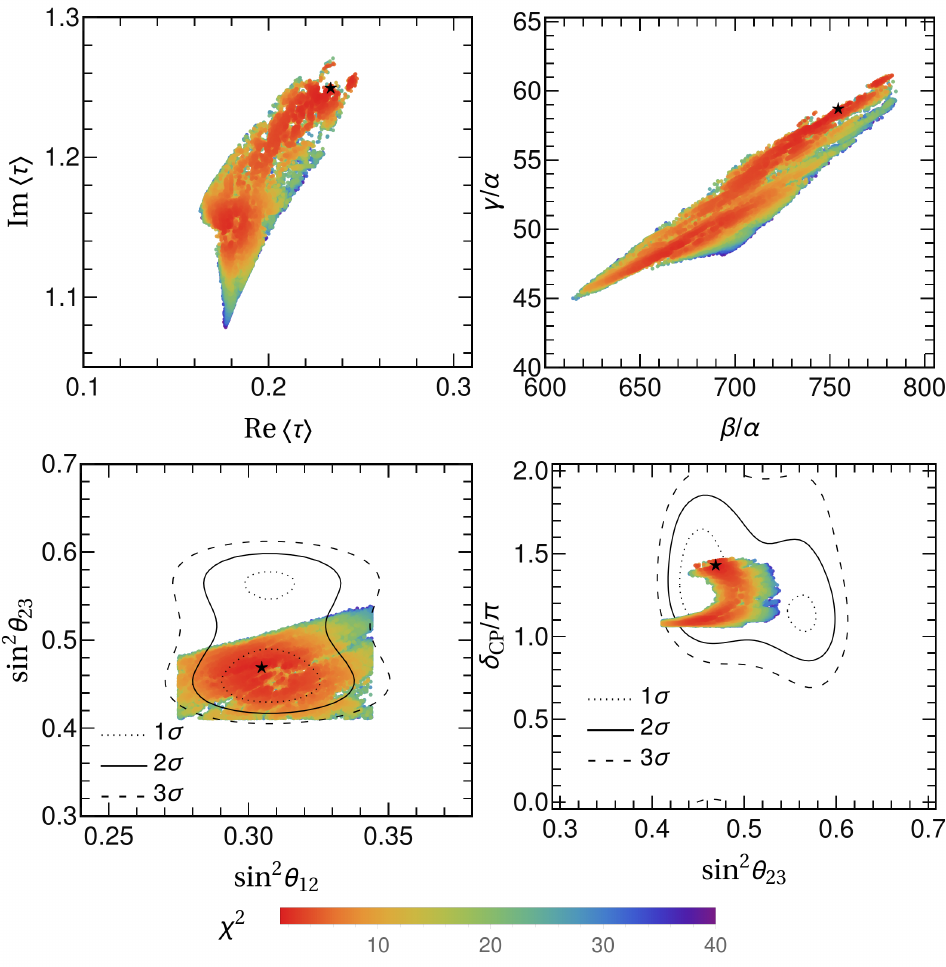}
\caption{\label{fig:model_lepton_7para_SSN2_WOCP} Predictions for correlations between the input free constant parameters, the neutrino mixing angles, CPV phases and neutrino masses in the lepton model where the neutrino masses are generated via the type I seesaw mechanism with  $N^{c}\sim {\bf 2}$ and have NO type of spectrum, and the gCP symmetry does not hold. The best fit values of the input parameters and lepton observables are indicated by black stars. }
\end{figure}

The agreement between predictions and experimental data can be achieved only for NO neutrino masses spectrum. In the case of IO neutrino masses spectrum, at the best fit point the prediction of the atmospheric mixing angle $\sin^{2}\theta_{23}=0.2224$ is outside the corresponding $3\sigma$ range $\sin^{2}\theta_{23}\in [0.412,0.611]$ as given in Table~\ref{tab:lepton-data}. For NO neutrino masses spectrum, the  best fit values of the input constant parameters and the lepton flavor observables are:
\begin{eqnarray}
\nonumber && \langle \tau \rangle =   0.2497 + 1.2685 i\,,~\beta/\alpha = 772.7069\,, ~\gamma/\alpha = 60.9505\,,  \\
\nonumber && \Big|\frac{\Lambda_{2}}{\Lambda_{1}}\Big| = 0.8925\,,~\text{arg}\left(\frac{\Lambda_{2}}{\Lambda_{1}}\right)=1.0901\,\pi \,,~\alpha v =  1.7245\times 10^{-3}\,{\rm GeV}\,, ~ \frac{g^{2}v^{2}}{\Lambda_{1}}= 32.9965\,{\rm meV}\,, \\
\nonumber && \sin^2\theta_{12}=0.306 \,,~ \sin^2\theta_{13}=0.02226\,,~ \sin^2\theta_{23}=0.456\,,~\delta_{CP}=1.460 \pi\,,~ \phi =0.301\pi\,,\\
&& m_1 = 0\,{\rm meV}\,,~ m_2 = 8.608\,{\rm meV}\,,~ m_3 = 50.073\,{\rm meV}\,,~ \chi^2_{\text{min}}=0.642\,. \label{eq:seesaw2}
\end{eqnarray}
In Figure~\ref{fig:model_lepton_7para_SSN2_WOCP}, we show correlations between the input free constant parameters, the neutrino masses, and neutrino mixing observables predicted in this model.

\begin{figure}[h!]
\centering
\includegraphics[width=0.98\textwidth]{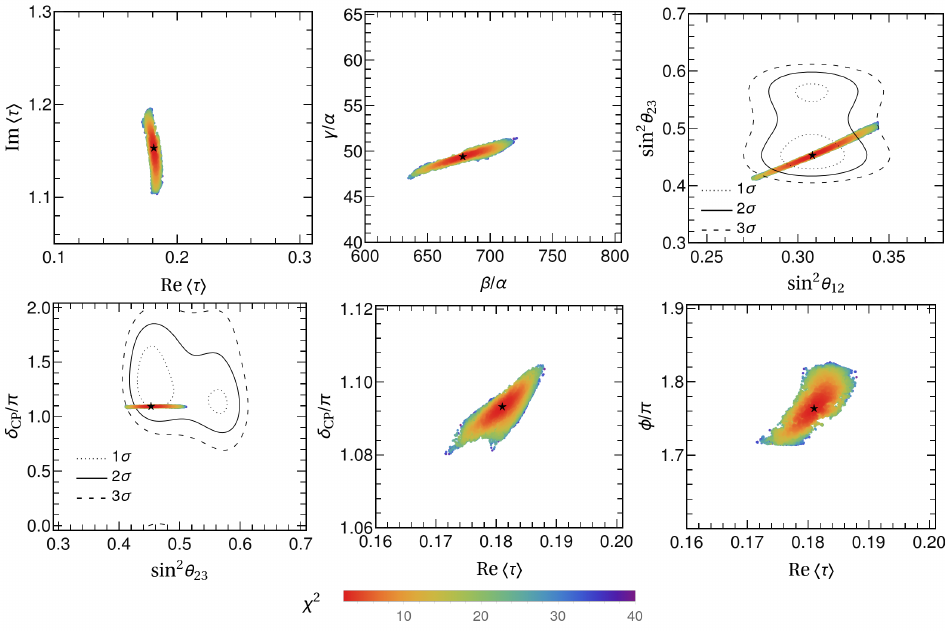}
\caption{\label{fig:model_lepton_7para_SSN2}
The same as in Figure \ref{fig:model_lepton_7para_SSN2_WOCP}
but with imposed gCP symmetry. }
\end{figure}

In the case that gCP symmetry is imposed, the parameter $\Lambda_{2}/\Lambda_1$ would be constrained to be real. Thus the model has 7 real parameters in this case: 3 real constants $\alpha$, $\beta$, $\gamma$ describing the 3 charged lepton masses and the remaining 2 real parameters $\Lambda_2/\Lambda_1$, $g^2/\Lambda_1$ and the complex modulus $\tau$ describing the 9 observables in the neutrino sector. We find that the experimental data of lepton masses and mixing angles can also be accommodated for NO neutrino mass spectrum in this case. The best fit values of the input parameters and lepton flavor observables are determined to be:
\begin{eqnarray}
\nonumber && \langle \tau \rangle =   0.1810 + 1.1528 i\,,~\beta/\alpha = 678.1592\,, ~\gamma/\alpha = 49.4148\,,  \\
\nonumber && \frac{\Lambda_{2}}{\Lambda_{1}} = -5.2401 \,,~\alpha v = 1.8203\,{\rm MeV}\,, ~ \frac{g^{2}v^{2}}{\Lambda_{1}}= 0.1746\,{\rm eV}\,, \\
\nonumber && \sin^2\theta_{12}=0.308 \,,~ \sin^2\theta_{13}=0.02223\,,~ \sin^2\theta_{23}=0.453\,,~ \delta_{CP}=1.093 \pi\,,~ \phi =1.764\pi\,,\\
&& m_1 = 0\,{\rm meV}\,, ~ m_2 = 8.608\,{\rm meV}\,,~ m_3 = 50.057\,{\rm meV}\,,~ \chi^2_{\text{min}}=2.000\,.\label{eq:seesaw2gCP}
\end{eqnarray}
In Figure~\ref{fig:model_lepton_7para_SSN2}, we show correlations between  some of the free constant parameters, the neutrino masses and neutrino mixing observables predicted in this model.

In the considered model the sum of neutrino masses $\sum_i m_i$,
the $J^{lep}_{CP}$ invariant and and the effective Majorana mass  $m_{\beta\beta}$, corresponding to the best fit
values of the relevant observables quoted
in Eq.~(\ref{eq:seesaw2}), and Eq.~(\ref{eq:seesaw2gCP}) are
given by:
\begin{equation}
\sum_i m_i = 58.681~(58.665)~{\rm meV}\,,
J^{lep}_{CP} = -\,0.0332~(-\,0.0097)\,,
m_{\beta\beta} = 1.861~(3.696){\rm meV}\,,
\label{eq:summbbseesaw2}
\end{equation}
where the values (values in brackets) correspond to the case
of not imposed (imposed) gCP symmetry.

Thus, the two versions of the model predict very different values
of $\delta_{CP}$ and of the $J^{lep}_{CP}$ invariant:
$\delta_{CP}= 1.460\pi$,   $J^{lep}_{CP} = -\,0.0332$
and $\delta_{CP}= 1.093\pi$, $J^{lep}_{CP} = -\,0.0097$.
Clearly, sufficient precise experimental measurements of
the Dirac CPV phase $\delta_{CP}$ and of the
$J^{lep}_{CP}$ factor could allow
to distinguish between the two versions without and with gCP symmetry of the model. Together with precise measurements of $\sum_i m_i$ and of
$\sin^2\theta_{23}$ they will provide a critical test of the model.

The predictions of the version of the model without gCP symmetry are very similar to those of the model considered in the preceding subsection. However, the values of $\delta_{CP}$ and  $J^{lep}_{CP}$ predicted in the version in which the gCP symmetry is imposed and of the model discussed in the preceding subsection differ significantly: $\delta_{CP}=1.460\pi$ (or $1.093\pi)$, $J^{lep}_{CP} =-\,0.0332$ (or $-\,0.0097$), to be compared with $\delta_{CP} = 1.385\pi$ (or $1.406\pi$), $J^{lep}_{CP} = -\,0.0313$ (or $-\,0.0319$). These differences in $\delta_{CP}$ and $J^{lep}_{CP}$ suggest that high precision measurements of these parameters could help distinguish between the two lepton flavor models.


\section{Conclusion\label{sec:conclusion}}

In the present study, we have explored the potential of the
non-supersymmetric modular invariance approach to the flavour problem
for lepton flavour model building. The approach is characterised by the
presence in the relevant formalism of the polyharmonic Maa{\ss} modular
forms of given level $N$, in addition to the standard modular forms of
the same level. For a fixed level $N$, the Yukawa coupling and fermion mass
matrices are expressed in terms of polyharmonic Maa{\ss} modular forms and
the standard modular forms of the level $N$ and a limited number of
constant parameters. The polyharmonic Maa{\ss} forms are non-holomorphic
modular forms. Non-trivial Maa{\ss} forms exist for zero, negative and
positive integer modular weights. The formalism of non-holomorphic modular
flavor symmetry in the framework of harmonic Maa{\ss} forms is introduced,
offering a novel avenue for understanding the flavor structure of fermions.

Using the finite modular group $S_4$ as a flavour symmetry group and assuming
that the three left-handed lepton doublets furnish a triplet irreducible
representation of $S_4$, we have constructed all possible lepton
flavour models in which the neutrino masses are generated either by the
Weinberg effective operator or by the type I seesaw mechanism and
in which Maa{\ss} forms of modular weights $-4\leq k_{Y}\leq 4$ can be
present. The independent charged lepton and neutrino masses matrices are
summarized in Table~\ref{tab:charged_lepton_model} and Tables~\ref{tab:Weinberg_operator_model}, \ref{tab:seesaw_neutrino_SSN2}, and \ref{tab:seesaw_neutrino_SSN11}. Focusing on models with minimal 7 (8) real constant parameters for the case with (without) gCP, we perform statistical analyses and
identified those that successfully describe the existing data on
the three neutrino mixing angles, the two neutrino mass squared differences
and the three charged lepton masses. We obtain predictions for each of these
viable models for the neutrino mass ordering, the absolute neutrino mass
scale, the Dirac and Majorana CP-violation phases and, correspondingly,
for the sum of neutrino masses and the neutrinoless double beta decay
effective Majorana mass. All the phenomenologically viable models as well as
the predictions for lepton observables are provided in
Tables~\ref{tab:lepton_res_par7_WO_NO},~\ref{tab:lepton_res_par7_SSN2_NO},~\ref{tab:lepton_res_par7_SSN2_IO},~\ref{tab:lepton_res_par7_SSN11_NO} and
Table~\ref{tab:lepton_res_par7_SSN11_IO}. On the basis of the predictions
thus obtained we have concluded, in particular, that:
i) a large number of the considered currently viable models would be ruled
out if it is definitely established that $\sin^2\theta_{23} \geq 0.5$ or
that $\sin^2\theta_{23} < 0.5$; a high precision determination of
$\sin^2\theta_{23}$ will further reduce the number of viable models;
ii) the very high precision measurement of $\sin^2\theta_{12}$ foreseen to be
performed by the JUNO expriment~\cite{JUNO:2022mxj}
would also reduce significantly the number of viable models;
iii) additional important tests of the models will be provided
by precision measurements of the Dirac CPV phase $\delta_{CP}$ and of
the $J^{lep}_{CP}$ factor as well as of the sum of the neutrino masses
$\sum_i m_i$. Approximately half of the models will be ruled out if
the neutrino mass spectrum is proven to be of the NO type (of the IO type).

To further illustrate our results, we have presented a very detailed
description and statistical analyses of three representative viable benchmark models: one in which neutrino masses originate from the Weinberg effective operator (Section~\ref{subsec:model-Weinberg}) and two in which they are generated by the type I seesaw mechanism
(Sections~\ref{subsec:model-seesaw_N11} and~\ref{subsec:model-seesaw_N2}).
We have considered two versions of the models: with gCP symmetry imposed
and without gCP symmetry. Each of these two version includes respectively
5 real, and 4 real and one complex, constant parameters in additon
to the complex value of the VEV of the modulus $\tau$. As in the case of
the general analysis, for each of these pairs of three models we derived
predictions for the neutrino mass ordering, the absolute neutrino mass scale,
the Dirac and Majorana CP-violation phases and, correspondingly, for the sum
of neutrino masses and the neutrinoless double beta decay effective Majorana
mass and discussed the possibility to test the models and to discriminate
between them experimentally. Given the fact that in the considered three pairs
of models the number of real parameters describing the neutrino sector
(four or five) is smaller than the number of the described nine observables
(three masses, three mixing angles and three CPV phases) and that
all observables depend on the VEV of the modulus $\tau$,
there are unusual correlations between some of the described or predicted
observables that varie with the model. We have shown graphically these
correlations in Figures.~\ref{fig:model_lepton_7para_WO},~\ref{fig:model_lepton_7para_WO_WOCP},~\ref{fig:model_lepton_7para_SSN11_WOCP},~\ref{fig:model_lepton_7para_SSN11},~\ref{fig:model_lepton_7para_SSN2_WOCP} and  Figure~\ref{fig:model_lepton_7para_SSN2}. In the case of the model with neutrino masses generated by
the Weinberg effective operator (Section~\ref{subsec:model-Weinberg}),
for example, there are rather strong correlations between $\sin^2\theta_{12}$
and $\sin^2\theta_{13}$ and between $\sin^2\theta_{23}$ and $\delta_{CP}$,
and weaker ones between $\sum_i m_i$ and $\sin^2\theta_{23}$ and
between $\sum_i m_i$ and $\delta_{CP}$
(Figures. \ref{fig:model_lepton_7para_WO} and
\ref{fig:model_lepton_7para_WO_WOCP}).

We foresee that eventually only a very few, if any, of the viable models
we have constructed would pass the test of the data from the upcoming high
precision: i) neutrino oscillation experiments (JUNO, T2HK+HK, DUNE),
ii) planned more precise neutrino mass experiments (KATRIN++, PROJRCT 8, etc.)
iii)  determination of the sum of neutrino masses using cosmological and
astrophysical data, and the test of the data, iv) from the next generation of
neutrinoless double beta decay experiments planned to be sensitive to
$m_{\beta\beta} \sim (5 - 10)$ meV. We look very much forward to the results of
these powerful tests of the models and of the whole non-supersymmetric modular
invariance approach to the flavour problem by the data.

\section*{Acknowledgements}

GJD and BYQ are supported by the National Natural Science Foundation of China under Grant No.~12375104. JNL is supported by the Grants No. NSFC-12147110 and the China Post-doctoral Science Foundation under Grant No. 2021M70. The work of S. T. P. was supported in part by the European Union's Horizon 2020 research and innovation programme under the Marie Sklodowska-Curie grant agreement No.~860881-HIDDeN, by the Italian
INFN program on Theoretical Astroparticle Physics and by the World Premier International Research Center Initiative (WPI Initiative, MEXT), Japan. GJD is grateful to the School of Physics, Northwest University for its hospitality during the completion of this work.

\clearpage

\section*{Appendix}

\setcounter{equation}{0}
\renewcommand{\theequation}{\thesection.\arabic{equation}}

\begin{appendix}

\section{\label{sec:S4_group}Modular group $\Gamma_4\cong S_4$ and polyharmonic Maa{\ss} forms of level $N=4$}
%
%

In the following, we shall present the finite modular group $\Gamma_4 \cong S_4$, along with the irreducible representations and Clebsch-Gordan coefficients of $S_4$. The Fourier expansions of the multiplets of level 4 polyharmonic Maa{\ss} forms are listed. All these results can be found in~\cite{Qu:2024rns}, but we include them here to be self-contained.

\subsection{The finite modular group $\Gamma_4\cong S_4$}

The inhomogeneous finite modular group $\Gamma_4$ is isomorphic to $S_4$ whose defining relations are
\begin{equation}
S_4=\left\{S,T|S^2=T^4=(ST)^3=1\right\}\,.
\end{equation}
%
The $S_4$ group has two singlet representations $\mathbf{1}$, $\mathbf{1}'$, a doublet representation $\mathbf{2}$, and two triplet representations $\mathbf{3}$, $\mathbf{3}'$. The generators $S$ and $T$ are represented by:
\begin{eqnarray}
\nonumber \mathbf{1}:~ & S=1,~&~T=1\,,\\
\nonumber \mathbf{1}':~ &S=-1,~&~T=-1\,,\\
\nonumber \mathbf{2}:~ &S=\dfrac{1}{2} \begin{pmatrix}
-1 ~& \sqrt{3}  \\  \sqrt{3} ~& 1
\end{pmatrix},~&~T=\begin{pmatrix}
1 ~& 0 \\ 0 ~& -1
\end{pmatrix} \,,\\
\nonumber \mathbf{3}:~ & S=\dfrac{1}{2}\begin{pmatrix}
0 ~& \sqrt{2} ~& \sqrt{2} \\
\sqrt{2} ~& -1 ~& 1 \\
\sqrt{2} ~& 1 ~& -1
\end{pmatrix},~&~T=\begin{pmatrix}
1 ~& 0 ~& 0 \\
0 ~& i ~& 0 \\
0 ~& 0 ~& -i
\end{pmatrix}\,,\\
\mathbf{3}':~ & S=-\dfrac{1}{2}\begin{pmatrix}
0 ~& \sqrt{2} ~& \sqrt{2} \\
\sqrt{2} ~& -1 ~& 1 \\
\sqrt{2} ~& 1 ~& -1
\end{pmatrix},~&~T=-\begin{pmatrix}
1 ~& 0 ~& 0 \\
0 ~& i ~& 0 \\
0 ~& 0 ~& -i
\end{pmatrix}\,. \label{eq:rep-basis-S4}
\end{eqnarray}
The tensor products of different $S_4$ multiplets are given by:
\begin{eqnarray}
\mathbf{2}\otimes\mathbf{2}=\mathbf{1}\oplus\mathbf{1}'\oplus\mathbf{2} ~&\text{with}&~\left\{
\begin{array}{l}
\mathbf{1}\sim\alpha_1\beta_1+\alpha_2\beta_2 \\ [0.1in]
\mathbf{1}'\sim\alpha_1\beta_2-\alpha_2\beta_1 \\ [0.1in]
\mathbf{2}\sim
\begin{pmatrix}
-\alpha_1\beta_1+\alpha_2\beta_2\\
\alpha_1\beta_2+\alpha_2\beta_1
\end{pmatrix}
\end{array}
\right. \\
\mathbf{2}\otimes\mathbf{3}=\mathbf{3}\oplus\mathbf{3}' ~&\text{with}&~\left\{
\begin{array}{l}
\mathbf{3}\sim
\begin{pmatrix}
2\alpha_1 \beta_1 \\
-\alpha_1 \beta_2 + \sqrt{3}\alpha_2 \beta_3 \\
-\alpha_1 \beta_3 + \sqrt{3}\alpha_2 \beta_2
\end{pmatrix} \\ [0.1in]
\mathbf{3}'\sim
\begin{pmatrix}
-2\alpha_2 \beta_1 \\
\sqrt{3}\alpha_1 \beta_3 + \alpha_2 \beta_2 \\
\sqrt{3}\alpha_1 \beta_2 + \alpha_2 \beta_3
\end{pmatrix} \\ [0.1in]
\end{array}
\right. \\
\mathbf{2}\otimes\mathbf{3}'=\mathbf{3}\oplus\mathbf{3}' ~&\text{with}&~\left\{
\begin{array}{l}
\mathbf{3}\sim
\begin{pmatrix}
-2\alpha_2 \beta_1 \\
\sqrt{3}\alpha_1 \beta_3 + \alpha_2 \beta_2 \\
\sqrt{3}\alpha_1 \beta_2 + \alpha_2 \beta_3
\end{pmatrix} \\ [0.1in]
\mathbf{3}'\sim
\begin{pmatrix}
2\alpha_1 \beta_1 \\
-\alpha_1 \beta_2 + \sqrt{3}\alpha_2 \beta_3 \\
-\alpha_1 \beta_3 + \sqrt{3}\alpha_2 \beta_2
\end{pmatrix} \\ [0.1in]
\end{array}
\right. \\
\mathbf{3}\otimes\mathbf{3}=\mathbf{3}'\otimes\mathbf{3}'=\mathbf{1}\oplus\mathbf{2}\oplus\mathbf{3}\oplus\mathbf{3}' ~&\text{with}&~\left\{
\begin{array}{l}
\mathbf{1} \sim\alpha_1 \beta_1 + \alpha_2 \beta_3 + \alpha_3 \beta_2 \\ [0.1in]
\mathbf{2}\sim\begin{pmatrix}
2\alpha_1 \beta_1 - \alpha_2 \beta_3 - \alpha_3 \beta_2 \\
\sqrt{3}\alpha_2 \beta_2 + \sqrt{3} \alpha_3 \beta_3
\end{pmatrix}\\ [0.1in]
\mathbf{3}\sim \begin{pmatrix}
\alpha_2 \beta_3 - \alpha_3 \beta_2 \\
\alpha_1 \beta_2 - \alpha_2 \beta_1 \\
\alpha_3 \beta_1 - \alpha_1 \beta_3
\end{pmatrix} \\ [0.1in]
\mathbf{3}'\sim
\begin{pmatrix}
\alpha_2 \beta_2 - \alpha_3 \beta_3 \\
-\alpha_1 \beta_3 - \alpha_3 \beta_1 \\
\alpha_1 \beta_2 + \alpha_2 \beta_1
\end{pmatrix} \\ [0.1in]
\end{array}
\right. \\
\label{eq:CG-S4}\mathbf{3}\otimes\mathbf{3}'=\mathbf{1}'\oplus\mathbf{2}\oplus\mathbf{3}\oplus\mathbf{3}' ~&\text{with}&~\left\{
\begin{array}{l}
\mathbf{1}' \sim\alpha_1 \beta_1 + \alpha_2 \beta_3 + \alpha_3 \beta_2 \\ [0.1in]
\mathbf{2}\sim\begin{pmatrix}
\sqrt{3}\alpha_2 \beta_2 + \sqrt{3} \alpha_3 \beta_3  \\
- 2\alpha_1 \beta_1+\alpha_2 \beta_3 + \alpha_3 \beta_2  \\
\end{pmatrix}\\ [0.1in]
\mathbf{3}\sim \begin{pmatrix}
\alpha_2 \beta_2 - \alpha_3 \beta_3 \\
-\alpha_1 \beta_3 - \alpha_3 \beta_1 \\
\alpha_1 \beta_2 + \alpha_2 \beta_1
\end{pmatrix}\\ [0.1in]
\mathbf{3}'\sim\begin{pmatrix}
\alpha_2 \beta_3 - \alpha_3 \beta_2 \\
\alpha_1 \beta_2 - \alpha_2 \beta_1 \\
\alpha_3 \beta_1 - \alpha_1 \beta_3
\end{pmatrix} \\ [0.1in]
\end{array}
\right.
\end{eqnarray}
Here $\alpha_i$ and $\beta_i$ stand for the elements of the first and second representations respectively.

%
\subsection{Polyharmonic Maa{\ss} form of level $N=4$}
%
%

The polyharmonic Maa{\ss} form of level $4$ can be organized into multiplets
of the finite modular group $\Gamma_4\cong S_4$~\cite{Qu:2024rns}. In the following, we list the expressions of the modular multiplets with weights
$k=-4, -2, 0, 2, 4, 6$, which are necessary in modular flavour model construction. The expressions of the polyharmonic Maa{\ss} forms involve the incomplete gamma function $\Gamma(s,x)$ which is defined as:
\begin{eqnarray}
\label{eq:incomplete-gamma}
\Gamma(s,x)=\int_x^{+\infty} e^{-t}t^{s-1}\,dt\,.
\end{eqnarray}
%
The incomplete gamma function has the following asymptotic behavior,
\begin{eqnarray}
\Gamma(s, x)\sim x^{s-1}\,e^{-x},~~~\text{as $|x|\rightarrow +\infty$}\,.
\end{eqnarray}
%
Moreover, it satisfies the following recurssion relation,
\begin{eqnarray}
\label{eq:Gamma-recursion}\Gamma(s+1,x)=s\Gamma(s,x) + x^s\,e^{-x}
\end{eqnarray}
%
For different low integer values of $s$ of interest for our analysis,
$\Gamma(s, x)$ is given by simple analytical expressions:
\begin{eqnarray}
\nonumber \Gamma(1,x)&=&e^{-x}\,, \\
\nonumber \Gamma(2,x)&=&(x+1)\,e^{-x}\,, \\
\nonumber \Gamma(3,x)&=&(x^2+2x+2)\,e^{-x}\,, \\
\nonumber \Gamma(4,x)&=&(x^3+3x^2+6x+6)\,e^{-x}\,, \\
\Gamma(5,x)&=&(x^4+4x^3+12x^2+24x+24)\,e^{-x}\,.
\end{eqnarray}

%

\begin{itemize}

\item{$k=-4$}

The weight $k=-4$ polyharmonic Maa{\ss} forms of level $4$ can be arranged into a singlet $\mathbf{1}$, a doublet $\mathbf{2}$ and a triplet $\mathbf{3}$ of $S_4$, i.e.
\begin{equation}
Y_{\mathbf{1}}^{(-4)}(\tau),~~~Y_{\mathbf{2}}^{(-4)}=\begin{pmatrix}
Y_{\mathbf{2},1}^{(-4)}(\tau) \\
Y_{\mathbf{2},2}^{(-4)}(\tau)
\end{pmatrix},~~~Y_{\mathbf{3}}^{(-4)}=\begin{pmatrix}
Y_{\mathbf{3},1}^{(-4)}(\tau) \\
Y_{\mathbf{3},2}^{(-4)}(\tau) \\
Y_{\mathbf{3},3}^{(-4)}(\tau)
\end{pmatrix}\,.
\end{equation}
The Fourier expansion of each component of the above modular multiplets reads as
\begin{eqnarray}
\nonumber Y_{\mathbf{1}}^{(-4)}(\tau)&=& \dfrac{y^5}{5} + \dfrac{63\Gamma(5,4\pi y)}{128\pi^5 q} + \dfrac{2079\Gamma(5,8\pi y)}{4096\pi^5 q^2} + \dfrac{427\Gamma(5,12\pi y)}{864\pi^5 q^3} + \dfrac{66591\Gamma(5,16\pi y)}{131072 \pi^5 q^4} + \cdots   \\
\nonumber &&+\dfrac{\pi}{80}\dfrac{\zeta(5)}{\zeta(6)} + \dfrac{189 q}{16\pi^5} + \dfrac{6237 q^2}{512 \pi^5} + \dfrac{427 q^3}{36 \pi^5} + \dfrac{199773 q^4}{16384 \pi^5} + \cdots \,, \\
\nonumber Y_{\mathbf{2},1}^{(-4)}(\tau) &=& \dfrac{y^5}{5} - \dfrac{ 33 \Gamma(5, 4 \pi y)}{128 \pi^5 q}- \dfrac{ 993 \Gamma(5, 8 \pi y)}{4096 \pi^5 q^2}- \dfrac{ 671 \Gamma(5, 12 \pi y)}{2592 \pi^5 q^3}- \dfrac{ 31713 \Gamma(5, 16 \pi y)}{131072 \pi^5 q^4}+ \cdots  \\
\nonumber && - \dfrac{\pi}{168}\dfrac{\zeta(5)}{\zeta(6)} - \dfrac{99 q}{16 \pi^5}- \dfrac{2979 q^2}{512 \pi^5}- \dfrac{671 q^3}{108 \pi^5}- \dfrac{95139 q^4}{16384 \pi^5}- \dfrac{154737 q^5}{25000 \pi^5} + \cdots \,, \\
\nonumber Y_{\mathbf{2},2}^{(-4)}(\tau)&=& \dfrac{\sqrt{3}q^{1/2}}{4\pi^5} \left( \dfrac{\Gamma(5, 2 \pi y)}{q}+ \dfrac{244 \Gamma(5, 6 \pi y)}{243 q^2}+ \dfrac{3126 \Gamma(5, 10 \pi y)}{3125 q^3}+ \dfrac{16808 \Gamma(5, 14 \pi y)}{16807 q^4} + \cdots  \right) \\
\nonumber && + \dfrac{6\sqrt{3}q^{1/2}}{\pi^5} \left( 1 + \dfrac{244 q}{243}+ \dfrac{3126 q^2}{3125}+ \dfrac{16808 q^3}{16807}+ \dfrac{59293 q^4}{59049}+ \dfrac{161052 q^5}{161051} + \cdots \right)\,, \\
\nonumber Y_{\mathbf{3},1}^{(-4)}(\tau)&=& \dfrac{y^5}{5} + \dfrac{\Gamma(5, 4 \pi y)}{128 \pi^5 q}- \dfrac{ 33 \Gamma(5, 8 \pi y)}{4096 \pi^5 q^2}+ \dfrac{61 \Gamma(5, 12 \pi y)}{7776 \pi^5 q^3}- \dfrac{ 993 \Gamma(5, 16 \pi y)}{131072 \pi^5 q^4}+ \dfrac{1563 \Gamma(5, 20 \pi y)}{200000 \pi^5 q^5} + \cdots  \\
\nonumber && - \dfrac{\pi}{5376}\dfrac{\zeta(5)}{\zeta(6)}  + \dfrac{3 q}{16 \pi^5}- \dfrac{99 q^2}{512 \pi^5}+ \dfrac{61 q^3}{324 \pi^5}- \dfrac{2979 q^4}{16384 \pi^5}+ \dfrac{4689 q^5}{25000 \pi^5} + \cdots \,, \\
\nonumber Y_{\mathbf{3},2}^{(-4)}(\tau)&=& \dfrac{61\sqrt{2}q^{1/4}}{243\pi^5} \left( \dfrac{\Gamma(5, 3 \pi y)}{q}+ \dfrac{1021086 \Gamma(5, 7 \pi y)}{1025227 q^2}+ \dfrac{9783909 \Gamma(5, 11 \pi y)}{9824111 q^3}+ \dfrac{3126 \Gamma(5, 15 \pi y)}{3125 q^4} + \cdots  \right)  \\
\nonumber && + \dfrac{6\sqrt{2}q^{1/4}}{\pi^5} \left( 1 + \dfrac{3126 q}{3125}+ \dfrac{59293 q^2}{59049}+ \dfrac{371294 q^3}{371293}+ \dfrac{1419858 q^4}{1419857}+ \dfrac{4101152 q^5}{4084101} + \cdots \right) \,, \\
\nonumber Y_{\mathbf{3},3}^{(-4)}(\tau)&=& \dfrac{q^{3/4}}{2\sqrt{2}\pi^5} \left( \dfrac{\Gamma(5, \pi y)}{q}+ \dfrac{3126 \Gamma(5, 5 \pi y)}{3125 q^2}+ \dfrac{59293 \Gamma(5, 9 \pi y)}{59049 q^3}+ \dfrac{371294 \Gamma(5, 13 \pi y)}{371293 q^4} + \cdots  \right) \\
&&+\dfrac{488\sqrt{2}q^{3/4}}{81\pi^5} \left( 1 + \dfrac{1021086 q}{1025227}+ \dfrac{9783909 q^2}{9824111}+ \dfrac{3126 q^3}{3125}+ \dfrac{150423075 q^4}{151042039} + \cdots \right)\,,
\end{eqnarray}
with $q=e^{2\pi i\tau}$ and $y$ is the imaginary part of the modulus $\tau$.

\item{$k=-2$}

Similar to previous case, one can organize the weight $k=-2$ polyharmonic Maa{\ss} forms of level $4$ into a trivial singlet $Y_{\mathbf{1}}^{(-2)}(\tau)$, a doublet $Y_{\mathbf{2}}^{(-2)}(\tau)$ and a triplet $Y_{\mathbf{3}}^{(-2)}(\tau)$ of $S_4$. Their Fourier expansion is given by
\begin{eqnarray}
\nonumber Y_{\mathbf{1}}^{(-2)}(\tau)&=& \dfrac{y^3}{3} - \dfrac{15\Gamma(3,4\pi y)}{4\pi^3 q} - \dfrac{135\Gamma(3,8\pi y)}{32\pi^3 q^2} - \dfrac{35\Gamma(3,12\pi y)}{9\pi^3 q^3} + \cdots \\
\nonumber&&-\dfrac{\pi}{12}\dfrac{\zeta(3)}{\zeta(4)} - \dfrac{15 q}{2\pi^3} - \dfrac{135 q^2}{16\pi^3} - \dfrac{70 q^3}{q\pi^3} - \dfrac{1095 q^4}{128\pi^3} - \dfrac{189 q^5}{25\pi^3} - \dfrac{35 q^6}{4\pi^3} + \cdots \,, \\
\nonumber Y_{\mathbf{2},1}^{(-2)}(\tau)&=& \dfrac{y^3}{3} + \dfrac{9 \Gamma(3, 4 \pi y)}{4 \pi^3 q}+ \dfrac{57 \Gamma(3, 8 \pi y)}{32 \pi^3 q^2}+ \dfrac{7 \Gamma(3, 12 \pi y)}{3 \pi^3 q^3}+ \dfrac{441 \Gamma(3, 16 \pi y)}{256 \pi^3 q^4} + \cdots  \\
\nonumber &&+\dfrac{\pi}{30} \dfrac{\zeta(3)}{\zeta(4)} + \dfrac{9 q}{2 \pi^3}+ \dfrac{57 q^2}{16 \pi^3}+ \dfrac{14 q^3}{3 \pi^3}+ \dfrac{441 q^4}{128 \pi^3}+ \dfrac{567 q^5}{125 \pi^3} + \cdots \,, \\
\nonumber Y_{\mathbf{2},2}^{(-2)}(\tau)&=& -\dfrac{2\sqrt{3}q^{1/2}}{\pi^3}\left( \dfrac{\Gamma(3, 2 \pi y)}{q}+ \dfrac{28 \Gamma(3, 6 \pi y)}{27 q^2}+ \dfrac{126 \Gamma(3, 10 \pi y)}{125 q^3}+ \dfrac{344 \Gamma(3, 14 \pi y)}{343 q^4} + \cdots  \right) \\
\nonumber && -\dfrac{4\sqrt{3}q^{1/2}}{\pi^3} \left( 1 + \dfrac{28 q}{27}+ \dfrac{126 q^2}{125}+ \dfrac{344 q^3}{343}+ \dfrac{757 q^4}{729}+ \dfrac{1332 q^5}{1331} + \cdots \right) \,, \\
\nonumber Y_{\mathbf{3},1}^{(-2)}(\tau)&=& \dfrac{y^3}{3} - \dfrac{ \Gamma(3, 4 \pi y)}{4 \pi^3 q}+ \dfrac{9 \Gamma(3, 8 \pi y)}{32 \pi^3 q^2}- \dfrac{ 7 \Gamma(3, 12 \pi y)}{27 \pi^3 q^3}+ \dfrac{57 \Gamma(3, 16 \pi y)}{256 \pi^3 q^4}- \dfrac{ 63 \Gamma(3, 20 \pi y)}{250 \pi^3 q^5} + \cdots  \\
\nonumber &&+ \dfrac{\pi}{240}\dfrac{\zeta(3)}{\zeta(4)} - \dfrac{q}{2 \pi^3}+ \dfrac{9 q^2}{16 \pi^3}- \dfrac{14 q^3}{27 \pi^3}+ \dfrac{57 q^4}{128 \pi^3}- \dfrac{63 q^5}{125 \pi^3} + \cdots \,, \\
\nonumber Y_{\mathbf{3},2}^{(-2)}(\tau)&=&-\dfrac{56\sqrt{2} q^{1/4}}{27\pi^3} \left( \dfrac{\Gamma(3, 3 \pi y)}{q}+ \dfrac{2322 \Gamma(3, 7 \pi y)}{2401 q^2}+ \dfrac{8991 \Gamma(3, 11 \pi y)}{9317 q^3}+ \dfrac{126 \Gamma(3, 15 \pi y)}{125 q^4} + \cdots  \right) \\
\nonumber && - \dfrac{4\sqrt{2} q^{1/4}}{\pi^3} \left( 1 + \dfrac{126 q}{125}+ \dfrac{757 q^2}{729}+ \dfrac{2198 q^3}{2197}+ \dfrac{4914 q^4}{4913}+ \dfrac{1376 q^5}{1323} + \cdots \right) \,, \\
\nonumber Y_{\mathbf{3},3}^{(-2)}(\tau)&=& -\dfrac{2\sqrt{2} q^{3/4}}{\pi^3} \left( \dfrac{\Gamma(3, \pi y)}{q}+ \dfrac{126 \Gamma(3, 5 \pi y)}{125 q^2}+ \dfrac{757 \Gamma(3, 9 \pi y)}{729 q^3}+ \dfrac{2198 \Gamma(3, 13 \pi y)}{2197 q^4} + \cdots  \right)  \\
&& -\dfrac{112\sqrt{2} q^{3/4}}{27\pi^3} \left( 1 + \dfrac{2322 q}{2401}+ \dfrac{8991 q^2}{9317}+ \dfrac{126 q^3}{125}+ \dfrac{6615 q^4}{6859}+ \dfrac{82134 q^5}{85169} + \cdots \right)\,.
\end{eqnarray}

\item{$k=0$}

At weight $k=0$ and level $N=4$, there are three linearly independent modular multiplets $Y_{\mathbf{1}}^{(0)}(\tau)$, $Y_{\mathbf{2}}^{(0)}(\tau)$ and  $Y_{\mathbf{3}}^{(0)}(\tau)$ of polyharmonic Maa{\ss} forms which are given by
\begin{eqnarray}
\nonumber Y_{\mathbf{1}}^{(0)}(\tau)&=&1\,,\\
\nonumber Y^{(0)}_{\mathbf{2},1}(\tau)&=& y - \dfrac{6\,e^{-4\pi y}}{\pi q}-\dfrac{3\,e^{-8\pi y}}{\pi q^2}-\dfrac{8\,e^{-12\pi y}}{\pi q^3}-\dfrac{3\,e^{-16\pi y}}{2\pi q^4}-\dfrac{36\,e^{-20\pi y}}{5\pi q^5} + \cdots \\
\nonumber&&-\frac{4\log 2}{\pi}-\dfrac{6q}{\pi}-\dfrac{3q^2}{\pi}-\dfrac{8q^3}{\pi}-\dfrac{3q^4}{2\pi}-\dfrac{36q^5}{5\pi}+\cdots \,, \\
\nonumber Y^{(0)}_{\mathbf{2},2}(\tau)&=&4\sqrt{3}\,q^{1/2}\left( \dfrac{e^{-2\pi y}}{\pi q}+\dfrac{4\,e^{-6\pi y}}{3\pi q^2}+\dfrac{6\,e^{-10\pi y}}{5\pi q^3}+\dfrac{8\,e^{-14\pi y}}{7\pi q^4}+\dfrac{13\,e^{-18\pi y}}{9\pi q^5}+\cdots \right) \\
\nonumber&&+\dfrac{4\sqrt{3}\,q^{1/2}}{\pi}\left(1+\dfrac{4}{3}q+\dfrac{6}{5}q^2+\dfrac{8}{7}q^3+\dfrac{13}{9}q^4+\dfrac{12}{11}q^5+\cdots \right)\,,\\
\nonumber Y_{\mathbf{3},1}^{(0)}(\tau)&=&y + \dfrac{2\,e^{-4\pi y}}{\pi q} - \dfrac{3\,e^{-8\pi y}}{\pi q^2} + \dfrac{8\, e^{-12\pi y}}{3\pi q^3} - \dfrac{3\,e^{-16\pi y}}{2\pi q^4} + \dfrac{12\,e^{-20\pi y}}{5\pi q^5} + \cdots \\
\nonumber &&-\frac{2\log 2}{\pi}+\dfrac{2q}{\pi}-\dfrac{3q^2}{\pi}+\dfrac{8q^3}{3\pi}-\dfrac{3q^4}{2\pi}+\dfrac{12q^5}{5\pi}-\dfrac{4q^6}{\pi}+\cdots\,, \\
\nonumber Y_{\mathbf{3},2}^{(0)}(\tau)&=&\dfrac{16\sqrt{2}q^{1/4}}{\pi}\left( \dfrac{e^{-3\pi y}}{3 q} + \dfrac{2\,e^{-7\pi y}}{7 q^2} + \dfrac{3\,e^{-11\pi y}}{11 q^3} + \dfrac{2\,e^{-15\pi y}}{5 q^4} + \dfrac{5\,e^{-19\pi y}}{19 q^5} + \cdots \right) \\
\nonumber &&+\dfrac{4\sqrt{2}q^{1/4}}{\pi}\left( 1 + \dfrac{6q}{5} + \dfrac{13q^2}{9} + \dfrac{14q^3}{13} + \dfrac{18q^4}{17} + \dfrac{32q^5}{21} + \dfrac{31 q^6}{25} + \cdots \right) \,, \\
\nonumber Y_{\mathbf{3},3}^{(0)}(\tau)&=&\dfrac{4\sqrt{2}q^{3/4}}{\pi} \left( \dfrac{e^{-\pi y}}{q} + \dfrac{6\,e^{-5\pi y}}{5q^2} + \dfrac{13\, e^{-9\pi y}}{9q^3} + \dfrac{14\,e^{-13\pi y}}{13q^4} + \dfrac{18\,e^{-17\pi y}}{14q^5} + \cdots \right) \,, \\
&&+\dfrac{16\sqrt{2}q^{3/4}}{\pi} \left( \dfrac{1}{3} + \dfrac{2q}{7} + \dfrac{3q^2}{11} + \dfrac{2 q^3}{5} + \dfrac{5q^4}{19} + \dfrac{6q^5}{23} + \dfrac{10q^6}{27} + \cdots \right)\,.
\end{eqnarray}
The expressions for $Y_{\bm{2},1}^{(0)}(\tau)$ and $Y_{\bm{2},2}^{(0)}(\tau)$ can be brought to the form:
\begin{eqnarray}
\nonumber
&&Y_{\bm{2},1}^{(0)}(\tau) = y -\,\dfrac{4}{\pi}\,\log 2
-\,\dfrac{2}{\pi} \left [6\,e^{-\,2\pi y}\,\cos(2x\pi)
+3 \,e^{-\,4\pi y}\,\cos(4x\pi) + 8\,e^{-\,6\pi y}\,\cos(6x\pi) +...\right]\,,
\\
&&Y_{\bm{2},2}^{(0)}(\tau) = \dfrac{8\sqrt{3}}{\pi}
\left[ e^{-\,\pi y}\,\cos(x\pi) +
\dfrac{4}{3} e^{-\,3\pi y}\,\cos(3x\pi) +
\dfrac{6}{5} e^{-\,5\pi y}\,\cos(5x\pi) + ...\right]\,.
\label{eq:Y02a}
\end{eqnarray}
It follows from these expressions that $Y_{\bm{2},1}^{(0)}(\tau)$ and $Y_{\bm{2},2}^{(0)}(\tau)$ are real functions. Using the fact that in the fundamental domain of the modular group $y\geq \sqrt{3}/2$ one can show further that for any $\tau =x+ i\,y$, up to corrections $\mathcal{O}(few\times 10^{-4}$) $Y_{\bm{2},1}^{(0)}(\tau)$ and $Y_{\bm{2},2}^{(0)}(\tau)$ are given by:
\begin{eqnarray}
\nonumber
&&Y_{\bm{2},1}^{(0)} \cong y -\,\dfrac{4}{\pi}\,\log 2
-\,\dfrac{12}{\pi}\,e^{-\,2\pi y}\,\cos(2x\pi) -\,\mathcal{O}(10^{-5})\,,\\
&&Y_{\bm{2},2}^{(0)} \cong \dfrac{8\sqrt{3}}{\pi}\,e^{-\,\pi y}\,\cos(x\pi)
+\mathcal{O}(10^{-4})\,.
\label{eq:Y02b}
\end{eqnarray}
Similarly the expressions for $Y_{\bm{3},i}^{(0)}(\tau)$, $i=1,2,3$,
can be simplified somewhat:
\begin{eqnarray}
\nonumber
&&Y_{\bm{3},1}^{(0)} = y -\,\dfrac{2}{\pi}\,\log2
+\,\dfrac{2}{\pi}
\left [ 2\,e^{-\,2\pi y}\,\cos(2x\pi)
-\,3 e^{-\,4\pi y}\,\cos(4x\pi) +
\dfrac{8}{3}\,e^{-\,6\pi y}\,\cos(6x\pi) + ...\right ]\,,\\
 \nonumber
&&Y_{\bm{3},2}^{(0)} = \dfrac{16\sqrt{2}}{\pi}\,
\left [\dfrac{1}{3} ( e^{i\,\frac{3\pi}{2} \tau} )^*
+ \dfrac{2}{7} ( e^{i\,\frac{7\pi}{2} \tau} )^*  +
\dfrac{3}{11} ( e^{i\,\frac{11\pi}{2} \tau} )^*+\ldots \right]\\
\nonumber
&&\qquad + \dfrac{4\sqrt{2}}{\pi}
\left [e^{i\,\frac{\pi}{2} \tau} + \dfrac{6}{5}e^{i\,\frac{5\pi}{2} \tau} +
\dfrac{13}{9}e^{i\,\frac{9\pi}{2} \tau} + ...\right]\,,\\
&&Y_{\bm{3},3}^{(0)} = \left( Y_{\bm{3},2}^{(0)} \right)^*\,.
\label{eq:Y03a}
\end{eqnarray}
Clearly, $Y_{\bm{3},1}^{(0)}$ is a real function.
Taking into account that  $y\geq \sqrt{3}/2$,
up to corrections $\mathcal{O}(10^{-4})$ we have:
\begin{eqnarray}
\nonumber
&&Y_{\bm{3},1}^{(0)} \cong y -\,\dfrac{2}{\pi}\,\log 2
+\,\dfrac{4}{\pi}\,e^{-\,2\pi y}\,\cos(2x\pi) -\,\mathcal{O}(10^{-5})\,,\\
\nonumber
&&Y_{\bm{3},2}^{(0)} \cong \dfrac{4\sqrt{2}}{\pi}
\left [
e^{i\,\frac{\pi}{2} \tau}\,\left ( 1 + \dfrac{6}{5}\, e^{i\,2\pi \tau} \right )
+ \dfrac{4}{3}\,\left ( e^{i\,\frac{3\pi}{2} \tau} \right)^*
\right ] + \mathcal{O}(10^{-4})\,,\\
&&Y_{\bm{3},3}^{(0)} =    \left( Y_{\bm{3},2}^{(0)}\right)^*              \,.
\label{eq:Y03b}
\end{eqnarray}

\item{$k=2$}

The weight 2 polyharmonic Maa{\ss} forms of level 4 are composed of the modified Eisenstein series $\widehat{E}_2(\tau)$ and the modular form multiplets of weight 2 and level 4~\cite{Penedo:2018nmg,Novichkov:2018ovf}
$Y^{(2)}_{\mathbf{2}}(\tau)$ and $Y^{(2)}_{\mathbf{3}}(\tau)$.
$\widehat{E}_2(\tau)$ forms a invariant singlet of $S_4$,
$Y^{(2)}_{\mathbf{2}}(\tau)$ and $Y^{(2)}_{\mathbf{3}}(\tau)$ can be expressed
in terms of Jacobi theta functions $\vartheta_1$ and $\vartheta_2$~
\cite{Novichkov:2020eep,Liu:2020akv,Liu:2020msy}:
\begin{eqnarray}
\nonumber Y_{\mathbf{1}}^{(2)}(\tau)&=&\widehat{E}_2(\tau)\,,\\
\nonumber Y_{\mathbf{2}}^{(2)}(\tau)&=&\begin{pmatrix}
Y_1 \\ Y_2
\end{pmatrix}=
\begin{pmatrix}
\vartheta_1^4 + \vartheta_2^4  \\
-2\sqrt{3} \vartheta_1^2 \vartheta_2^2
\end{pmatrix} \,, \\
Y_{\mathbf{3}}^{(2)}(\tau)&=&\begin{pmatrix}
Y_3 \\ Y_4  \\ Y_5
\end{pmatrix}=
\begin{pmatrix}
\vartheta_1^4 - \vartheta_2^4  \\
2\sqrt{2} \vartheta_1^3 \vartheta_2 \\
2\sqrt{2} \vartheta_1 \vartheta_2^3
\end{pmatrix}\,,
\end{eqnarray}
%
where
\begin{eqnarray}
\nonumber \vartheta_1(\tau) &=& \sum_{m\in \mathbb{Z}} e^{2\pi i \tau m^2} = 1 + 2 q + 2 q^4 + 2 q^9 + 2 q^{16} + \cdots \,, \\
\vartheta_2(\tau) &=& - \sum_{m\in \mathbb{Z}} e^{2\pi i \tau (m+1/2)^2} = - 2 q^{1/4} ( 1+ q^2 + q^6 + q^{12} + \cdots ) \,.
\end{eqnarray}
%
$Y_{\mathbf{1}}^{(2)}(\tau)$ is a non-holomorphic
function of $\tau$ with the following
series expansion:
\begin{equation}
 Y_{\mathbf{1}}^{(2)}(\tau) = 1- \frac{3}{\pi y}-24q -72q^{2} -168q^{4}
-144q^{5}+ \cdots\,.
\label{eq:Y21serexp}
\end{equation}
%
Both $Y^{(2)}_{\mathbf{2}}(\tau)$ and $Y^{(2)}_{\mathbf{3}}(\tau)$ are holomorphic
functions of $\tau$
\footnote{
Note that the definition of  $Y^{(2)}_{\mathbf{3}}(\tau)$ employed by us
corresponds to $Y^{(2)}_{\mathbf{3'}}(\tau)$ defined in
\cite{Penedo:2018nmg,Novichkov:2018ovf,Novichkov:2020eep}.
}
whose $q$-expansions are given by \cite{Penedo:2018nmg,Novichkov:2020eep}:
\begin{eqnarray}
\nonumber
Y_{\mathbf{2}}^{(2)}(\tau)&=&\begin{pmatrix}
Y_1 \\ Y_2
\end{pmatrix}=\begin{pmatrix}
1 + 24 q + 24 q^2 + 96 q^3 + 24 q^4 + 144 q^5 \cdots \\
-8\sqrt{3}\, q^{1/2} (1 + 4 q + 6 q^2 +8 q^3 + 13 q^4 + 12 q^5 + \cdots)
\end{pmatrix}\,, \\
Y_{\mathbf{3}}^{(2)}(\tau)&=&\begin{pmatrix}
Y_3 \\ Y_4  \\ Y_5
\end{pmatrix}=\begin{pmatrix}
1 - 8 q + 24 q^2 - 32 q^3 + 24 q^4 - 48 q^5 + \cdots  \\
-4\sqrt{2}\, q^{1/4} (1 + 6 q + 13 q^2 + 14 q^3 + 18 q^4 + 32 q^5 + \cdots) \\
-16\sqrt{2} \,q^{3/4} (1 + 2 q + 3 q^2 + 6 q^3 + 5 q^4 + 6 q^5 + \cdots)
\end{pmatrix}\,.
\end{eqnarray}
%

\item{$k=4$}

The weight $4$ polyharmonic Maa{\ss} forms coincide with the modular forms,
and they can be obtained from the tensor products of weight 2 modular forms
\cite{Penedo:2018nmg,Novichkov:2018ovf,Novichkov:2020eep,Liu:2020akv,Liu:2020msy}:
\begin{eqnarray}
\nonumber Y^{(4)}_{\mathbf{1}}(\tau) &=& \left(Y_{\mathbf{2}}^{(2)}Y_{\mathbf{2}}^{(2)}\right)_{\mathbf{1}}=Y_1^2 + Y_2^2 \,, \\
\nonumber Y^{(4)}_{\mathbf{2}}(\tau)&=&-\left(Y_{\mathbf{2}}^{(2)}Y_{\mathbf{2}}^{(2)}\right)_{\mathbf{2}}= \begin{pmatrix}
Y_1^2 - Y_2^2 \\
-2Y_1 Y_2
\end{pmatrix} \,, \\
\nonumber Y^{(4)}_{\mathbf{3}}(\tau) &=& \frac{1}{2}\left(Y_{\mathbf{2}}^{(2)}Y_{\mathbf{3}}^{(2)}\right)_{\mathbf{3}}= \begin{pmatrix}
Y_1 Y_3 \\
-\dfrac{1}{2}Y_1 Y_4 + \dfrac{\sqrt{3}}{2}Y_2 Y_5 \\
-\dfrac{1}{2}Y_1 Y_5 + \dfrac{\sqrt{3}}{2}Y_2 Y_4
\end{pmatrix}\,, \\
Y^{(4)}_{\mathbf{3}'}(\tau) &=& \frac{1}{2}\left(Y_{\mathbf{2}}^{(2)}Y_{\mathbf{3}}^{(2)}\right)_{\mathbf{3}'}= \begin{pmatrix}
-Y_2 Y_3 \\
\dfrac{\sqrt{3}}{2}Y_1 Y_5 + \dfrac{1}{2}Y_2 Y_4 \\
\dfrac{\sqrt{3}}{2}Y_1 Y_4 + \dfrac{1}{2}Y_2 Y_5
\end{pmatrix}\,.
\end{eqnarray}

\item{$k=6$}

The weight $6$ polyharmonic Maa{\ss} forms of level $4$ can be organized into
six multiplets of $S_4$: $Y^{(6)}_{\mathbf{1}}(\tau)$, $Y^{(6)}_{\mathbf{1}'}(\tau)$, $Y^{(6)}_{\mathbf{2}}(\tau)$, $Y^{(6)}_{\mathbf{3}I}(\tau)$, $Y^{(6)}_{\mathbf{3}II}(\tau)$ and $Y^{(6)}_{\mathbf{3}'}(\tau)$ with
\begin{eqnarray}
\nonumber Y^{(6)}_{\mathbf{1}}(\tau)&=&(Y_{\mathbf{2}}^{(2)}Y_{\mathbf{2}}^{(4)})_{\mathbf{1}} =Y_1^3 - 3Y_1 Y_2^2 \,, \\
\nonumber Y^{(6)}_{\mathbf{1}'}(\tau)&=&(Y_{\mathbf{2}}^{(2)}Y_{\mathbf{2}}^{(4)})_{\mathbf{1}'} = Y_2^3 -3Y_1^2 Y_2\,, \\
\nonumber Y^{(6)}_{\mathbf{2}}(\tau)&=&(Y_{\mathbf{2}}^{(2)}Y_{\mathbf{1}}^{(4)})_{\mathbf{2}} =\begin{pmatrix}
 Y_1(Y_1^2+Y_2^2)  \\
 Y_2 (Y_1^2+Y_2^2)
\end{pmatrix} \,, \\
\nonumber Y^{(6)}_{\mathbf{3}I}(\tau) &=& \frac{1}{2} (Y_{\mathbf{3}}^{(2)}Y_{\mathbf{2}}^{(4)})_{\mathbf{3}} =\begin{pmatrix}
(Y_1^2-Y_2^2) Y_3  \\
\dfrac{1}{2}(Y_2^2-Y_1^2)Y_4 - \sqrt{3}Y_1 Y_2 Y_5 \\
\dfrac{1}{2}(Y_2^2-Y_1^2)Y_5 - \sqrt{3}Y_1 Y_2 Y_4 \\
\end{pmatrix} \,, \\
\nonumber Y^{(6)}_{\mathbf{3}II}(\tau)&=&(Y_{\mathbf{3}}^{(2)}Y_{\mathbf{1}}^{(4)})_{\mathbf{3}} = \begin{pmatrix}
Y_3(Y_1^2+ Y_2^2)  \\
Y_4(Y_1^2+ Y_2^2) \\
Y_5(Y_1^2+ Y_2^2)
\end{pmatrix} \,, \\
Y^{(6)}_{\mathbf{3}'}(\tau)&=& (Y_{\mathbf{3}}^{(2)}Y_{\mathbf{2}}^{(4)})_{\mathbf{3}'} =\begin{pmatrix}
4 Y_1 Y_2 Y_3 \\
\sqrt{3}(Y_1^2 - Y_2^2) Y_5 - 2 Y_1 Y_2 Y_4 \\
\sqrt{3}(Y_1^2 - Y_2^2) Y_4- 2 Y_1 Y_2 Y_5
\end{pmatrix}\,.
\end{eqnarray}
%
The above multiplets of polyharmonic Maa{\ss} forms at level 4 are summarized in table~\ref{Tab:LeveL4_MM}.

\end{itemize}

\begin{table}[t!]
\centering
\begin{tabular}{|c|c|}
\hline  \hline

Weight $k_Y$ & Polyharmonic Maa{\ss} forms $Y^{(k_Y)}_{\mathbf{r}}$ \\ \hline

$k_Y=-4$ & $Y^{(-4)}_{\mathbf{1}}$,\; $Y^{(-4)}_{\mathbf{2}}$,\; $Y^{(-4)}_{\mathbf{3}}$\\ \hline

$k_Y=-2$ & $Y^{(-2)}_{\mathbf{1}}$,\; $Y^{(-2)}_{\mathbf{2}}$,\; $Y^{(-2)}_{\mathbf{3}}$\\ \hline

$k_Y=0$ & $Y^{(0)}_{\mathbf{1}}$,\; $Y^{(0)}_{\mathbf{2}}$,\; $Y^{(0)}_{\mathbf{3}}$\\ \hline

$k_Y=2$ & $Y^{(2)}_{\mathbf{1}}$,\; $Y^{(2)}_{\mathbf{2}}$,\; $Y^{(2)}_{\mathbf{3}}$\\ \hline

$k_Y=4$ & $Y^{(4)}_{\mathbf{1}}$,\; $Y^{(4)}_{\mathbf{2}}$,\; $Y^{(4)}_{\mathbf{3}}$,\; $Y^{(4)}_{\mathbf{3}'}$\\ \hline

$k_Y=6$ & $Y^{(6)}_{\mathbf{1}}$, \; $Y^{(6)}_{\mathbf{1}'}$, \; $Y^{(6)}_{\mathbf{2}}$, \; $Y^{(6)}_{\mathbf{3I}}$, \; $Y^{(6)}_{\mathbf{3II}}$,\; $Y^{(6)}_{\mathbf{3}'}$\\
\hline \hline
\end{tabular}\caption{\label{Tab:LeveL4_MM}Summary of polyharmonic Maa{\ss} form of level $N=4$ at weights $k_{Y}=-4,-2,0,2,4,6$.}
\end{table}

%
\section{\label{sec:effective_parameters}The number of effective parameters in $M_{\nu}$}
%
%

In section~\ref{sec:two_right_handed_neutrinos}, we mentioned that in the case of
$N^{c}\sim \mathbf{1}^{j_{1}}\oplus \mathbf{1}^{j_{2}}$, there are 3 and 5 real
parameters in the effective light neutrino mass matrix $M_{\nu}$ for
$[j_{1}+j_{2}]=0$ and $[j_{1}+j_{2}]=1$ respectively. In the following, we will
clarify the above conclusion in detail. Given $L\sim \mathbf{3}^{i}$ and
$N^{c}\sim \mathbf{1}^{j_{1}}\oplus \mathbf{1}^{j_{2}}$, the Dirac neutrino mass
matrix $M_{\nu_{D}}$ is given by Eq.~\eqref{eq:N11_MD}.
For $[j_{1}+j_{2}]=0$, the heavy neutrino Majorana
mass matrix $M_{N^{c}}$
is determined by Eq.~\eqref{eq:MN_1}. Using the seesaw expression
in Eq.~\eqref{eq:seesaw}, we can obtain the explicit form of $M_{\nu}$:
\begin{equation}
  M_{\nu}=-\frac{\beta_{1}^{2}v^{2}}{\Lambda g_{1}Y_{\mathbf{1}}^{(2k_{N^c_1})}}M_{\nu}^{I}-\frac{\beta_{2}^{2}v^{2}}{\Lambda g_{2}Y_{\mathbf{1}}^{(2k_{N^c_2})}}M_{\nu}^{II}\,,
\end{equation}
%
with
\begin{eqnarray}\small \nonumber
  M_{\nu}^{I}&=&\left(\begin{matrix}
Y_{\mathbf{3}^{[j_{1}+i]},1}^{(k_{N^{c}_{1}}+k_{L})}Y_{\mathbf{3}^{[j_{1}+i]},1}^{(k_{N^{c}_{1}}+k_{L})}~&~Y_{\mathbf{3}^{[j_{1}+i]},1}^{(k_{N^{c}_{1}}+k_{L})}Y_{\mathbf{3}^{[j_{1}+i]},3}^{(k_{N^{c}_{1}}+k_{L})}~&~Y_{\mathbf{3}^{[j_{1}+i]},1}^{(k_{N^{c}_{1}}+k_{L})}Y_{\mathbf{3}^{[j_{1}+i]},2}^{(k_{N^{c}_{1}}+k_{L})}\\
Y_{\mathbf{3}^{[j_{1}+i]},3}^{(k_{N^{c}_{1}}+k_{L})}Y_{\mathbf{3}^{[j_{1}+i]},1}^{(k_{N^{c}_{1}}+k_{L})}~&~Y_{\mathbf{3}^{[j_{1}+i]},3}^{(k_{N^{c}_{1}}+k_{L})}Y_{\mathbf{3}^{[j_{1}+i]},3}^{(k_{N^{c}_{1}}+k_{L})}~&~Y_{\mathbf{3}^{[j_{1}+i]},3}^{(k_{N^{c}_{1}}+k_{L})}Y_{\mathbf{3}^{[j_{1}+i]},2}^{(k_{N^{c}_{1}}+k_{L})}\\
Y_{\mathbf{3}^{[j_{1}+i]},2}^{(k_{N^{c}_{1}}+k_{L})}Y_{\mathbf{3}^{[j_{1}+i]},1}^{(k_{N^{c}_{1}}+k_{L})}~&~Y_{\mathbf{3}^{[j_{1}+i]},2}^{(k_{N^{c}_{1}}+k_{L})}Y_{\mathbf{3}^{[j_{1}+i]},3}^{(k_{N^{c}_{1}}+k_{L})}~&~Y_{\mathbf{3}^{[j_{1}+i]},2}^{(k_{N^{c}_{1}}+k_{L})}Y_{\mathbf{3}^{[j_{1}+i]},2}^{(k_{N^{c}_{1}}+k_{L})}\\
\end{matrix}\right)\,,\\
\label{eq:MI_MII}
  M_{\nu}^{II}&=&\left(\begin{matrix}
Y_{\mathbf{3}^{[j_{2}+i]},1}^{(k_{N^{c}_{2}}+k_{L})}Y_{\mathbf{3}^{[j_{2}+i]},1}^{(k_{N^{c}_{2}}+k_{L})}~&~Y_{\mathbf{3}^{[j_{2}+i]},1}^{(k_{N^{c}_{2}}+k_{L})}Y_{\mathbf{3}^{[j_{2}+i]},3}^{(k_{N^{c}_{2}}+k_{L})}~&~Y_{\mathbf{3}^{[j_{2}+i]},1}^{(k_{N^{c}_{2}}+k_{L})}Y_{\mathbf{3}^{[j_{2}+i]},2}^{(k_{N^{c}_{2}}+k_{L})}\\
Y_{\mathbf{3}^{[j_{2}+i]},3}^{(k_{N^{c}_{2}}+k_{L})}Y_{\mathbf{3}^{[j_{2}+i]},1}^{(k_{N^{c}_{2}}+k_{L})}~&~Y_{\mathbf{3}^{[j_{2}+i]},3}^{(k_{N^{c}_{2}}+k_{L})}Y_{\mathbf{3}^{[j_{2}+i]},3}^{(k_{N^{c}_{2}}+k_{L})}~&~Y_{\mathbf{3}^{[j_{2}+i]},3}^{(k_{N^{c}_{2}}+k_{L})}Y_{\mathbf{3}^{[j_{2}+i]},2}^{(k_{N^{c}_{2}}+k_{L})}\\
Y_{\mathbf{3}^{[j_{2}+i]},2}^{(k_{N^{c}_{2}}+k_{L})}Y_{\mathbf{3}^{[j_{2}+i]},1}^{(k_{N^{c}_{2}}+k_{L})}~&~Y_{\mathbf{3}^{[j_{2}+i]},2}^{(k_{N^{c}_{2}}+k_{L})}Y_{\mathbf{3}^{[j_{2}+i]},3}^{(k_{N^{c}_{2}}+k_{L})}~&~Y_{\mathbf{3}^{[j_{2}+i]},2}^{(k_{N^{c}_{2}}+k_{L})}Y_{\mathbf{3}^{[j_{2}+i]},2}^{(k_{N^{c}_{2}}+k_{L})}\\
\end{matrix}\right)\,.
\end{eqnarray}
%
We see that there are two effective constant parameters $\frac{\beta_{1}^{2}v^{2}}{\Lambda g_{1}}$ and
$\frac{\beta_{2}^{2}v^{2}}{\Lambda g_{2}}$ in $M_{\nu}$,
$\frac{\beta_{1}^{2}v^{2}}{\Lambda g_{1}}$ can be real
while $\frac{\beta_{2}^{2}v^{2}}{\Lambda g_{2}}$ is complex.

In the case of $[j_{1}+j_{2}]=1$, the general form of $M_{N^{c}}$ is presented in
Eq.~\eqref{eq:MN_2}. We find that the effective light neutrino mass matrix
$M_{\nu}$ takes following form:
\begin{eqnarray}
  M_{\nu}=\frac{v^{2}}{\Lambda}\frac{\left[ -\beta_{1}^{2}g_{2}Y_{\mathbf{1}}^{(2k_{N^c_2})}M_{\nu}^{I}-\beta_{2}^{2}g_{1}Y_{\mathbf{1}}^{(2k_{N^c_1})}M_{\nu}^{II} + \beta_{1}\beta_{2}g_{3}Y_{\mathbf{1}^{[j_1+j_2]}}^{(k_{N^c_1}+k_{N^c_2})}M_{\nu}^{III}\right]}{g_{1}g_{2}Y_{\mathbf{1}}^{(2k_{N^c_1})}Y_{\mathbf{1}}^{(2k_{N^c_2})}-\left(g_{3}Y_{\mathbf{1}^{[j_1+j_2]}}^{(k_{N^c_1}+k_{N^c_2})} \right)^{2}}\,.
\end{eqnarray}
%
The matrices  $M_{\nu}^{I}$, $M_{\nu}^{II}$ are defined
in Eq.~\eqref{eq:MI_MII}, while $M_{\nu}^{III}$
is given by:
\begin{eqnarray}
\small \nonumber
  M_{\nu}^{III}&=&\left(\begin{matrix}
2Y_{\mathbf{3}^{[j_{1}+i]},1}^{(k_{N^{c}_{1}}+k_{L})}Y_{\mathbf{3}^{[j_{2}+i]},1}^{(k_{N^{c}_{2}}+k_{L})}~&~Y_{\mathbf{3}^{[j_{1}+i]},3}^{(k_{N^{c}_{1}}+k_{L})}Y_{\mathbf{3}^{[j_{2}+i]},1}^{(k_{N^{c}_{2}}+k_{L})}~&~Y_{\mathbf{3}^{[j_{1}+i]},2}^{(k_{N^{c}_{1}}+k_{L})}Y_{\mathbf{3}^{[j_{2}+i]},1}^{(k_{N^{c}_{2}}+k_{L})}\\
Y_{\mathbf{3}^{[j_{1}+i]},3}^{(k_{N^{c}_{1}}+k_{L})}Y_{\mathbf{3}^{[j_{2}+i]},1}^{(k_{N^{c}_{2}}+k_{L})}~&~2Y_{\mathbf{3}^{[j_{1}+i]},3}^{(k_{N^{c}_{1}}+k_{L})}Y_{\mathbf{3}^{[j_{2}+i]},3}^{(k_{N^{c}_{2}}+k_{L})}~&~Y_{\mathbf{3}^{[j_{1}+i]},3}^{(k_{N^{c}_{1}}+k_{L})}Y_{\mathbf{3}^{[j_{2}+i]},2}^{(k_{N^{c}_{2}}+k_{L})}\\
Y_{\mathbf{3}^{[j_{1}+i]},2}^{(k_{N^{c}_{1}}+k_{L})}Y_{\mathbf{3}^{[j_{2}+i]},1}^{(k_{N^{c}_{2}}+k_{L})}~&~Y_{\mathbf{3}^{[j_{1}+i]},3}^{(k_{N^{c}_{1}}+k_{L})}Y_{\mathbf{3}^{[j_{2}+i]},2}^{(k_{N^{c}_{2}}+k_{L})}~&~2Y_{\mathbf{3}^{[j_{1}+i]},2}^{(k_{N^{c}_{1}}+k_{L})}Y_{\mathbf{3}^{[j_{2}+i]},2}^{(k_{N^{c}_{2}}+k_{L})}\\
\end{matrix}\right)\\
&~&+\left(\begin{matrix}
0 &Y_{\mathbf{3}^{[j_{1}+i]},1}^{(k_{N^{c}_{1}}+k_{L})}Y_{\mathbf{3}^{[j_{2}+i]},3}^{(k_{N^{c}_{2}}+k_{L})}~&~Y_{\mathbf{3}^{[j_{1}+i]},1}^{(k_{N^{c}_{1}}+k_{L})}Y_{\mathbf{3}^{[j_{2}+i]},2}^{(k_{N^{c}_{2}}+k_{L})}\\
Y_{\mathbf{3}^{[j_{1}+i]},1}^{(k_{N^{c}_{1}}+k_{L})}Y_{\mathbf{3}^{[j_{2}+i]},3}^{(k_{N^{c}_{2}}+k_{L})}~&~ 0 ~&~Y_{\mathbf{3}^{[j_{1}+i]},2}^{(k_{N^{c}_{1}}+k_{L})}Y_{\mathbf{3}^{[j_{2}+i]},3}^{(k_{N^{c}_{2}}+k_{L})}\\
Y_{\mathbf{3}^{[j_{1}+i]},1}^{(k_{N^{c}_{1}}+k_{L})}Y_{\mathbf{3}^{[j_{2}+i]},2}^{(k_{N^{c}_{2}}+k_{L})}~&~Y_{\mathbf{3}^{[j_{1}+i]},2}^{(k_{N^{c}_{1}}+k_{L})}Y_{\mathbf{3}^{[j_{2}+i]},3}^{(k_{N^{c}_{2}}+k_{L})}~&~ 0\\
\end{matrix}\right)\,.
\end{eqnarray}
%
In this case there are three effective constant parameters in $M_{\nu}$:
\begin{eqnarray} \nonumber
 && \frac{v^{2}}{\Lambda}\frac{-\beta_{1}^{2}g_{2}Y_{\mathbf{1}}^{(2k_{N^c_2})}}{g_{1}g_{2}Y_{\mathbf{1}}^{(2k_{N^c_1})}Y_{\mathbf{1}}^{(2k_{N^c_2})}-\left(g_{3}Y_{\mathbf{1}^{[j_1+j_2]}}^{(k_{N^c_1}+k_{N^c_2})} \right)^{2}}\,,\\ \nonumber
&& \frac{v^{2}}{\Lambda}\frac{-\beta_{2}^{2}g_{1}Y_{\mathbf{1}}^{(2k_{N^c_1})}}{g_{1}g_{2}Y_{\mathbf{1}}^{(2k_{N^c_1})}Y_{\mathbf{1}}^{(2k_{N^c_2})}-\left(g_{3}Y_{\mathbf{1}^{[j_1+j_2]}}^{(k_{N^c_1}+k_{N^c_2})} \right)^{2}}\,,\\
&& \frac{v^{2}}{\Lambda}\frac{\beta_{1}\beta_{2}g_{3}Y_{\mathbf{1}^{[j_1+j_2]}}^{(k_{N^c_1}+k_{N^c_2})}}{g_{1}g_{2}Y_{\mathbf{1}}^{(2k_{N^c_1})}Y_{\mathbf{1}}^{(2k_{N^c_2})}-\left(g_{3}Y_{\mathbf{1}^{[j_1+j_2]}}^{(k_{N^c_1}+k_{N^c_2})} \right)^{2}}\,,
\end{eqnarray}
%
where the first one can be real and the remain two parameters are complex.

%
\section{Viable lepton flavor models\label{sec:app_viable_models}}
%
%

In this section, we provide phenomenologically viable
lepton models with $7$ (8) real input parameters in the case where gCP symmetry is (not) imposed, and the predictions for the best fit values of the lepton mass and mixing observables will be listed in the following. We impose the bound on the neutrino mass sum $m_1+m_2+m_3<0.12$ eV from the Planck collaboration~\cite{Planck:2018vyg}.

For neutrino masses generated via the Weinberg operator, in the case of a NO neutrino mass spectrum, only 16 out of 80 lepton models are compatible with experimental data, regardless of whether gCP symmetry is imposed. In contrast, for an IO neutrino mass spectrum, none of the models align with experimental results. The viable models and their corresponding best-fit results for lepton observables (obtained with gCP) are summarized in Table~\ref{tab:lepton_res_par7_WO_NO}.

If neutrino masses are generated through the minimal type-I seesaw mechanism with two right-handed neutrinos, we consider two distinct assignments for the right-handed neutrinos: \( N^{c} \sim \mathbf{2} \) and \( N^{c} \sim \mathbf{1}^{j_{1}} \oplus \mathbf{1}^{j_{2}} \). For \( N^{c} \sim \mathbf{2} \), 56 (116) out of 400 models are consistent with experimental data for NO (IO) when gCP symmetry is imposed. When gCP symmetry is not imposed, an additional real free parameter emerges in the models, resulting in 32 (20) extra viable models for NO (IO) beyond those obtained with gCP. The viable models for NO and IO neutrino mass spectra are listed in Tables~\ref{tab:lepton_res_par7_SSN2_NO} and~\ref{tab:lepton_res_par7_SSN2_IO}, respectively. For \( N^{c} \sim \mathbf{1}^{j_{1}} \oplus \mathbf{1}^{j_{2}} \), among the 180 ``minimal'' models, 16 (51) models can account for the experimental data at the \( 3\sigma \) level for the NO (IO) neutrino mass spectrum with gCP symmetry. Without gCP symmetry, there are an additional 20 (48) viable models for the NO (IO) neutrino mass spectrum. The detailed results are presented in Table~\ref{tab:lepton_res_par7_SSN11_NO} and Table~\ref{tab:lepton_res_par7_SSN11_IO}. Note that in Tables \ref{tab:lepton_res_par7_WO_NO} - \ref{tab:lepton_res_par7_SSN11_IO}, the best-fit predictions for lepton observables are provided for the case with gCP symmetry if the model is viable under both gCP and non-gCP scenarios.

It follows from Tables
\ref{tab:lepton_res_par7_WO_NO} -
\ref{tab:lepton_res_par7_SSN11_IO},
in particular, that \\
i) a large number of the currently viable models would be ruled out if it is definitely established that
$\sin^2\theta_{23} \geq 0.5$ or that $\sin^2\theta_{23} < 0.5$,
a high precision determination of $\sin^2\theta_{23}$
will further reduce the number of viable models;
\\
ii) the very high precision measurement of
$\sin^2\theta_{12}$ forseen to be performed by the JUNO
expriment \cite{JUNO:2022mxj}
would also reduce significantly the number of
viable models;\\
iii) additional important tests of the models will be provided
by precision measurements of the Dirac CPV phase
$\delta_{CP}$ and of the $J^{lep}_{CP}$ factor
as well as of the sum of the neutrino masses $\sum_i m_i$.
It follows also from the results in Tables \ref{tab:lepton_res_par7_WO_NO} -
\ref{tab:lepton_res_par7_SSN11_IO}
that all  models whose predictions are listed in Tables
12, 13, 16 and 17 (Tables 9, 10, 11, 14 and 15)
will be ruled out if the neutrino mass spectrum is proven to be
of NO type (of IO type).

We foresee  that eventually only a very few models, if any,
would pass the test of the data from the upcoming high precision
i) neutrino oscillation experiments (JUNO, T2HK+HK, DUNE),
ii) planned more precise neutrino mass experiments (KATRIN++, PROJECT 8, etc.)
iii)  determination of the sum of neutrino masses
using cosmological and astrophysical data,
and the test of the data iv) from the next generation of
neutrinoless double beta decay experiments planned to be sensitive
to $m_{\beta\beta} \sim (5 - 10)$ meV.

\begin{table}[ht!]
 \centering \resizebox{1.0\textwidth}{!}{
\begin{tabular}{|c|c|c|c|c|c|c|c|c|c|c|c|c|} \hline \hline
\multicolumn{12}{|c|}{Weinberg operator with/without gCP (NO)} \\ \hline
Model & $\sin^2\theta_{12}$ &$\sin^2\theta_{13}$ &$\sin^2\theta_{23}$&$\delta_{CP}/\pi$ & $\alpha_{21}/\pi$  &$\alpha_{31}/\pi$ & $m_1$/meV & $m_2$/meV & $m_3$/meV & $m_{\beta\beta}$/meV & $\chi^{2}_{\text{min}}$ \\ \hline $\mathcal{C}_{20}-\mathcal{W}_{4}$ & $0.305$ & $0.02241$ & $0.411$ & $1.245$ & $0.243$ & $1.145$ & $7.640$ & $11.509$ & $50.630$ & $7.896$ & $7.287$\\ \hline$\mathcal{C}_{20}-\mathcal{W}_{3}$ & $0.305$ & $0.02241$ & $0.411$ & $1.245$ & $0.234$ & $1.904$ & $3.620$ & $9.338$ & $50.181$ & $4.298$ & $7.287$\\ \hline$\mathcal{C}_{20}-\mathcal{W}_{2}$ & $0.305$ & $0.02241$ & $0.411$ & $1.245$ & $0.200$ & $0.655$ & $6.083$ & $10.541$ & $50.418$ & $8.027$ & $7.287$\\ \hline$\mathcal{C}_{20}-\mathcal{W}_{1}$ & $0.305$ & $0.02241$ & $0.411$ & $1.245$ & $0.367$ & $1.736$ & $10.684$ & $13.720$ & $51.178$ & $8.616$ & $7.287$\\ \hline$\mathcal{C}_{10}-\mathcal{W}_{4}$ & $0.305$ & $0.02241$ & $0.411$ & $1.245$ & $0.243$ & $1.145$ & $7.641$ & $11.510$ & $50.630$ & $7.896$ & $7.282$\\ \hline$\mathcal{C}_{10}-\mathcal{W}_{3}$ & $0.305$ & $0.02241$ & $0.411$ & $1.245$ & $0.234$ & $1.904$ & $3.619$ & $9.338$ & $50.181$ & $4.298$ & $7.282$\\ \hline$\mathcal{C}_{10}-\mathcal{W}_{2}$ & $0.305$ & $0.02241$ & $0.411$ & $1.245$ & $0.200$ & $0.655$ & $6.084$ & $10.541$ & $50.418$ & $8.028$ & $7.282$\\ \hline$\mathcal{C}_{10}-\mathcal{W}_{1}$ & $0.305$ & $0.02241$ & $0.411$ & $1.245$ & $0.367$ & $1.736$ & $10.685$ & $13.721$ & $51.178$ & $8.616$ & $7.282$\\ \hline$\mathcal{C}_{9}-\mathcal{W}_{4}$ & $0.305$ & $0.02241$ & $0.411$ & $1.245$ & $0.243$ & $1.145$ & $7.640$ & $11.510$ & $50.630$ & $7.896$ & $7.285$\\ \hline$\mathcal{C}_{9}-\mathcal{W}_{3}$ & $0.305$ & $0.02241$ & $0.411$ & $1.245$ & $0.234$ & $1.904$ & $3.621$ & $9.339$ & $50.181$ & $4.299$ & $7.285$\\ \hline$\mathcal{C}_{9}-\mathcal{W}_{2}$ & $0.305$ & $0.02241$ & $0.411$ & $1.245$ & $0.200$ & $0.655$ & $6.083$ & $10.541$ & $50.418$ & $8.027$ & $7.285$\\ \hline$\mathcal{C}_{9}-\mathcal{W}_{1}$ & $0.305$ & $0.02241$ & $0.411$ & $1.245$ & $0.367$ & $1.736$ & $10.684$ & $13.721$ & $51.178$ & $8.616$ & $7.285$\\ \hline$\mathcal{C}_{6}-\mathcal{W}_{4}$ & $0.305$ & $0.02241$ & $0.411$ & $1.245$ & $0.243$ & $1.145$ & $7.641$ & $11.510$ & $50.630$ & $7.896$ & $7.281$\\ \hline$\mathcal{C}_{6}-\mathcal{W}_{3}$ & $0.305$ & $0.02241$ & $0.411$ & $1.245$ & $0.234$ & $1.904$ & $3.619$ & $9.338$ & $50.181$ & $4.297$ & $7.281$\\ \hline$\mathcal{C}_{6}-\mathcal{W}_{2}$ & $0.305$ & $0.02241$ & $0.411$ & $1.245$ & $0.200$ & $0.655$ & $6.084$ & $10.541$ & $50.418$ & $8.028$ & $7.281$\\ \hline$\mathcal{C}_{6}-\mathcal{W}_{1}$ & $0.305$ & $0.02241$ & $0.411$ & $1.245$ & $0.367$ & $1.736$ & $10.685$ & $13.722$ & $51.178$ & $8.616$ & $7.281$\\ \hline \hline \end{tabular}}
\caption{\label{tab:lepton_res_par7_WO_NO}The best fit values of the lepton observables at the minimum of the $\chi^2$ for the lepton models under the assumption of NO neutrino masses. Here we assume that neutrino masses are generated via Weinberg operator.  We give the predictions for neutrino mixing angles $\theta_{12}$, $\theta_{13}$, $\theta_{23}$, and Dirac CP violating phase $\delta_{CP}$ as well as Majorana CP violating phases $\alpha_{21}$, $\alpha_{31}$, and the light neutrino masses $m_{1,2,3}$ and the effective mass $m_{\beta\beta}$ in neutrinoless double beta decay.}
\end{table}

\begin{small}
\begin{landscape}
\setlength\LTcapwidth{\textwidth}
\setlength\LTleft{1.1in}            
\setlength\LTright{0pt}           
\begin{longtable}{|c|c|c|c|c|c|c|c|c|c|c|}
\caption{\label{tab:lepton_res_par7_SSN2_NO}The best fit values of lepton observables at the minimum of the $\chi^2$ under the assumption of NO neutrino masses. Here we assume that the neutrino mass are described by the type-I seesaw mechanism with two right-handed neutrinos $N^{c}\sim \mathbf{2}$. We give the predictions for neutrino mixing angles $\theta_{12}$, $\theta_{13}$, $\theta_{23}$, and Dirac CP violating phase $\delta_{CP}$ as well as Majorana CP violating phase $\phi$, and the light neutrino masses $m_{1,2,3}$ and the effective mass $m_{\beta\beta}$ in neutrinoless double beta decay.} \\
\hline
\multicolumn{11}{|c|}{Seesaw mechanism with/without gCP (NO) ($N^{c}\sim \mathbf{2}$)} \\ \hline
Model & $\sin^2\theta_{12}$ &$\sin^2\theta_{13}$ &$\sin^2\theta_{23}$&$\delta_{CP}/\pi$ & $\phi/\pi$ & $m_1$/meV & $m_2$/meV & $m_3$/meV & $m_{\beta\beta}$/meV & $\chi^{2}_{\text{min}}$ \\ \hline
\endfirsthead

\multicolumn{11}{c}%
   {{\bfseries \tablename\ \thetable{} -- continued from previous page}} \\
\hline
Model & $\sin^2\theta_{12}$ &$\sin^2\theta_{13}$ &$\sin^2\theta_{23}$&$\delta_{CP}/\pi$ & $\phi/\pi$ & $m_1$/meV & $m_2$/meV & $m_3$/meV & $m_{\beta\beta}$/meV & $\chi^{2}_{\text{min}}$ \\ \hline
\endhead

\hline \multicolumn{11}{|r|}{{continues on next page}} \\ \hline
\endfoot

\hline \hline
\endlastfoot

$\mathcal{C}_{20}-\mathcal{D}_{3}-\mathcal{N}_{5}$ & $0.280$ & $0.02216$ & $0.433$ & $1.838$ & $1.638$ & $0$ & $8.608$ & $49.881$ & $1.977$ & $14.337$\\ \hline$\mathcal{C}_{20}-\mathcal{D}_{3}-\mathcal{N}_{3}$ & $0.307$ & $0.02221$ & $0.507$ & $1.343$ & $0.964$ & $0$ & $8.608$ & $50.050$ & $3.242$ & $7.833$\\ \hline$\mathcal{C}_{20}-\mathcal{D}_{1}-\mathcal{N}_{4}$ & $0.300$ & $0.02219$ & $0.492$ & $1.400$ & $1.592$ & $0$ & $8.608$ & $50.150$ & $3.081$ & $4.669$\\ \hline$\mathcal{C}_{20}-\mathcal{D}_{1}-\mathcal{N}_{5}$ & $0.305$ & $0.02222$ & $0.475$ & $1.615$ & $0.878$ & $0$ & $8.608$ & $50.114$ & $3.639$ & $3.516$\\ \hline$\mathcal{C}_{20}-\mathcal{D}_{1}-\mathcal{N}_{1}$ & $0.308$ & $0.02223$ & $0.453$ & $1.093$ & $1.763$ & $0$ & $8.608$ & $50.057$ & $3.696$ & $2.000$\\ \hline$\mathcal{C}_{10}-\mathcal{D}_{3}-\mathcal{N}_{5}$ & $0.280$ & $0.02216$ & $0.433$ & $1.838$ & $1.638$ & $0$ & $8.608$ & $49.881$ & $1.977$ & $14.339$\\ \hline$\mathcal{C}_{10}-\mathcal{D}_{1}-\mathcal{N}_{4}$ & $0.300$ & $0.02219$ & $0.492$ & $1.399$ & $1.592$ & $0$ & $8.608$ & $50.150$ & $3.081$ & $4.674$\\ \hline$\mathcal{C}_{10}-\mathcal{D}_{1}-\mathcal{N}_{5}$ & $0.305$ & $0.02222$ & $0.475$ & $1.615$ & $0.878$ & $0$ & $8.608$ & $50.114$ & $3.640$ & $3.517$\\
\hline
\end{longtable}
\end{landscape}
\end{small}

\begin{small}
\begin{landscape}
\setlength\LTcapwidth{\textwidth}
\setlength\LTleft{1.1in}            
\setlength\LTright{0pt}           
 \begin{longtable}{|c|c|c|c|c|c|c|c|c|c|c|}
\caption{\label{tab:lepton_res_par7_SSN2_IO}The same as in Table~\ref{tab:lepton_res_par7_SSN2_NO} but for IO neutrino mass spectrum.} \\
\hline
\multicolumn{11}{|c|}{Seesaw mechanism with/without gCP (IO) ($N^{c}\sim \mathbf{2}$)} \\ \hline
Model & $\sin^2\theta_{12}$ &$\sin^2\theta_{13}$ &$\sin^2\theta_{23}$&$\delta_{CP}/\pi$ & $\phi/\pi$ & $m_1$/meV & $m_2$/meV & $m_3$/meV & $m_{\beta\beta}$/meV & $\chi^{2}_{\text{min}}$ \\ \hline
\endfirsthead

\multicolumn{11}{c}%
   {{\bfseries \tablename\ \thetable{} -- continued from previous page}} \\
\hline
Model & $\sin^2\theta_{12}$ &$\sin^2\theta_{13}$ &$\sin^2\theta_{23}$&$\delta_{CP}/\pi$ & $\phi/\pi$ & $m_1$/meV & $m_2$/meV & $m_3$/meV & $m_{\beta\beta}$/meV & $\chi^{2}_{\text{min}}$ \\ \hline
\endhead

\hline \multicolumn{11}{|r|}{{continues on next page}} \\ \hline
\endfoot

\hline \hline
\endlastfoot

$\mathcal{C}_{20}-\mathcal{D}_{4}-\mathcal{N}_{4}$ & $0.307$ & $0.02146$ & $0.469$ & $1.555$ & $0.241$ & $49.124$ & $49.872$ & $0$ & $45.391$ & $23.931$\\ \hline$\mathcal{C}_{20}-\mathcal{D}_{4}-\mathcal{N}_{5}$ & $0.307$ & $0.02146$ & $0.469$ & $1.471$ & $1.862$ & $49.122$ & $49.870$ & $0$ & $47.325$ & $23.864$\\ \hline$\mathcal{C}_{20}-\mathcal{D}_{4}-\mathcal{N}_{2}$ & $0.307$ & $0.02146$ & $0.469$ & $1.534$ & $0.209$ & $49.126$ & $49.874$ & $0$ & $46.089$ & $23.967$\\ \hline$\mathcal{C}_{20}-\mathcal{D}_{4}-\mathcal{N}_{1}$ & $0.307$ & $0.02146$ & $0.469$ & $1.505$ & $1.996$ & $49.124$ & $49.872$ & $0$ & $48.293$ & $23.857$\\ \hline$\mathcal{C}_{20}-\mathcal{D}_{3}-\mathcal{N}_{5}$ & $0.307$ & $0.02221$ & $0.499$ & $1.503$ & $0.070$ & $49.121$ & $49.870$ & $0$ & $48.008$ & $10.886$\\ \hline$\mathcal{C}_{20}-\mathcal{D}_{3}-\mathcal{N}_{3}$ & $0.307$ & $0.02222$ & $0.498$ & $1.446$ & $0.023$ & $49.127$ & $49.876$ & $0$ & $48.233$ & $11.215$\\ \hline$\mathcal{C}_{20}-\mathcal{D}_{3}-\mathcal{N}_{1}$ & $0.307$ & $0.02221$ & $0.499$ & $1.475$ & $1.972$ & $49.122$ & $49.871$ & $0$ & $48.215$ & $10.971$\\ \hline$\mathcal{C}_{10}-\mathcal{D}_{4}-\mathcal{N}_{4}$ & $0.307$ & $0.02146$ & $0.470$ & $1.555$ & $0.241$ & $49.124$ & $49.872$ & $0$ & $45.392$ & $23.812$\\
\hline

\end{longtable}
\end{landscape}
\end{small}

\begin{table}[ht!]
 \centering \resizebox{1.0\textwidth}{!}{
\begin{tabular}{|c|c|c|c|c|c|c|c|c|c|c|c|} \hline \hline
\multicolumn{11}{|c|}{Seesaw mechanism with/without gCP (NO) ($N^{c}\sim \mathbf{1}\oplus \mathbf{1}'$)} \\ \hline
  Model & $\sin^2\theta_{12}$ &$\sin^2\theta_{13}$ &$\sin^2\theta_{23}$&$\delta_{CP}/\pi$ & $\phi/\pi$ & $m_1$/meV & $m_2$/meV & $m_3$/meV & $m_{\beta\beta}$/meV & $\chi^{2}_{\text{min}}$\\ \hline
$\mathcal{C}_{9}-\mathfrak{D}_{8}-\mathfrak{N}_{13}$ & $0.313$ & $0.02225$ & $0.460$ & $1.889$ & $1.795$ & $0$ & $8.608$ & $50.045$ & $3.083$ & $7.991$\\ \hline$\mathcal{C}_{17}-\mathfrak{D}_{8}-\mathfrak{N}_{13}$ & $0.313$ & $0.02225$ & $0.460$ & $1.889$ & $1.795$ & $0$ & $8.608$ & $50.045$ & $3.082$ & $7.990$\\ \hline$\mathcal{C}_{8}-\mathfrak{D}_{8}-\mathfrak{N}_{13}$ & $0.313$ & $0.02225$ & $0.460$ & $1.889$ & $1.795$ & $0$ & $8.608$ & $50.045$ & $3.081$ & $7.985$\\ \hline$\mathcal{C}_{2}-\mathfrak{D}_{9}-\mathfrak{N}_{12}$ & $0.303$ & $0.02227$ & $0.455$ & $1.406$ & $0.827$ & $0$ & $8.608$ & $50.044$ & $3.186$ & $0.409$\\ \hline$\mathcal{C}_{2}-\mathfrak{D}_{9}-\mathfrak{N}_{11}$ & $0.301$ & $0.02229$ & $0.456$ & $1.404$ & $0.866$ & $0$ & $8.608$ & $50.042$ & $3.256$ & $0.552$\\ \hline$\mathcal{C}_{2}-\mathfrak{D}_{9}-\mathfrak{N}_{10}$ & $0.303$ & $0.02227$ & $0.455$ & $1.405$ & $0.831$ & $0$ & $8.608$ & $50.044$ & $3.195$ & $0.421$\\
\hline \hline \end{tabular}}
\caption{\label{tab:lepton_res_par7_SSN11_NO}The same as in Table~\ref{tab:lepton_res_par7_SSN2_NO} but for $N^{c}\sim \mathbf{1}\oplus \mathbf{1}'$.}
\end{table}

\begin{center}
\begin{small}
\begin{landscape}
\setlength\LTcapwidth{\textwidth}
\setlength\LTleft{1.2in}            
\setlength\LTright{0pt}           
 \begin{longtable}{|c|c|c|c|c|c|c|c|c|c|c|}
\caption{\label{tab:lepton_res_par7_SSN11_IO}The same as in Table~\ref{tab:lepton_res_par7_SSN11_NO} but for IO neutrino mass spectrum.} \\
\hline
\multicolumn{11}{|c|}{Seesaw mechanism with/without gCP (IO) ($N^{c}\sim \mathbf{1}\oplus \mathbf{1}'$)} \\ \hline
Model & $\sin^2\theta_{12}$ &$\sin^2\theta_{13}$ &$\sin^2\theta_{23}$&$\delta_{CP}/\pi$ & $\phi/\pi$ & $m_1$/meV & $m_2$/meV & $m_3$/meV & $m_{\beta\beta}$/meV & $\chi^{2}_{\text{min}}$ \\ \hline
\endfirsthead

\multicolumn{11}{c}%
   {{\bfseries \tablename\ \thetable{} -- continued from previous page}} \\
\hline
Model & $\sin^2\theta_{12}$ &$\sin^2\theta_{13}$ &$\sin^2\theta_{23}$&$\delta_{CP}/\pi$ & $\phi/\pi$ & $m_1$/meV & $m_2$/meV & $m_3$/meV & $m_{\beta\beta}$/meV & $\chi^{2}_{\text{min}}$ \\ \hline
\endhead

\hline \multicolumn{11}{|r|}{{continues on next page}} \\ \hline
\endfoot

\hline \hline
\endlastfoot

$\mathcal{C}_{6}-\mathcal{D}_{10}-\mathcal{N}_{8}$ & $0.307$ & $0.02160$ & $0.542$ & $1.525$ & $0.041$ & $49.120$ & $49.868$ & $0$ & $48.196$ & $2.707$\\ \hline$\mathcal{C}_{6}-\mathcal{D}_{10}-\mathcal{N}_{9}$ & $0.307$ & $0.02160$ & $0.542$ & $1.525$ & $0.041$ & $49.120$ & $49.868$ & $0$ & $48.196$ & $2.707$\\ \hline$\mathcal{C}_{6}-\mathcal{D}_{10}-\mathcal{N}_{7}$ & $0.307$ & $0.02160$ & $0.542$ & $1.525$ & $0.041$ & $49.120$ & $49.868$ & $0$ & $48.196$ & $2.707$\\ \hline$\mathcal{C}_{6}-\mathcal{D}_{10}-\mathcal{N}_{6}$ & $0.307$ & $0.02160$ & $0.542$ & $1.525$ & $0.041$ & $49.120$ & $49.868$ & $0$ & $48.196$ & $2.707$\\ \hline$\mathcal{C}_{6}-\mathcal{D}_{10}-\mathcal{N}_{5}$ & $0.307$ & $0.02160$ & $0.542$ & $1.525$ & $0.041$ & $49.120$ & $49.868$ & $0$ & $48.196$ & $2.707$\\ \hline$\mathcal{C}_{6}-\mathcal{D}_{8}-\mathcal{N}_{13}$ & $0.307$ & $0.02237$ & $0.566$ & $1.610$ & $1.957$ & $49.142$ & $49.890$ & $0$ & $48.171$ & $0.551$\\ \hline$\mathcal{C}_{13}-\mathcal{D}_{10}-\mathcal{N}_{8}$ & $0.307$ & $0.02161$ & $0.542$ & $1.525$ & $0.041$ & $49.120$ & $49.868$ & $0$ & $48.196$ & $2.645$\\ \hline$\mathcal{C}_{13}-\mathcal{D}_{10}-\mathcal{N}_{9}$ & $0.307$ & $0.02161$ & $0.542$ & $1.525$ & $0.041$ & $49.120$ & $49.868$ & $0$ & $48.196$ & $2.645$\\
\hline

\end{longtable}
\end{landscape}
\end{small}
\end{center}

\end{appendix}



\begin{thebibliography}{10}

\bibitem{ParticleDataGroup:2024cfk}
{\bfseries Particle Data Group} Collaboration, S.~Navas {\em et~al.}, ``{Review
  of particle physics},''
  \href{http://dx.doi.org/10.1103/PhysRevD.110.030001}{{\em Phys. Rev. D}
  {\bfseries 110} no.~3, (2024) 030001}.

\bibitem{King:2017guk}
S.~F. King, ``{Unified Models of Neutrinos, Flavour and CP Violation},''
  \href{http://dx.doi.org/10.1016/j.ppnp.2017.01.003}{{\em Prog. Part. Nucl.
  Phys.} {\bfseries 94} (2017) 217--256},
\href{http://arxiv.org/abs/1701.04413}{{\ttfamily arXiv:1701.04413 [hep-ph]}}.

\bibitem{Petcov:2017ggy}
S.~T. Petcov, ``{Discrete Flavour Symmetries, Neutrino Mixing and Leptonic CP
  Violation},'' \href{http://dx.doi.org/10.1140/epjc/s10052-018-6158-5}{{\em
  Eur. Phys. J. C} {\bfseries 78} no.~9, (2018) 709},
  \href{http://arxiv.org/abs/1711.10806}{{\ttfamily arXiv:1711.10806
  [hep-ph]}}.

\bibitem{Feruglio:2019ybq}
F.~Feruglio and A.~Romanino, ``{Lepton flavor symmetries},''
  \href{http://dx.doi.org/10.1103/RevModPhys.93.015007}{{\em Rev. Mod. Phys.}
  {\bfseries 93} no.~1, (2021) 015007},
  \href{http://arxiv.org/abs/1912.06028}{{\ttfamily arXiv:1912.06028
  [hep-ph]}}.

\bibitem{Xing:2020ijf}
Z.-z. Xing, ``{Flavor structures of charged fermions and massive neutrinos},''
  \href{http://dx.doi.org/10.1016/j.physrep.2020.02.001}{{\em Phys. Rept.}
  {\bfseries 854} (2020) 1--147},
  \href{http://arxiv.org/abs/1909.09610}{{\ttfamily arXiv:1909.09610
  [hep-ph]}}.

\bibitem{Ding:2024ozt}
G.-J. Ding and J.~W.~F. Valle, ``{The symmetry approach to quark and lepton
  masses and mixing},'' \href{http://arxiv.org/abs/2402.16963}{{\ttfamily
  arXiv:2402.16963 [hep-ph]}}.

\bibitem{Feruglio:2017spp}
F.~Feruglio, \href{http://dx.doi.org/10.1142/9789813238053_0012}{``{Are
  neutrino masses modular forms?},''} in {\em From My Vast Repertoire ...:
  Guido Altarelli's Legacy}, A.~Levy, S.~Forte, and G.~Ridolfi, eds.,
  pp.~227--266.
\newblock 2019.
\newblock
\href{http://arxiv.org/abs/1706.08749}{{\ttfamily arXiv:1706.08749 [hep-ph]}}.
\newblock

\bibitem{Kobayashi:2023zzc}
T.~Kobayashi and M.~Tanimoto, ``{Modular flavor symmetric models},''
\newblock 7, 2023.
\newblock \href{http://arxiv.org/abs/2307.03384}{{\ttfamily arXiv:2307.03384
  [hep-ph]}}.

\bibitem{Ding:2023htn}
G.-J. Ding and S.~F. King, ``{Neutrino mass and mixing with modular
  symmetry},'' \href{http://dx.doi.org/10.1088/1361-6633/ad52a3}{{\em Rept.
  Prog. Phys.} {\bfseries 87} no.~8, (2024) 084201},
  \href{http://arxiv.org/abs/2311.09282}{{\ttfamily arXiv:2311.09282
  [hep-ph]}}.

\bibitem{Ding:2020zxw}
G.-J. Ding, F.~Feruglio, and X.-G. Liu, ``{Automorphic Forms and Fermion
  Masses},'' \href{http://dx.doi.org/10.1007/JHEP01(2021)037}{{\em JHEP}
  {\bfseries 01} (2021) 037}, \href{http://arxiv.org/abs/2010.07952}{{\ttfamily
  arXiv:2010.07952 [hep-th]}}.

\bibitem{Qu:2024rns}
B.-Y. Qu and G.-J. Ding, ``{Non-holomorphic modular flavor symmetry},''
  \href{http://dx.doi.org/10.1007/JHEP08(2024)136}{{\em JHEP} {\bfseries 08}
  (2024) 136}, \href{http://arxiv.org/abs/2406.02527}{{\ttfamily
  arXiv:2406.02527 [hep-ph]}}.

\bibitem{Novichkov:2019sqv}
P.~P. Novichkov, J.~T. Penedo, S.~T. Petcov, and A.~V. Titov, ``{Generalised CP
  Symmetry in Modular-Invariant Models of Flavour},''
  \href{http://dx.doi.org/10.1007/JHEP07(2019)165}{{\em JHEP} {\bfseries 07}
  (2019) 165},
\href{http://arxiv.org/abs/1905.11970}{{\ttfamily arXiv:1905.11970 [hep-ph]}}.

\bibitem{Baur:2019kwi}
A.~Baur, H.~P. Nilles, A.~Trautner, and P.~K.~S. Vaudrevange, ``{Unification of
  Flavor, CP, and Modular Symmetries},''
  \href{http://dx.doi.org/10.1016/j.physletb.2019.03.066}{{\em Phys. Lett. B}
  {\bfseries 795} (2019) 7--14},
  \href{http://arxiv.org/abs/1901.03251}{{\ttfamily arXiv:1901.03251
  [hep-th]}}.

\bibitem{Nomura:2024atp}
T.~Nomura and H.~Okada, ``{Type-II seesaw of a non-holomorphic modular $A_4$
  symmetry},'' \href{http://arxiv.org/abs/2408.01143}{{\ttfamily
  arXiv:2408.01143 [hep-ph]}}.

\bibitem{Penedo:2018nmg}
J.~Penedo and S.~Petcov, ``{Lepton Masses and Mixing from Modular $S_4$
  Symmetry},'' \href{http://dx.doi.org/10.1016/j.nuclphysb.2018.12.016}{{\em
  Nucl. Phys. B} {\bfseries 939} (2019) 292--307},
  \href{http://arxiv.org/abs/1806.11040}{{\ttfamily arXiv:1806.11040
  [hep-ph]}}.

\bibitem{Novichkov:2018ovf}
P.~Novichkov, J.~Penedo, S.~Petcov, and A.~Titov, ``{Modular S$_{4}$ models of
  lepton masses and mixing},''
  \href{http://dx.doi.org/10.1007/JHEP04(2019)005}{{\em JHEP} {\bfseries 04}
  (2019) 005}, \href{http://arxiv.org/abs/1811.04933}{{\ttfamily
  arXiv:1811.04933 [hep-ph]}}.

\bibitem{deMedeirosVarzielas:2019cyj}
I.~de~Medeiros~Varzielas, S.~F. King, and Y.-L. Zhou, ``{Multiple modular
  symmetries as the origin of flavor},''
  \href{http://dx.doi.org/10.1103/PhysRevD.101.055033}{{\em Phys. Rev. D}
  {\bfseries 101} no.~5, (2020) 055033},
  \href{http://arxiv.org/abs/1906.02208}{{\ttfamily arXiv:1906.02208
  [hep-ph]}}.

\bibitem{King:2019vhv}
S.~F. King and Y.-L. Zhou, ``{Trimaximal TM$_1$ mixing with two modular $S_4$
  groups},'' \href{http://dx.doi.org/10.1103/PhysRevD.101.015001}{{\em Phys.
  Rev. D} {\bfseries 101} no.~1, (2020) 015001},
  \href{http://arxiv.org/abs/1908.02770}{{\ttfamily arXiv:1908.02770
  [hep-ph]}}.

\bibitem{Criado:2019tzk}
J.~C. Criado, F.~Feruglio, and S.~J. King, ``{Modular Invariant Models of
  Lepton Masses at Levels 4 and 5},''
  \href{http://dx.doi.org/10.1007/JHEP02(2020)001}{{\em JHEP} {\bfseries 02}
  (2020) 001}, \href{http://arxiv.org/abs/1908.11867}{{\ttfamily
  arXiv:1908.11867 [hep-ph]}}.

\bibitem{Ding:2019gof}
G.-J. Ding, S.~F. King, X.-G. Liu, and J.-N. Lu, ``{Modular S$_{4}$ and A$_{4}$
  symmetries and their fixed points: new predictive examples of lepton
  mixing},'' \href{http://dx.doi.org/10.1007/JHEP12(2019)030}{{\em JHEP}
  {\bfseries 12} (2019) 030}, \href{http://arxiv.org/abs/1910.03460}{{\ttfamily
  arXiv:1910.03460 [hep-ph]}}.

\bibitem{Wang:2019ovr}
X.~Wang and S.~Zhou, ``{The minimal seesaw model with a modular S$_{4}$
  symmetry},'' \href{http://dx.doi.org/10.1007/JHEP05(2020)017}{{\em JHEP}
  {\bfseries 05} (2020) 017}, \href{http://arxiv.org/abs/1910.09473}{{\ttfamily
  arXiv:1910.09473 [hep-ph]}}.

\bibitem{Zhao:2021jxg}
Y.~Zhao and H.-H. Zhang, ``{Adjoint SU(5) GUT model with modular $S_{4}$
  symmetry},'' \href{http://dx.doi.org/10.1007/JHEP03(2021)002}{{\em JHEP}
  {\bfseries 03} (2021) 002}, \href{http://arxiv.org/abs/2101.02266}{{\ttfamily
  arXiv:2101.02266 [hep-ph]}}.

\bibitem{King:2021fhl}
S.~F. King and Y.-L. Zhou, ``{Twin modular S$_{4}$ with SU(5) GUT},''
  \href{http://dx.doi.org/10.1007/JHEP04(2021)291}{{\em JHEP} {\bfseries 04}
  (2021) 291}, \href{http://arxiv.org/abs/2103.02633}{{\ttfamily
  arXiv:2103.02633 [hep-ph]}}.

\bibitem{Ding:2021zbg}
G.-J. Ding, S.~F. King, and C.-Y. Yao, ``{Modular $S_4\times SU(5)$ GUT},''
  \href{http://dx.doi.org/10.1103/PhysRevD.104.055034}{{\em Phys. Rev. D}
  {\bfseries 104} no.~5, (2021) 055034},
  \href{http://arxiv.org/abs/2103.16311}{{\ttfamily arXiv:2103.16311
  [hep-ph]}}.

\bibitem{Qu:2021jdy}
B.-Y. Qu, X.-G. Liu, P.-T. Chen, and G.-J. Ding, ``{Flavor mixing and CP
  violation from the interplay of an S4 modular group and a generalized CP
  symmetry},'' \href{http://dx.doi.org/10.1103/PhysRevD.104.076001}{{\em Phys.
  Rev. D} {\bfseries 104} no.~7, (2021) 076001},
  \href{http://arxiv.org/abs/2106.11659}{{\ttfamily arXiv:2106.11659
  [hep-ph]}}.

\bibitem{Nomura:2021ewm}
T.~Nomura and H.~Okada, ``{Linear seesaw model with a modular $S_4$ flavor
  symmetry},'' \href{http://arxiv.org/abs/2109.04157}{{\ttfamily
  arXiv:2109.04157 [hep-ph]}}.

\bibitem{deMedeirosVarzielas:2023ujt}
I.~de~Medeiros~Varzielas, S.~F. King, and M.~Levy, ``{A modular SU (5) littlest
  seesaw},'' \href{http://dx.doi.org/10.1007/JHEP05(2024)203}{{\em JHEP}
  {\bfseries 05} (2024) 203}, \href{http://arxiv.org/abs/2309.15901}{{\ttfamily
  arXiv:2309.15901 [hep-ph]}}.

\bibitem{deAdelhartToorop:2011re}
R.~de~Adelhart~Toorop, F.~Feruglio, and C.~Hagedorn, ``{Finite Modular Groups
  and Lepton Mixing},''
  \href{http://dx.doi.org/10.1016/j.nuclphysb.2012.01.017}{{\em Nucl. Phys.}
  {\bfseries B858} (2012) 437--467},
\href{http://arxiv.org/abs/1112.1340}{{\ttfamily arXiv:1112.1340 [hep-ph]}}.

\bibitem{Li:2021buv}
C.-C. Li, X.-G. Liu, and G.-J. Ding, ``{Modular symmetry at level 6 and a new
  route towards finite modular groups},''
  \href{http://dx.doi.org/10.1007/JHEP10(2021)238}{{\em JHEP} {\bfseries 10}
  (2021) 238}, \href{http://arxiv.org/abs/2108.02181}{{\ttfamily
  arXiv:2108.02181 [hep-ph]}}.

\bibitem{Ding:2020msi}
G.-J. Ding, S.~F. King, C.-C. Li, and Y.-L. Zhou, ``{Modular Invariant Models
  of Leptons at Level 7},''
  \href{http://dx.doi.org/10.1007/JHEP08(2020)164}{{\em JHEP} {\bfseries 08}
  (2020) 164}, \href{http://arxiv.org/abs/2004.12662}{{\ttfamily
  arXiv:2004.12662 [hep-ph]}}.

\bibitem{Bilenky:1980cx}
S.~M. Bilenky, J.~Hosek, and S.~T. Petcov, ``{On Oscillations of Neutrinos with
  Dirac and Majorana Masses},''
  \href{http://dx.doi.org/10.1016/0370-2693(80)90927-2}{{\em Phys. Lett. B}
  {\bfseries 94} (1980) 495--498}.

\bibitem{Krastev:1988yu}
P.~I. Krastev and S.~T. Petcov, ``{Resonance Amplification and t Violation
  Effects in Three Neutrino Oscillations in the Earth},''
  \href{http://dx.doi.org/10.1016/0370-2693(88)90404-2}{{\em Phys. Lett. B}
  {\bfseries 205} (1988) 84--92}.

\bibitem{Jarlskog:1985ht}
C.~Jarlskog, ``{Commutator of the Quark Mass Matrices in the Standard
  Electroweak Model and a Measure of Maximal CP Nonconservation},''
  \href{http://dx.doi.org/10.1103/PhysRevLett.55.1039}{{\em Phys. Rev. Lett.}
  {\bfseries 55} (1985) 1039}.

\bibitem{Pascoli:2005zb}
S.~Pascoli, S.~T. Petcov, and T.~Schwetz, ``{The Absolute neutrino mass scale,
  neutrino mass spectrum, majorana CP-violation and neutrinoless double-beta
  decay},'' \href{http://dx.doi.org/10.1016/j.nuclphysb.2005.11.003}{{\em Nucl.
  Phys. B} {\bfseries 734} (2006) 24--49},
  \href{http://arxiv.org/abs/hep-ph/0505226}{{\ttfamily arXiv:hep-ph/0505226}}.

\bibitem{Xing:2007fb}
Z.-z. Xing, H.~Zhang, and S.~Zhou, ``{Updated Values of Running Quark and
  Lepton Masses},'' \href{http://dx.doi.org/10.1103/PhysRevD.77.113016}{{\em
  Phys. Rev. D} {\bfseries 77} (2008) 113016},
  \href{http://arxiv.org/abs/0712.1419}{{\ttfamily arXiv:0712.1419 [hep-ph]}}.

\bibitem{Esteban:2020cvm}
I.~Esteban, M.~Gonzalez-Garcia, M.~Maltoni, T.~Schwetz, and A.~Zhou, ``{The
  fate of hints: updated global analysis of three-flavor neutrino
  oscillations},'' \href{http://dx.doi.org/10.1007/JHEP09(2020)178}{{\em JHEP}
  {\bfseries 09} (2020) 178}, \href{http://arxiv.org/abs/2007.14792}{{\ttfamily
  arXiv:2007.14792 [hep-ph]}}.

\bibitem{minuit}
\url{https://seal.web.cern.ch/seal/snapshot/work-packages/mathlibs/minuit/}.

\bibitem{Planck:2018vyg}
{\bfseries Planck} Collaboration, N.~Aghanim {\em et~al.}, ``{Planck 2018
  results. VI. Cosmological parameters},''
  \href{http://dx.doi.org/10.1051/0004-6361/201833910}{{\em Astron. Astrophys.}
  {\bfseries 641} (2020) A6}, \href{http://arxiv.org/abs/1807.06209}{{\ttfamily
  arXiv:1807.06209 [astro-ph.CO]}}. [Erratum: Astron.Astrophys. 652, C4
  (2021)].

\bibitem{Feroz:2007kg}
F.~Feroz and M.~P. Hobson, ``{Multimodal nested sampling: an efficient and
  robust alternative to MCMC methods for astronomical data analysis},''
  \href{http://dx.doi.org/10.1111/j.1365-2966.2007.12353.x}{{\em Mon. Not. Roy.
  Astron. Soc.} {\bfseries 384} (2008) 449},
\href{http://arxiv.org/abs/0704.3704}{{\ttfamily arXiv:0704.3704 [astro-ph]}}.

\bibitem{Feroz:2008xx}
F.~Feroz, M.~P. Hobson, and M.~Bridges, ``{MultiNest: an efficient and robust
  Bayesian inference tool for cosmology and particle physics},''
  \href{http://dx.doi.org/10.1111/j.1365-2966.2009.14548.x}{{\em Mon. Not. Roy.
  Astron. Soc.} {\bfseries 398} (2009) 1601--1614},
\href{http://arxiv.org/abs/0809.3437}{{\ttfamily arXiv:0809.3437 [astro-ph]}}.

\bibitem{KamLAND-Zen:2024eml}
{\bfseries KamLAND-Zen} Collaboration, S.~Abe {\em et~al.}, ``{Search for
  Majorana Neutrinos with the Complete KamLAND-Zen Dataset},''
  \href{http://arxiv.org/abs/2406.11438}{{\ttfamily arXiv:2406.11438
  [hep-ex]}}.

\bibitem{LEGEND:2021bnm}
{\bfseries LEGEND} Collaboration, N.~Abgrall {\em et~al.}, ``{The Large
  Enriched Germanium Experiment for Neutrinoless $\beta\beta$ Decay}:
  {LEGEND-1000 Preconceptual Design Report},''
  \href{http://arxiv.org/abs/2107.11462}{{\ttfamily arXiv:2107.11462
  [physics.ins-det]}}.

\bibitem{nEXO:2021ujk}
{\bfseries nEXO} Collaboration, G.~Adhikari {\em et~al.}, ``{nEXO: neutrinoless
  double beta decay search beyond 10$^{28}$ year half-life sensitivity},''
  \href{http://dx.doi.org/10.1088/1361-6471/ac3631}{{\em J. Phys. G} {\bfseries
  49} no.~1, (2022) 015104}, \href{http://arxiv.org/abs/2106.16243}{{\ttfamily
  arXiv:2106.16243 [nucl-ex]}}.

\bibitem{bbonuReview:Neutrino2024}
R.~Guenette, {\em {Other present and future 0vDBD experiments, talk given at
  the XXXI International Conference on Neutrino Physics and Astrophysics, June
  17 - 22, 2024, Milano, Italy.}}
\newblock
  \url{https://agenda.infn.it/event/37867/contributions/233915/attachments/121855/177755/Guenette_0nbb_NEUTRINO_2024.pdf}.

\bibitem{Katrin:2024tvg}
{\bfseries Katrin} Collaboration, M.~Aker {\em et~al.}, ``{Direct neutrino-mass
  measurement based on 259 days of KATRIN data},''
  \href{http://arxiv.org/abs/2406.13516}{{\ttfamily arXiv:2406.13516
  [nucl-ex]}}.

\bibitem{KATRIN:2021dfa}
{\bfseries KATRIN} Collaboration, M.~Aker {\em et~al.}, ``{The design,
  construction, and commissioning of the KATRIN experiment},''
  \href{http://dx.doi.org/10.1088/1748-0221/16/08/T08015}{{\em JINST}
  {\bfseries 16} no.~08, (2021) T08015},
  \href{http://arxiv.org/abs/2103.04755}{{\ttfamily arXiv:2103.04755
  [physics.ins-det]}}.

\bibitem{Project8:2022wqh}
{\bfseries Project 8} Collaboration, A.~A. Esfahani {\em et~al.}, ``{The
  Project 8 Neutrino Mass Experiment},'' in {\em {Snowmass 2021}}.
\newblock 3, 2022.
\newblock \href{http://arxiv.org/abs/2203.07349}{{\ttfamily arXiv:2203.07349
  [nucl-ex]}}.

\bibitem{Hyper-Kamiokande:2018ofw}
{\bfseries Hyper-Kamiokande} Collaboration, K.~Abe {\em et~al.},
  ``{Hyper-Kamiokande Design Report},''
  \href{http://arxiv.org/abs/1805.04163}{{\ttfamily arXiv:1805.04163
  [physics.ins-det]}}.

\bibitem{DUNE:2020ypp}
{\bfseries DUNE} Collaboration, B.~Abi {\em et~al.}, ``{Deep Underground
  Neutrino Experiment (DUNE), Far Detector Technical Design Report, Volume II:
  DUNE Physics},'' \href{http://arxiv.org/abs/2002.03005}{{\ttfamily
  arXiv:2002.03005 [hep-ex]}}.

\bibitem{Alekou:2022emd}
A.~Alekou {\em et~al.}, ``{The European Spallation Source neutrino super-beam
  conceptual design report},''
  \href{http://dx.doi.org/10.1140/epjs/s11734-022-00664-w}{{\em Eur. Phys. J.
  ST} {\bfseries 231} no.~21, (2022) 3779--3955},
  \href{http://arxiv.org/abs/2206.01208}{{\ttfamily arXiv:2206.01208
  [hep-ex]}}. [Erratum: Eur.Phys.J.ST 232, 15--16 (2023)].

\bibitem{T2K:2023smv}
{\bfseries T2K} Collaboration, K.~Abe {\em et~al.}, ``{Measurements of neutrino
  oscillation parameters from the T2K experiment using $3.6\times 10^{21}$
  protons on target},''
  \href{http://dx.doi.org/10.1140/epjc/s10052-023-11819-x}{{\em Eur. Phys. J.
  C} {\bfseries 83} no.~9, (2023) 782},
  \href{http://arxiv.org/abs/2303.03222}{{\ttfamily arXiv:2303.03222
  [hep-ex]}}.

\bibitem{NOvA:2021nfi}
{\bfseries NOvA} Collaboration, M.~A. Acero {\em et~al.}, ``{Improved
  measurement of neutrino oscillation parameters by the NOvA experiment},''
  \href{http://dx.doi.org/10.1103/PhysRevD.106.032004}{{\em Phys. Rev. D}
  {\bfseries 106} no.~3, (2022) 032004},
  \href{http://arxiv.org/abs/2108.08219}{{\ttfamily arXiv:2108.08219
  [hep-ex]}}.

\bibitem{JUNO:2022mxj}
{\bfseries JUNO} Collaboration, A.~Abusleme {\em et~al.}, ``{Sub-percent
  precision measurement of neutrino oscillation parameters with JUNO},''
  \href{http://dx.doi.org/10.1088/1674-1137/ac8bc9}{{\em Chin. Phys. C}
  {\bfseries 46} no.~12, (2022) 123001},
  \href{http://arxiv.org/abs/2204.13249}{{\ttfamily arXiv:2204.13249
  [hep-ex]}}.

\bibitem{Novichkov:2020eep}
P.~P. Novichkov, J.~T. Penedo, and S.~T. Petcov, ``{Double cover of modular
  $S_4$ for flavour model building},''
  \href{http://dx.doi.org/10.1016/j.nuclphysb.2020.115301}{{\em Nucl. Phys. B}
  {\bfseries 963} (2021) 115301},
  \href{http://arxiv.org/abs/2006.03058}{{\ttfamily arXiv:2006.03058
  [hep-ph]}}.

\bibitem{Liu:2020akv}
X.-G. Liu, C.-Y. Yao, and G.-J. Ding, ``{Modular invariant quark and lepton
  models in double covering of $S_4$ modular group},''
  \href{http://dx.doi.org/10.1103/PhysRevD.103.056013}{{\em Phys. Rev. D}
  {\bfseries 103} no.~5, (2021) 056013},
  \href{http://arxiv.org/abs/2006.10722}{{\ttfamily arXiv:2006.10722
  [hep-ph]}}.

\bibitem{Liu:2020msy}
X.-G. Liu, C.-Y. Yao, B.-Y. Qu, and G.-J. Ding, ``{Half-integral weight modular
  forms and application to neutrino mass models},''
  \href{http://dx.doi.org/10.1103/PhysRevD.102.115035}{{\em Phys. Rev. D}
  {\bfseries 102} no.~11, (2020) 115035},
  \href{http://arxiv.org/abs/2007.13706}{{\ttfamily arXiv:2007.13706
  [hep-ph]}}.

\end{thebibliography}


\providecommand{\href}[2]{#2}\begingroup\raggedright\endgroup

\end{document}